\documentclass[a4paper,UKenglish,cleveref,autoref,thm-restate]{lipics-v2021}

\usepackage{amsmath,nccmath}
\usepackage{amsfonts,amssymb}
\usepackage{bussproofs}
\usepackage{tikz-cd}
\usepackage[all]{xy} \SelectTips{cm}{}  %
\usepackage{color, ifthen}
\usepackage{comment}
\usepackage{mathtools}
\usepackage{cite}

\DeclareMathOperator*{\colim}{colim}
\newcommand{\clat}[0]{\mathbf{CLat}_{\wedge}}
\newcommand{\Pf}[0]{\mathcal{P} }
\newcommand{\dotP}[0]{\dot{\mathcal{P}} }
\newcommand{\T}[0]{\mathcal{T} }
\newcommand{\dotT}[0]{\dot{\mathcal{T}} }
\newcommand{\Coalg}[1]{\mathbf{Coalg}(#1)}
\newcommand{\Alg}[1]{\mathbf{Alg}(#1)}
\newcommand{\Kl}[1]{\mathbf{Kl}(#1)}
\newcommand{\relto}{-{\kern-1.5ex}\raisebox{-.1pt}{\mbox{$\shortmid$}}{\kern-2ex}\to}
\newcommand{\longrelto}{-{\kern-0ex}\raisebox{-.1pt}{\mbox{$\shortmid$}}{\kern-2.2ex}\longrightarrow}
\newcommand{\asdf}{\alpha^*\dot{F}}
\newcommand{\cost}{IF/I coincidence}
\newcommand{\CoSt}{IF/I Coincidence}
\newcommand{\congrightarrow}{\mathrel{\stackrel{
           \raisebox{.5ex}{$\scriptstyle\cong\,$}}{
           \raisebox{0ex}[0ex][0ex]{$\rightarrow$}}}}
\newcommand{\iso}{\congrightarrow}
\newcommand{\tr}{\Sigma}
\newcommand{\ptr}{\mathrm{ptr}}
\newcommand{\Set}{\mathbf{Set}}
\newcommand{\ok}{\checkmark}
\newcommand{\notok}{?}

\newcommand{\mT}{\mathcal{T}}
\newcommand{\dT}{\dot{\mathcal{T}}}

\newcommand{\myparagraph}[1]{\noindent \textbf{\sffamily  #1}\quad}

\newcommand{\rloop}[2][-]{\save \POS!R(.7) \ar@(ru,rd)^#1{#2} \restore}
\newcommand{\lloop}[2][-]{\save \POS!L(.7) \ar@(lu,ld)_#1{#2} \restore}

\specialcomment{auxproof}
{\mbox{}\newline\textbf{BEGIN: AUX-PROOF}\dotfill\newline}
{\mbox{}\newline\textbf{END: AUX-PROOF}\dotfill\newline}

\newif\ifdraft\draftfalse

\ifdraft
\usepackage{todonotes}

\newcommand{\conf}[1]{}
\excludecomment{auxproof}
\else
\usepackage[disable]{todonotes}

\newcommand{\conf}[1]{}
\excludecomment{auxproof}
\fi

\theoremstyle{plain}
\newtheorem{mylemma}{Lemma}[section]
\newtheorem{myproposition}[mylemma]{Proposition}
\newtheorem{mytheorem}[mylemma]{Theorem}
\newtheorem{mycorollary}[mylemma]{Corollary}

\theoremstyle{definition}
\newtheorem{mydefinition}[mylemma]{Definition}
\newtheorem{myremark}[mylemma]{Remark}

\newtheorem{myexample}[mylemma]{Example}

\newtheorem{mynotation}[mylemma]{Notation}

\theoremstyle{remark}
\newtheorem*{myproof}{Proof}
\def\myqed{\qed}

\usepackage{wrapfig}
\usepackage{cutwin, lipsum}
\opencutright

\nolinenumbers

\title{Fibrational Initial Algebra-Final Coalgebra Coincidence over Initial Algebras:\\ Turning Verification Witnesses Upside Down}

\titlerunning{Fibrational Initial Algebra-Final Coalgebra Coincidence over Initial Algebras} %

\author{Mayuko Kori}{The Graduate University for Advanced Studies (SOKENDAI), Hayama, Japan \and National Institute of Informatics, Tokyo, Japan}{mkori@nii.ac.jp}{https://orcid.org/0000-0002-8495-5925}{}%

\author{Ichiro Hasuo}{The Graduate University for Advanced Studies (SOKENDAI), Hayama, Japan \and National Institute of Informatics, Tokyo, Japan}{hasuo@nii.ac.jp}{https://orcid.org/0000-0002-8300-4650}{}

\author{Shin-ya Katsumata}{National Institute of Informatics, Tokyo, Japan}{s-katsumata@nii.ac.jp}{https://orcid.org/0000-0001-7529-5489}{}

\authorrunning{M. Kori, I. Hasuo and S. Katsumata} %

\Copyright{Mayuko Kori, Ichiro Hasuo and Shin-ya Katsumata} %



\category{} %

\EventEditors{John Q. Open and Joan R. Access}
\EventNoEds{2}
\EventLongTitle{42nd Conference on Very Important Topics (CVIT 2016)}
\EventShortTitle{CVIT 2016}
\EventAcronym{CVIT}
\EventYear{2016}
\EventDate{December 24--27, 2016}
\EventLocation{Little Whinging, United Kingdom}
\EventLogo{}
\SeriesVolume{42}
\ArticleNo{23}

\begin{document}

\maketitle

\begin{abstract}
  The coincidence between initial algebras (IAs) and final coalgebras (FCs) is a phenomenon that  underpins various important results in theoretical computer science. In this paper, we identify a general fibrational condition for the IA-FC coincidence, namely in the fiber over an initial algebra in the base category. Identifying (co)algebras  in a fiber as (co)inductive predicates, our fibrational IA-FC coincidence allows one to use \emph{coinductive} witnesses (such as invariants) for  verifying \emph{inductive} properties (such as liveness). Our general fibrational theory features the technical condition of stability of chain colimits; we  extend the framework to the presence of a monadic effect, too, restricting to fibrations of complete lattice-valued predicates. Practical benefits of our categorical theory are exemplified by  new ``upside-down'' witness notions for three verification problems: probabilistic liveness, and acceptance and model-checking with respect to bottom-up tree automata.
\end{abstract}

\section{Introduction\conf{50}}
\myparagraph{Categorical Algebras and Coalgebras}
Categorical algebras and coalgebras are omnipresent in theoretical computer science. For a category $\mathbb{C}$ and an endofunctor $F\colon\mathbb{C}\to\mathbb{C}$, an $F$-algebra is a $\mathbb{C}$-morphism $a\colon FX\to X$, while an $F$-coalgebra is $c\colon X\to FX$. These structures occur in many different settings with different $\mathbb{C}$ and $F$; the identification of such (co)algebras has
yielded a number of concrete benefits, such as rigorous system/program semantics, verification methods, and programming language constructs.

One principal use of categorical (co)algebras is as models of \emph{data structures} such as terms and state-based systems.
Examples
 include modeling of inductive datatypes by initial algebras~\cite{BirdM97}, and the theory of coalgebras~\cite{Rutten00a,Jacobs16coalgBook} that captures state-based behaviors.
Here, the base category $\mathbb{C}$ is typically that of (structured) sets and (structure-preserving) maps.
 (In this paper, such a category will constitute a \emph{base category} of a fibration).

Another principal use of (co)algebras is as logical \emph{recursive specifications}. Here the base category $\mathbb{C}$ is typically a complete lattice of truth values (such as $\mathbf{2}=\{\bot, \top\}$) and the functor $F\colon \mathbb{C}\to\mathbb{C}$ is identified with a monotone function. Liveness properties are modeled by least fixed points (lfp's); safety properties are greatest ones (gfp's); and by the classic Knaster--Tarski theorem, these are identified with initial algebras and final coalgebras, respectively.
(In this paper, such a category will appear as a \emph{fibre category} of a fibration).

\myparagraph{Initial Algebras and Final Coalgebras}
In the above variety of occurrences of (co)algebras, \emph{initial algebras} and \emph{final coalgebras} play key roles. Their definition is by suitable universality: $\beta\colon F(\mu F)\to \mu F$ is initial if there is a unique algebra morphism from $\beta$ to an arbitrary algebra $a\colon FX\to X$; and dually for final coalgebras.
\begin{displaymath}
  \vcenter{\xymatrix@R=.6em{
  {F(\mu F)}
   \ar[d]^-{\cong}_-{\text{init.}}
   \ar@{.>}[r]
  &
  FX \ar[d]
  \\
  \mu F \ar@{.>}[r]
  &
  X
  }}
\;
  \vcenter{\xymatrix@R=.6em{
     FX \ar@{.>}[r]
  &
   {F(\nu F)}
  \\
  X \ar@{.>}[r]
  \ar[u]
  &
  \nu F
   \ar[u]^-{\cong}_-{\text{final}}
  }}
\end{displaymath}
Their (co)algebra structures are isomorphisms by the Lambek lemma. The latter extends the Knaster--Tarski theorem from lattices to categories.

In many occurrences of (co)algebras in computer science, initial algebras represent \emph{finitary} entities while final coalgebras represent \emph{infinitary} entities. For example, when $\mathbb{C}=\Set$ (the category of sets and functions) and $F$ is a functor that models a datatype constructor, the carrier $\mu F$ of an initial algebra represents the \emph{inductive datatype}---collecting all finite trees ``of shape $F$''---while the carrier $\nu F$ of a final coalgebra is for the \emph{coinductive datatype} and collects all (finite and infinite) trees. This intuition is found also in the logical (co)algebras: liveness properties (initial algebras) can be witnessed within finitely many steps, while safety properties (final coalgebras) are verified only after infinitely many steps.

\vspace{.3em}

\myparagraph{Initial Algebra-Final Coalgebra Coincidence}
In this paper, we are interested in the coincidence of an initial algebra and a final coalgebra (the \emph{IA-FC coincidence}). While it may sound unlikely in view of the contrast between finitary and infinitary, the coincidence has been found in different areas in computer science, underpinning  fundamental results.

One  example is in \emph{domain theory}:  cpo-enrichment yields the IA-FC coincidence, which is used to  solve recursive domain equations of mixed variance~\cite{SmythP82,Freyd90,Fiore96b,Zamdzhiev19}. Another example is in \emph{process semantics}: specifically, in the coalgebraic characterization of finite trace semantics~\cite{HasuoJS07b}, the IA-FC coincidence in some Kleisli categories $\Kl{T}$ has been observed.

\vspace{.3em}

\myparagraph{Contribution: the Fibrational IF/I Coincidence and Application to Verification Witnesses}
In this paper, we identify a general \emph{fibrational} condition for the IA-FC coincidence: under mild assumptions, we have the IA-FC coincidence
 in the fiber over an initial algebra in the base category (the \emph{\cost{}}). Identifying the base IA as a datatype, and the fibre IA/FC as lfp/gfp specifications, the \cost{} implies the coincidence between \emph{induction} and \emph{coinduction} as reasoning principles, assuming they are over a (finitary) algebraic datatype.

This coincidence allows us to \emph{turn witness notions upside down}, that is, to use coinductive witness notions for establishing inductive properties. In general, inductive witness notions for lfp properties (such as ranking functions) tend to be more complex than coinductive witness notions for gfp properties (such as invariants). When we have the \cost{}, the latter can now be used for lfp properties.

Our technical contributions are as follows. We work with a  \emph{fibration} $p\colon\mathbb{E}\to\mathbb{B}$, where $\mathbb{B}$ is intuitively a category of sets and functions, and $\mathbb{E}$ equips these sets with predicates.
\begin{itemize}
  \item We identify a general fibrational framework for what we call the \emph{\cost{}}---the coincidence of IAs (lfp predicates) and FCs (gfp predicates) in the fiber over an initial algebra in $\mathbb{B}$ (an inductive datatype).
The \cost{} relies only on mild fibrational assumptions,
notable among which are \emph{fibredness} of functors and \emph{stability} of certain colimits.
Although we restrict fibrations to posetal ones in the main text (\S{}\ref{sec:ip_for_clat}),
a similar result for general fibrations can be shown (Appendix~\ref{subsec:ip}).
  \item
    As a notable class of examples, in \S{}\ref{sec:general}
    we show that the fibration of  $\Omega$-valued predicates exhibits the \cost{}  (where $\Omega$ is an arbitrary  complete lattice for truth values).

   Furthermore, we study the \cost{} in the presence of monadic effects~\cite{Moggi91a}, building on the fibrational framework from~\cite{AguirreK20}.
  \item
  These theoretical results are used to obtain coinductive (invariant-like) witness notions for inductive (lfp, liveness) properties. Specifically, we present new witness notions for probabilistic verification (\S{}\ref{sec:live}) and verification with tree automata as specifications (\S{}\ref{sec:treeUnified}).
\end{itemize}

\vspace{.3em}

\myparagraph{Related Work}
Many works are discussed in the technical sections;  we discuss some others.

The work~\cite{Pitts96} shows uniqueness of fixed points above what is called a \emph{minimal invariant};
the latter corresponds to the lifting of a morphism which is both an initial algebra and a final coalgebra.
Our \cost{} can yield such lifting under some assumptions (see Thm.~\ref{thm:cost_lift}).
The proof in~\cite{Pitts96} relies on homset enrichment,  unlike our fibrational framework.

One of our main ideas is to use the IA-FC coincidence for novel proof methods for recursive specifications (\S{}\ref{sec:live}--\ref{sec:treeUnified}), mixing lfp's and gfp's. This is pursued also in~\cite{UrabeHH17, CaprettaUV09} where \emph{corecursive algebras} induce the lfp-gfp coincidence.

\vspace{.3em}

\myparagraph{Organization}
After recalling fibrations and the chain construction of initial algebras in~\S{}\ref{sec:pre},
we formulate our
\cost{} in \S{}\ref{sec:ip}
and present sufficient conditions for the coincidence in \S{}\ref{sec:ip_for_clat}.
In~\S{}\ref{sec:general}, these results are specialized to fibrations of $\Omega$-valued predicates, where we additionally include monadic effects. This paves the way to the concrete applications in~\S{}\ref{sec:live}--\ref{sec:treeUnified}, where we present seemingly new verification techniques for probabilistic liveness and witnesses of tree automata. We defer many proofs  to the appendix.

\section{Preliminaries\conf{70}} \label{sec:pre}
\subsection{Fibrations}\label{subsec:fibPrelim}
A fibration $p: \mathbb{E} \to \mathbb{B}$ is a functor that models indexing and substitution.
That is,
a functor $p: \mathbb{E} \to \mathbb{B}$ can be seen as a family of categories $(\mathbb{E}_X)_{X \in \mathbb{B}}$ that is equipped with substitution functors that change the index $X$.

In our examples, the base category $\mathbb{B}$ is that of sets and (potentially effectful) functions; and the total category $\mathbb{E}$ models ``predicates'' over sets in $\mathbb{B}$.
We review a minimal set of definitions and results on fibrations. See~\cite{CLTT} for  details.

\begin{mydefinition}[fibre, fibration]
  Let $p: \mathbb{E} \to \mathbb{B}$ be a functor.

      For each $X\in\mathbb{B}$, the \emph{fibre category} (or simply \emph{fibre}) $\mathbb{E}_X$ over $X$ is the category with
    objects $P \in \mathbb{E}$ such that $pP = X$ and morphisms $f: P \to Q$ such that $pf=\mathrm{id}_{X}$.
    An object $P \in \mathbb{E}_X$ is said to be \emph{above} $X$
    and a morphism $f \in \mathbb{E}_X$ is said to be \emph{vertical}.

    A morphism $f: P \to Q$ in $\mathbb{E}$ is \emph{cartesian}
      if it satisfies the following universality:
      for each $g: R \to Q$ in $\mathbb{E}$ and $k: pR \to pP$ in $\mathbb{B}$ with $pg = pf \circ k$,
      there exists a unique morphism $h: R \to P$ satisfying $g = f \circ h$ and $ph = k$. See the diagram below.
    \begin{displaymath}
      \vcenter{\xymatrix@R=.1em@C-1.7em{
          \mathbb{E} \ar[ddd]_-{p}
          & R
          \ar@/^/[rrd]^-{g}
          \ar@{.>}[rd]_-{h}
          \\
          &
          &
          P \ar[r]_-{f}
          &
          Q
          &
          l^{*}Q
          \ar[r]^-{\overline{l}}
          &
          Q
          \\
          & pR
          \ar@/^/[rrd]^-{pg}
          \ar@{->}[rd]_-{k}
          \\
          \mathbb{B}
          &
          &
          pP \ar[r]_-{pf}
          &
          pQ
          &
          X
          \ar[r]_-{l}
          &
          pQ
      }}
    \end{displaymath}

    The functor   $p: \mathbb{E} \to \mathbb{B}$ is a \emph{fibration}
      if, for each $Q \in \mathbb{E}$ and each $l: X \to pQ$ in $\mathbb{B}$,
      there exists $l^*Q\in \mathbb{E}$ and a morphism $\overline{l}: l^*Q \to Q$ such that $p\overline{l}=l$ and $\overline{l}$ is cartesian.

    The functor
 $p: \mathbb{E} \to \mathbb{B}$ is an \emph{opfibration} if $p^{\mathrm{op}}: \mathbb{E}^{\mathrm{op}} \to \mathbb{B}^{\mathrm{op}}$ is a fibration.
      A functor that is both a fibration and an opfibration is called a \emph{bifibration}.
\end{mydefinition}
When $p$ is a fibration, the correspondence from $Q$ to $l^* Q$ described above
induces the \emph{substitution functor} $l^*: \mathbb{E}_Y \to \mathbb{E}_X$ which replaces the index.
The following characterization of bifibrations is useful for us: a fibration $p$ is a bifibration if and only if each substitution functor $l^*: \mathbb{E}_Y \to \mathbb{E}_X$ (often called a \emph{pullback}) has a left adjoint $l_*: \mathbb{E}_X \to \mathbb{E}_Y$ (often called a \emph{pushforward}).

We are interested in reasoning over algebraic datatypes, that is in categorical terms, predicates in $\mathbb{E}_{\mu F}$ over the carrier $\mu F$ of the initial algebra for $F\colon \mathbb{B}\to\mathbb{B}$.
For this purpose
we often consider a tuple $(p, F, \dot{F})$ in the following definition.
\begin{mydefinition}[(fibred) lifting]\label{def:lifting}
  Let $p: \mathbb{E} \to \mathbb{B}$ be a functor and $F$ be an endofunctor on $\mathbb{B}$.
  We say that an endofunctor $\dot{F}$ on $\mathbb{E}$ is a \emph{lifting} of $F$ along $p$
  if $p \circ \dot{F} = F \circ p$ (see above).

  Assuming that $p$ is a fibration,
  a lifting $\dot{F}$ is \emph{fibred}
  if $\dot{F}$ preserves cartesian morphisms.
\end{mydefinition}

In this paper, we focus on a certain class of posetal fibrations called $\clat$-fibrations.
They can be seen as \emph{topological functors}~\cite{Herrlich74} whose fibres are posets.
This class abstracts treatment of spacial and logical structures.

\begin{mydefinition}[$\clat$-fibration] \label{def:clat}
  A \emph{$\clat$-fibration} is a fibration $p\colon\mathbb{E}\to\mathbb{B}$ where each fibre $\mathbb{E}_{X}$
is a complete lattice and each substitution $f^{*}\colon \mathbb{E}_{Y}\to \mathbb{E}_{X}$ preserves all meets $\bigwedge$.

  In each fibre $\mathbb{E}_X$, the order is denoted by $\leq_X$ or $\leq$.
Its least and greatest elements are denoted by $\bot_X$ and $\top_X$;
  its join and meet are denoted by $\bigvee$ and $\bigwedge$.
\end{mydefinition}
The above simple axioms of $\clat$-fibrations
induce many useful structures~\cite{Komorida19, SprungerKDH18}.
One of them is that a $\clat$-fibration is always a bifibration
whose pushforwards $f_{*}$ arise essentially by Freyd's adjoint functor theorem.
Another one is that $\clat$-fibrations lift colimits.
This is proved by~\cite[Prop.~9.2.2 and Exercise~9.2.4]{CLTT}.
\begin{myproposition} \label{prop:clat}
  Let $p: \mathbb{E} \to \mathbb{B}$ be a $\clat$-fibration.
  \begin{enumerate}
    \item $p$ is a bifibration.
    \item If $\mathbb{B}$ is (co)complete then $\mathbb{E}$ is also (co)complete and $p$ strictly preserves (co)limits.
  \myqed
  \end{enumerate}
\end{myproposition}

\begin{myexample}[$\clat$-fibration]\label{eg:clat}
\begin{itemize}
  \item ($\mathbf{Pre} \to \Set$, $\mathbf{Pred} \to \Set$) These forgetful functors
  are $\clat$-fibrations.
      Here
      $\mathbf{Pre}$ is the category of preordered sets
       $(X, \leq_X)$ and order-preserving functions between them.
      $\mathbf{Pred}$ is that of predicates:
      objects are $P \subseteq X$, and morphisms $f: (P \subseteq X) \to (Q \subseteq Y)$ are functions $f: X \to Y$ satisfying $f(P) \subseteq Q$.

    \item
      \begin{minipage}[t]{.75\textwidth}
      ($\mathbf{ERel} \to \Set$) The functor $\mathbf{ERel} \to \Set$
  defined by the change-of-base~\cite{CLTT}, as shown in the right,
   is a $\clat$-fibration. Concretely,
      $\mathbf{ERel}$ is the category of sets with binary relations $(X,R\subseteq X\times X)$ as objects, and relation-preserving maps as morphisms.
      \end{minipage}
      \hfill
      \begin{minipage}[t]{.18\textwidth}
        \begin{math}
        \xymatrix@R=1em@C=1em{
          {\mathbf{ERel}} \ar[r] \ar[d]
         &{\mathbf{Pred}} \ar[d]
         \\
          {\Set} \ar[r]^-{(-)^2}
         &{\Set}}
      \end{math}
\end{minipage}

    \item
\begin{minipage}[t]{.75\textwidth}
        (Domain fibration $d^{\Omega}: \Set/\Omega \to \Set$)
        For each complete lattice $\Omega$,
        we introduce a $\clat$-fibration
        $d^{\Omega}: \Set/\Omega \to \Set$ defined as follows.
\end{minipage}
\hfill
\begin{minipage}[t]{.18\textwidth}
      \begin{math}
        \xymatrix@R=1em@C=0.5em{
          {X} \ar[rr]^-{h}_-{\leq_X}
              \ar[rd]_-{f}
         &&
          {Y}
              \ar[ld]^-{g}
         \\
         &\Omega&}
      \end{math}
      \end{minipage}

      Here, $\Set/\Omega$  is a lax slice category
        defined as follows:
        objects of $\Set/\Omega$ are pairs $(X,f: X \to \Omega)$ of a set and a function (an ``$\Omega$-valued predicate on $X$''); we shall often write simply $f: X \to \Omega$ for the pair $(X,f)$.
  Its morphisms from $f: X\to \Omega$ to $g: Y \to \Omega$ are functions $h: X \to  Y$ such that $f \leq_X g \circ h$, as shown above.

  Then $d_\Omega$ is the evident forgetful functor, extracting the upper part of the above triangle.

  The order $\leq_X$ used there is the pointwise order between functions of the type $X \to \Omega$; the same order $\leq_{X}$ defines the order in each fiber $(\Set/\Omega)_{X}=\Set(X,\Omega)$.
    Following \cite[Def.~4.1]{AguirreK20}, we call $d^\Omega$
    a \emph{domain fibration} (from the lax slice category).
\end{itemize}
\end{myexample}

\subsection{Chain Construction of Initial Algebras} \label{subsec:adamek}
\begin{mydefinition}[chain-cocomplete category]
  A category $\mathbb{C}$ is \emph{chain-cocomplete}
  if $\mathbb{C}$ has a colimit of every chain.
  We write $0$ for a colimit of the empty chain (i.e.\ an initial object).
\end{mydefinition}
Noteworthy is that chain-cocompleteness is equivalent to existence of an initial object and filtered colimits, see \cite[Cor.~1.7]{Adamek94} for further details.

\begin{mydefinition}[initial chain{\cite{Adamek74}, \cite[Def.~3.2]{AdamekMM18}}] \label{def:init_chain}
  Let $\mathbb{C}$ be a chain-cocomplete category, and $F: \mathbb{C} \to \mathbb{C}$ be an endofunctor.
 The \emph{initial chain} of $F$ is the following diagram:
\begin{equation}\label{eq:initialChain}
  0 \xrightarrow{\alpha_{0, 1}} F0 \xrightarrow{\alpha_{1, 2}} \cdots
  \longrightarrow F^{\lambda}0 \xrightarrow{\alpha_{\lambda, \lambda+1}} \cdots.
\end{equation}
This consists of the following.
  \begin{itemize}
    \item (Objects) It has objects $F^{i}0$ for each $i\in\mathrm{Ord}$ (where $\mathrm{Ord}$ is the category of ordinals), defined by $F^{0}0 = 0$, $F^{i+1}0 = F(F^{i}0)$, and for a limit ordinal $i$,  $F^{i}0 = \colim_{j < i} F^{j}0$.
    \item (Morphisms) It has morphisms $\alpha_{i,j}\colon F^{i}0\to F^{j}0$ for all ordinals $i,j$ such that $i\le j$, defined inductively on $i$.
        (Base case) $\alpha_{0, j}: 0 \to F^{j}0$ is the unique morphism.
        (Step case) $\alpha_{i+1, j+1}$ is $F\alpha_{i, j}$; for a limit ordinal $j$, $\alpha_{i+1,j}$ is from the colimiting cocone for $F^j 0$.
        (Limit case) When $i$ is a limit ordinal,
          $\alpha_{i, j}$ is induced by universality of $F^i 0 = \colim_{k < i} F^k 0$.
  \end{itemize}
  If $\alpha_{\lambda, \lambda+1}$ is an isomorphism, then we say that the initial chain of $F$ \emph{converges} in $\lambda$ steps.
\end{mydefinition}

\begin{myproposition}[{from~\cite{Adamek74}, \cite[Thm.~3.5]{AdamekMM18}}] \label{prop:converge_to_initial}
  In the setting of Def.~\ref{def:init_chain},
assume that the initial chain
converges in $\lambda$ steps.
Then $\alpha_{\lambda, \lambda+1}^{-1}: F^{\lambda+1} 0 \iso F^{\lambda}0$ is an initial $F$-algebra. \myqed
\end{myproposition}

The dual of the initial chain in Def.~\ref{def:init_chain} is called \emph{the final chain}.
This also satisfies the dual of Prop.~\ref{prop:converge_to_initial} (yielding final coalgebras), see~\cite[Def.~3.20 and Thm.~3.21]{AdamekMM18}.

The converse of Prop.~\ref{prop:converge_to_initial} holds if we restrict to $\Set$.

\begin{myproposition}[from {\cite{Trnkova75}, \cite[Cor.~3.16]{AdamekMM18}}] \label{prop:initial_to_converge}
A set functor has an initial algebra if and only if the initial chain converges.
\myqed
\end{myproposition}
\noindent We often write $\mu F$ for the carrier of an initial algebra of $F$.

The next basic lemma is important for us. Its dual (for coalgebras) is observed e.g.\ in~\cite{HasuoCKJ13}.
\begin{mylemma} \label{lem:init}
Assume that $p: \mathbb{E} \to \mathbb{B}$ is a fibration,  that both $\mathbb{E}$ and $\mathbb{B}$ are chain-cocomplete, and that $p$ strictly preserves chain colimits. Let $\dot{F}$ be a lifting of $F\colon \mathbb{B}\to\mathbb{B}$ along $p$.

Consider the following initial chains.
\begin{displaymath}
  \vcenter{\xymatrix@R=.6em@C-.5em{
  {\mathbb{E}}
  \ar[d]_{p}
  &
  {0}
  \ar[r]^-{\dot{\alpha}_{0,1}}
  &
  {\dot{F}0}
  \ar[r]^-{\dot{\alpha}_{1,2}}
  &
  \cdots
   \ar[r]
  &
  {\dot{F}^{\lambda}0}
  \ar[r]^-{\dot{\alpha}_{\lambda,\lambda+1}}
  &
  \cdots
  \\
  {\mathbb{B}}
  &
  {0}
  \ar[r]^-{\alpha_{0,1}}
  &
  {F0}
  \ar[r]^-{\alpha_{1,2}}
  &
  \cdots
   \ar[r]
  &
  {F^{\lambda}0}
  \ar[r]^-{\alpha_{\lambda,\lambda+1}}
  &
  \cdots
  }}
\end{displaymath}
\begin{enumerate}
  \item We have $\alpha_{i, j} = p\dot{\alpha}_{i, j}$ for all ordinals $i,j$ with $i<j$.
  \item Moreover, if the upper initial chain for $\dot{F}$ converges and yield an initial $\dot{F}$-algebra $\dot{\alpha}\colon \dot{F}(\mu\dot{F})\to \mu\dot{F}$, then $p\dot{\alpha}\colon Fp(\mu\dot{F})\to p(\mu\dot{F})$ is an initial $F$-algebra. \myqed
\end{enumerate}
\end{mylemma}

\section{Initial Algebra-Final Coalgebra Coincidence over Initial Algebras} \label{sec:ip}
In this section,
we formulate our target coincidence called the \emph{\cost{}}.
It is a fibrational IA-FC coincidence over an initial algebra.
\begin{mydefinition}[\cost{}] \label{def:cost}
  Let $p: \mathbb{E} \to \mathbb{B}$ be a fibration, and
  $\dot{F}$ be a lifting of $F$. We say
that the tuple
\begin{math}
   \bigl(\,p\colon \mathbb{E} \to \mathbb{B} ,\;
        F\colon \mathbb{B}\to\mathbb{B},\;
       \dot{F}\colon\mathbb{E}\to\mathbb{E}\, \bigr)
\end{math}
satisfies the \emph{IA-FC coincidence over an initial algebra (\cost{}, for short)} if the following is satisfied.

\noindent
  \begin{enumerate}
  \item\label{item:baseInitAlg} There is an initial $F$-algebra $\beta\colon F(\mu F)\iso\mu F$.
  \item\label{item:totalInitAlg} There is an initial $\dot{F}$-algebra $\dot{\beta}\colon \dot{F}(\mu \dot{F})\iso\mu \dot{F}$ above $\beta$.
  \item\label{item:totalFinalCoalgToo}
    Moreover, $\dot{\beta}^{-1}$ is final over $\beta^{-1}$ in the following sense: for each $\dot{F}$-coalgebra $\gamma$ above $\beta^{-1}$ (shown below diagram on the left),
   there exists a unique vertical coalgebra morphism $f$ from $\gamma$ to $\dot{\beta}^{-1}$ (below diagram on the right, where vertical means $pf = \mathrm{id}_{\mu F}$).
  \end{enumerate}
\begin{displaymath}
  \vcenter{\xymatrix@R=.7em@C-1.5em{
   {\mathbb{E}} \ar@(ul, dl)_{\dot{F}}
  \ar[dd]_{p}
  &
  &
  &&
  {\dot{F}(\mu\dot{F})}
  &
  {\mu\dot{F}}
  \ar[l]_-{\dot{\beta}^{-1}}^-{\cong}
  \\
  &
  {\dot{F}P}
  &
  {P}
  \ar[l]_-{\gamma}
  &
  \Longrightarrow
  &
  {\dot{F}P}
  \ar@{.>}[u]^{\dot{F}f}
  &
  {P}
  \ar[l]_-{\gamma}
  \ar@{.>}[u]_{f}
  \\
  {\mathbb{B}} \ar@(ul, dl)_{F}
  &
  {F(\mu F)}
  &
  {\mu F\mathrlap{;}}
  \ar[l]_-{\beta^{-1}}^-{\cong}
  &&
  {F(\mu F)}
  &
  {\mu F\mathrlap{;}}
  \ar[l]_-{\beta^{-1}}^-{\cong}
}}
\end{displaymath}
\end{mydefinition}

\vspace{.3em}

\myparagraph{\CoSt{} in Fibrations of (Co)algebras}
The \cost{} in Def.~\ref{def:cost} is nicely organized in terms of \emph{fibrations of (co)algebras}: the last two conditions in Def.~\ref{def:cost} can be stated succinctly in advanced fibrational terms.

Given a functor  $F: \mathbb{B} \to \mathbb{B}$,
 $\Alg{F}$ is the category of \emph{$F$-algebras}, where an object is a pair $(X\in \mathbb{B}, a\colon FX\to X)$ and a morphism from $(X,a)$ to $ (Y,b)$ is $f\colon X\to Y$ such that $b\circ Ff=f\circ a$. Dually,   $\Coalg{F}$ is the category of \emph{$F$-coalgebras}, where an object is $(X\in \mathbb{B}, c\colon X\to FX)$ and a morphism from $(X,c)$ to $(Y,d)$ is $f$ such that $d\circ f=Ff\circ c$.

Then a fibration $p$ and a fibred lifting $\dot{F}$   yield fibrations of (co)algebras.
\begin{myproposition}[from~{\cite[Prop.~4.1]{HasuoCKJ13}}] \label{prop:alg_coalg}
A lifting $\dot{F}$ of $F$ along a fibration $p$ induces functors
$\Alg{p}: \Alg{\dot{F}} \to \Alg{F}$ and
$\Coalg{p}: \Coalg{\dot{F}} \to \Coalg{F}$,
given by
\begin{align*}
 \Alg{p}&: (\dot{F}P \xrightarrow{q} P) \longmapsto (FpP = p\dot{F}P \xrightarrow{pq} pP), \\
 \Coalg{p}&: (P \xrightarrow{r} \dot{F}P) \longmapsto (pP \xrightarrow{pr} p\dot{F}P = FpP).
\end{align*}
The functor $\Alg{p}$ is a fibration. If additionally $\dot{F}$ is a fibred lifting, then $\Coalg{p}$ is a fibration, too.
For an opfibration $p$,
we have a result dual to the above: $\Coalg{p}$ is an opfibration; so is $\Alg{p}$ if $\dot{F}$ is an opfibred lifting (preserving cocartesian morphisms).
\myqed
\end{myproposition}

The functor $\Coalg{p}\colon \Coalg{\dot{F}} \to \Coalg{F}$
in Prop.~\ref{prop:alg_coalg} plays an important role in the following development. It
is thought of as a functor where
\begin{itemize}
  \item (following the coalgebraic tradition)
  state-based transition systems $c\colon X\to FX$ and behavior-preserving morphisms between them populate the base category  $\Coalg{F}$,
 and
  \item
  \emph{invariants}---i.e.\ predicates $P\in \mathbb{E}_{X}$ over $X$ that are \emph{preserved} by transitions---populate the total category $\Coalg{\dot{F}}$. The arrows in $\Coalg{\dot{F}}$ are logical implication of invariants.
\end{itemize}
The following reformulation is proved in Appendix~\ref{ap:cost_fibre}, together with technical remarks.
\begin{myproposition} \label{prop:cost_fibre}
   The following is equivalent to Cond.~\ref{item:totalInitAlg} and~\ref{item:totalFinalCoalgToo} in Def.~\ref{def:cost}, respectively.
  \begin{enumerate}
   \renewcommand{\labelenumi}{\arabic{enumi}'.}
   \setcounter{enumi}{1}
   \item
    There is an initial object $\dot{\beta}$ in the fiber $\Alg{\dot{F}}_{\beta}$.
  \item
   $\dot{\beta}^{-1}$ is a final object in the fiber $\Coalg{\dot{F}}_{\beta^{-1}}$.
   \myqed
  \end{enumerate}
  \end{myproposition}
\vspace{.3em}

\myparagraph{\CoSt{} in a $\clat$-fibration}
Here we shall rewrite conditions in Def.~\ref{def:cost} for $\clat$-fibrations.
But first, we need the following investigation of these conditions.

An initial $\dot{F}$-algebra lying above an initial $F$-algebra is a norm (Cond.~\ref{item:baseInitAlg}--\ref{item:totalInitAlg}; cf.\ Lem.~\ref{lem:init}).
What is special is the finality of the initial $\dot{F}$-algebra (Cond.~\ref{item:totalFinalCoalgToo}).
The intuition of the latter is the following:
\begin{quote}
  \emph{an lfp and a gfp coincide, in the fiber over the base initial algebra.}
\end{quote}
Intuitively, $P$ with $(\gamma: P \to \dot{F}P) \in \Coalg{\dot{F}}_{\beta^{-1}}$ is an \emph{invariant}---it is a predicate
  that is preserved by the transition $\beta^{-1}$.
  Indeed, the morphism $\gamma$ is equivalent to a morphism
  \begin{align*}
& \gamma^{\dagger}\colon P\longrightarrow (\beta^{-1})^{*}\dot{F}P \quad\text{over $\mathrm{id}_{\mu F}$, that is,}\\
& %
  P\le (\beta^{-1})^{*}\dot{F}P \quad\text{if the fibration $p\colon\mathbb{E}\to\mathbb{B}$ is posetal,}
  \end{align*}
  by pulling back along $\beta^{-1}$.
The latter inequality signifies that $P$ is an invariant.
This equivalence is formulated as follows.
  \begin{mylemma} \label{lem:iso_alg_coalg}
  Let $p: \mathbb{E} \to \mathbb{B}$ be a fibration and $\dot{F}$ be a lifting of $F$ along $p$.
  For any isomorphism $\alpha: X \iso FX$ in $\mathbb{B}$,
  $\Alg{\dot{F}}_{\alpha^{-1}} \cong \Alg{\asdf}$
  and
  $\Coalg{\dot{F}}_\alpha \cong \Coalg{\asdf}$.
  \myqed
\end{mylemma}

Therefore, Cond.~\ref{item:totalFinalCoalgToo} requires that $\dot{\beta}^{-1}$ gives a greatest invariant.
In view of the Knaster--Tarski theorem (that a greatest post-fixed point is a greatest \emph{fixed} point), this means that $\dot{\beta}^{-1}$ is a gfp if $p$ is a $\clat$-fibration. Symmetrically, Cond.~\ref{item:totalInitAlg} (rephrased as Cond.~\ref{item:totalInitAlg}') requires that $\dot{\beta}$ is an lfp.
Therefore, the \cost{} yields a coincidence between an lfp and a gfp.
This plays an important role in the next section.
\begin{myproposition} \label{prop:cost_clat}
If $p$ is a $\clat$-fibration
then Cond.~\ref{item:totalInitAlg} and~\ref{item:totalFinalCoalgToo} in Def.~\ref{def:cost} are equivalent to the following condition:
there is a unique fixed-point $\mu \dot{F}$ of $(\beta^{-1})^*\dot{F}: \mathbb{E}_{\mu F} \to \mathbb{E}_{\mu F}$.
\myqed
\end{myproposition}

\vspace{.3em}

\myparagraph{\CoSt{} over Base IA-FC Coincidence} The IF/I coincidence (Def.~\ref{def:cost}) allows a simpler formulation in the special case where the IA-FC coincidence is \emph{already there in the base category}. In this case, $\dot{\beta}^{-1}$ is final not only in a suitable fiber (Cond.~3 of Def.~\ref{def:cost}; cf.\ Prop.~\ref{prop:cost_fibre}), but also \emph{globally} in the total category $\mathbb{E}$. See Appendix~\ref{ap:cost_lift} for details.

This special setting (the base IA-FC coincidence) is known to hold in domain-theoretic settings~\cite{SmythP82,Freyd90}. We use this  setting (specifically the IA-FC coincidence in a Kleisli category~\cite{HasuoJS07b}) in one of our applications (\S{}\ref{sec:treeUnified}).

\begin{mytheorem}[\cost{} over the base coincidence] \label{thm:cost_lift}
  Let $p: \mathbb{E} \to \mathbb{B}$ be a bifibration
  and $(p, F, \dot{F})$ be a tuple satisfying the \cost{}.
  If there exists initial $F$-algebra $\beta: F(\mu F) \cong \mu F$ (in $ \mathbb{B}$) such that $\beta^{-1}$ is a final $F$-coalgebra,
  then there exists an initial $\dot{F}$-algebra $\dot{\beta}: \dot{F}(\mu \dot{F}) \cong \mu \dot{F}$  (in $ \mathbb{E}$) above $\beta$ such that $\dot{\beta}^{-1}$ is a final $\dot{F}$-coalgebra. \qed
\end{mytheorem}

\section{\CoSt{} from Stable Chain Colimits} \label{sec:ip_for_clat}
We now present our main observation, namely that the \cost{} is a general phenomenon that relies only on a few mild assumptions. These assumptions include 1) that $\dot{F}$ is fibred (Def.~\ref{def:lifting}) and 2) \emph{stability} of chain colimits (Def.~\ref{def:stable_chain_colim}).

Here in \S{}\ref{sec:ip_for_clat}, we restrict the underlying fibration $p\colon\mathbb{E}\to\mathbb{B}$ to a $\clat$-fibration over $\Set$ (Def.~\ref{def:clat}). This restriction simplifies proofs and technical developments. Nevertheless, we have a general coincidence theorem for not necessarily posetal fibrations; it is found in Appendix~\ref{subsec:ip}. The general proof hinges on stable chain colimits, too.

The following is a key assumption. It is a fibrational adaptation of \emph{pullback-stable colimit}, a notion studied in (higher) topos theory and categorical logic~\cite{Lurie09,CarboniLW93,HeindelS09}.
\begin{mydefinition}[stable chain colimits] \label{def:stable_chain_colim}
  We say that a fibration $p: \mathbb{E} \to \mathbb{B}$ has \emph{stable chain colimits}
  if the following condition holds: for each $\lambda \in \mathrm{Ord}$ and each diagram $D: \mathrm{Ord}_{<\lambda} \to \mathbb{B}$,
  \begin{enumerate}
  \item $\mathbb{B}$ has a colimit of $D$.
    The $i$-th cocone component is denoted by $\kappa_{i}\colon Di\to \colim D$.
  \item Moreover, for each object $P \in \mathbb{E}_{\colim D}$ above $\colim D$, we have $P \cong \colim \kappa_i^*P$, with the cartesian liftings $(\kappa_i^*P \to P)_{i \in \mathrm{Ord}_{< \lambda}}$ forming a colimiting cocone.
  \begin{displaymath}
   \vcenter{\xymatrix@R=.6em@C=1.2em{
        \mathbb{E} \ar[d]^p
        &\kappa_0^*P \ar[r] &\kappa_1^*P \ar[r] &\cdots \ar[r] &P \mathrlap{\;(\cong \colim \kappa_i^*P)} \\
        \mathbb{B}
        &D0 \ar[r] &D1 \ar[r] &\cdots \ar[r] &\colim D
  }}
\end{displaymath}
  \end{enumerate}
\end{mydefinition}
The first condition is equivalent to chain-cocompleteness.
The situation of the second condition is illustrated as the above diagram. Stability requires that the upper cocone is colimiting.
In the diagram, we note that morphisms $\kappa_i^*P \to \kappa_j^* P$ above $D(i \to j)$ are well-defined (where $i \leq j \leq \lambda$); they are induced by universality of the cartesian liftings $\kappa_j^*P \to P$.

Letting $\lambda=0$ in Def.~\ref{def:stable_chain_colim} yields the following property.
\begin{mylemma} \label{lem:init_obj}
Let $p\colon \mathbb{E}\to\mathbb{B}$ be a fibration with stable chain colimits. Then,
  all objects in $\mathbb{E}_0$ are initial in $\mathbb{E}$. \myqed
\end{mylemma}

\begin{myexample} \label{eg:stable chain colimits}
The fibrations in Example~\ref{eg:clat}---$\mathbf{Pre} \to \Set$, $\mathbf{Pred} \to \Set$, $\mathbf{ERel} \to \Set$, and the domain fibration $d^{\Omega}$ for any complete lattice $\Omega$---all have stable chain colimits.
Non-examples are deferred to Appendix~\ref{ap:non_example}.
\end{myexample}

\begin{mytheorem}[Main result] \label{thm:cost_clat}
  Let $p: \mathbb{E} \to \Set$ be a $\clat$-fibration
  and $\dot{F}$ be a lifting of $F$ along $p$.
  Assume further the following conditions:
  \begin{enumerate}
    \item there exists an initial $F$-algebra; \label{item:cost_clat1}
    \item $\dot{F}$ is a fibred lifting of $F$; \label{item:cost_clat2}
    \item $p$ has stable chain colimits. \label{item:cost_clat3}
  \end{enumerate}
  Then $(p, F, \dot{F})$ satisfies the \cost{} (Def.~\ref{def:cost}).
\end{mytheorem}

We prove the theorem in the rest of the section.
Due to Prop.~\ref{prop:cost_clat}, it suffices to show that  $(\beta^{-1})^*\dot{F}\colon \mathbb{E}_{\mu F} \to \mathbb{E}_{\mu F}$ has a unique fixed point,
 where $\beta$ is an initial $F$-algebra.
Cond.~\ref{item:cost_clat1} in Thm.~\ref{thm:cost_clat} yields that the initial chain of $F$ converges
and  gives an initial $F$-algebra (Prop.~\ref{prop:converge_to_initial} and~\ref{prop:initial_to_converge}).

We analyze the initial chains of $F$ and $\dot{F}$, which is shown on the below.
 \begin{displaymath}
  \vcenter{\xymatrix@R=.8em@C=2em{
  {\mathbb{E}}
  \ar[d]_{p}
  &
  {0}
  \ar[r]^-{\dot{\alpha}_{0,1}}
  &
  {\dot{F}0}
  \ar[r]^-{\dot{\alpha}_{1,2}}
  &
  \cdots
   \ar[r]
  &
  {\dot{F}^{\lambda}0}
  \ar@{->}[r]^-{\dot{\alpha}_{\lambda,\lambda+1}}
  &
  {\dot{F}^{\lambda+1}0}
  \ar[r]
  &
  \cdots
  \\
  {\Set}
  &
  {0}
  \ar[r]^-{\alpha_{0,1}}
  &
  {F0}
  \ar[r]^-{\alpha_{1,2}}
  &
  \cdots
   \ar[r]
  &
  {F^{\lambda}0}
  \ar@{->}[r]^-{\alpha_{\lambda,\lambda+1}}
  &
  {\dot{F}^{\lambda+1}0}
  \ar[r]
  &
  \cdots
  }}
 \end{displaymath}
  Prop.~\ref{prop:clat} and Lem.~\ref{lem:init} ensure that each chain morphism $\dot{\alpha}_{i, i+1}$ is above $\alpha_{i, i+1}$. Then, assuming that the initial chain of $F$ converges in $\lambda$ steps, the functor $(\beta^{-1})^*\dot{F}$ of our interest is equal to $\alpha_{\lambda,\lambda+1}^*\dot{F}$.

Fig.~\ref{fig:clat} is the key diagram about a unique fixed-point of $\alpha_{\lambda, \lambda+1}^*\dot{F}$.
For simplicity, we write $\alpha$ for $\alpha_{\lambda, \lambda+1}$.
We find  the initial chain of $\dot{F}$ as its middle row; the initial chain of $\alpha^*\dot{F}$ as the bottom half of the last column; and the final chain of $\alpha^*\dot{F}$ as the top half.
The other objects in the diagram are obtained by applying substitution to the last column.

\begin{figure*}[h] \footnotesize
  \begin{math}
  \vcenter{\xymatrix@R=1em@C=2em{
  \alpha_{0, \lambda}^*\top \ar[r] &\alpha_{1, \lambda}^*\top \ar[r] &\alpha_{2, \lambda}^*\top \ar[r] &\cdots \ar[r] &\top \\
  \alpha_{0, \lambda}^*\alpha^*\dot{F}\top \ar[r] \ar@{=}[u] &\alpha_{1, \lambda}^*\alpha^*\dot{F}\top \ar[r] \ar[u] &\alpha_{2, \lambda}^*\alpha^*\dot{F}\top \ar[r] \ar[u] &\cdots \ar[r] &\alpha^*\dot{F}\top \ar[u] \\
  \alpha_{0, \lambda}^*(\alpha^*\dot{F})^2\top \ar[r] \ar@{=}[u] \ar@{=}[d] &\alpha_{1, \lambda}^*(\alpha^*\dot{F})^2\top \ar[r] \ar@{=}[u] \ar@{=}[d] &\alpha_{2, \lambda}^*(\alpha^*\dot{F})^2\top \ar[r] \ar[u]  \ar@{=}[d] &\cdots \ar[r] &(\alpha^*\dot{F})^2\top \ar[u] \ar@{<-}[d] \\
   & & & & \\
  0 \ar@{}[u]|-(1){\vdots} \ar@{}[d]|-{\vdots} \ar[r]^{\dot{\alpha}_{0, 1}} &\dot{F}0 \ar@{}[d]|-{\vdots} \ar[r]^{\dot{\alpha}_{1, 2}} \ar@{}[u]|-(1){\vdots} &\dot{F}^20 \ar[r] \ar@{}[d]|-{\vdots} \ar@{}[u]|-(1){\vdots} &\cdots \ar[r] &\dot{F}^\lambda 0 \ar@{}[d]|-{\vdots} \ar@{}[u]|-(1){\vdots} \mathrlap{\text{\quad (the initial $\dot{F}$-chain)}}\\
   & & & & \\
  \alpha_{0, \lambda}^*(\alpha^*\dot{F})^2\bot \ar[r] \ar@{=}[u] &\alpha_{1, \lambda}^*(\alpha^*\dot{F})^2\bot \ar[r] \ar@{=}[u] &\alpha_{2, \lambda}^*(\alpha^*\dot{F})^2\bot \ar[r] \ar@{=}[u] &\cdots \ar[r] &(\alpha^*\dot{F})^2\bot \ar[u] \\
  \alpha_{0, \lambda}^*\alpha^*\dot{F}\bot \ar[r] \ar@{=}[u] &\alpha_{1, \lambda}^*\alpha^*\dot{F}\bot \ar[r] \ar@{=}[u] &\alpha_{2, \lambda}^*\alpha^*\dot{F}\bot \ar[r] \ar[u] &\cdots \ar[r] &\alpha^*\dot{F}\bot \ar[u] \\
  \alpha_{0, \lambda}^*\bot \ar[r] \ar@{=}[u] &\alpha_{1, \lambda}^*\bot \ar[r] \ar[u] &\alpha_{2, \lambda}^*\bot \ar[r] \ar[u] &\cdots \ar[r] &\bot \ar[u]
\\
  0 \ar[r]^{\alpha_{0, 1}} &F0 \ar[r]^{\alpha_{1, 2}} &F^20 \ar[r] &\cdots \ar[r] &F^\lambda 0 \mathrlap{\text{\quad (the initial $F$-chain)}}
}
}
\end{math}
\caption{IA-FC coincidence for $\clat$-fibrations, in Prop.~\ref{prop:coinci_clat}. Here we write  $\alpha$  for $\alpha_{\lambda,\lambda+1}$; the choice of $\lambda$ is arbitrary  (the initial $\dot{F}$-chain may not have stabilized).}
\label{fig:clat}
\end{figure*}

The next result is the key technical observation. It says 1) the upper rows become closer to the initial $\dot{F}$-chain as we go below; and 2) symmetrically, the lower rows become closer to the same as we go up. Its proof is by transfinite induction; the stability assumption is crucially used in its limit case.
\begin{myproposition} \label{prop:coinci_clat}
  Consider the setting of Thm.~\ref{thm:cost_clat}. Let $\lambda$ be an arbitrary ordinal.
  We write $\alpha, \dot{\alpha}$ for $\alpha_{\lambda, \lambda+1}, \dot{\alpha}_{\lambda, \lambda+1}$
  and $\top, \bot$ for the maximum and minimum of the complete lattice $\mathbb{E}_{F^\lambda 0}$.
  For each ordinal $i$, the objects
  $(\alpha^*\dot{F})^i \bot$ and $(\alpha^*\dot{F})^i \top$ above $F^{\lambda}0$ are defined by
  the initial chain and the final chain of $\alpha^*\dot{F}$ (the last column of Fig.~\ref{fig:clat}).

  Then we have  $\alpha_{i, \lambda}^*(\alpha^*\dot{F})^i \bot = \dot{F}^i 0 = \alpha_{i, \lambda}^*(\alpha^*\dot{F})^i \top$ for each $i$ with $i\le \lambda$.
\end{myproposition}
\begin{myproof}[{sketch; a full proof is in Appendix~\ref{ap:clat_ip}}]
The proof is by transfinite induction on $i$.
  The base case is clear because $\mathbb{E}_0$ includes only one object by Lemma~\ref{lem:init_obj} and the posetal assumption on $p$.

  In the step case, fibredness of the lifting $\dot{F}$ lifts the equality for $i$, which is $\alpha^{*}_{i, \lambda}(\alpha^*\dot{F})^i\bot = \dot{F}^i0 = \alpha^{*}_{i, \lambda}(\alpha^*\dot{F})^i\top$ (the induction hypothesis), to the desired equality for $i+1$.

  The limit case is less obvious than the other cases.
  We rewrite the target objects (e.g. $\alpha_{i, \lambda}^*(\alpha^*\dot{F})^i \bot$) to chain colimits (e.g. $\colim_{j < i} \alpha_{j, \lambda}^*(\alpha^*\dot{F})^i \bot$)
  by stability of chain colimits,
  and use the fact that colimits of diagonal elements (e.g. $\colim_{j < i} \alpha_{j, \lambda}^*(\alpha^*\dot{F})^j \bot$) are equal to $\dot{F}^i 0$ by the induction hypothesis.
  See Appendix~\ref{ap:clat_ip} for a full proof.
  \myqed
\end{myproof}

Letting $i=\lambda$ in
Prop.~\ref{prop:coinci_clat} yields that
$(\alpha^*\dot{F})^\lambda \bot = \dot{F}^\lambda 0 = (\alpha^*\dot{F})^\lambda \top$.
Therefore,
both the initial  and  final chains of $\alpha^*\dot{F}$ (the last column in Fig.~\ref{fig:clat}) converge in $\lambda$ steps.
We conclude that $\dot{F}^\lambda 0$ is both the lfp and gfp for $\alpha^*\dot{F}\colon \mathbb{E}_{F^{\lambda}0}\to\mathbb{E}_{F^{\lambda}0}$, hence is
its unique fixed point.

Here are some consequences of the proposition. In the next result, note that the number of converging steps of $F$ and that of $\dot{F}$ are not the same in general.
See Appendix~\ref{ap:non_example} for an example.
\begin{mycorollary} \label{cor:same_converge}
  Let $p: \mathbb{E} \to \Set$ be a $\clat$-fibration
  and $\dot{F}$ be a fibred lifting of $F$ along $p$.
  Assume $p$ has stable chain colimits.
  Then, the initial chain of $F$ converges in $\lambda$ steps
  if and only if that of $\dot{F}$ converges in $\lambda$ steps.
  \myqed
\end{mycorollary}

\begin{mycorollary} \label{cor:init_alg_search}
  In the setting of Cor.~\ref{cor:same_converge},
  if $F$ has an initial algebra $\alpha$,
  then any isomorphism $\dot{F}P \to P$ above $\alpha$ is an initial algebra of $\dot{F}$.
  \myqed
\end{mycorollary}

We are finally in a position to prove our main theorem.
\begin{myproof}[Thm.~\ref{thm:cost_clat}]
  Using Prop.~\ref{prop:initial_to_converge} and Cor.~\ref{cor:same_converge},
  Cond.~\ref{item:cost_clat1} ensures the existence of an ordinal $\lambda$ such that both the initial chains of $F$ and $\dot{F}$ converges in the steps.
  Then $\alpha_{\lambda, \lambda+1}^{-1}$ is an initial $F$-algebra by Prop.~\ref{prop:converge_to_initial} and $\dot{F}^\lambda 0$ is a unique fixed-point of $\alpha_{\lambda, \lambda+1}^*\dot{F}$ by Prop.~\ref{prop:coinci_clat}.
  Prop.~\ref{prop:cost_clat} concludes the proof.
  \myqed
\end{myproof}

\section{Coincidence for \texorpdfstring{$\Omega$}{Omega}-Valued Predicates, Pure and Effectful\conf{70}} \label{sec:general}
 We instantiate the above categorical results  to an important family of examples, namely \emph{$\Omega$-valued predicates} (Example~\ref{eg:clat}).
In this setting, a functor lifting $\dot{F}$  (\S{}\ref{sec:ip}) has a concrete presentation as an $F$-algebra, an observation that helps identification of many examples.

Besides the ``pure'' setting modeled by the fibration $\Set/\Omega\to\Set$, we also consider the ``effectful'' setting $\Kl{\dot{\mathcal{T}}}\to\Kl{\mathcal{T}}$, where effects are modeled by a monad $\mT$~\cite{Moggi91a} with its lifting $\dT$ along $d^\Omega$, and the base category is the Kleisli category for $\mathcal{T}$. The categorical construction of the fibration $\Kl{\dot{\mathcal{T}}}\to\Kl{\mathcal{T}}$ is described later in~\S{}\ref{subsec:effectful}; the construction builds upon the recent results in~\cite{AguirreK20}.

The theoretical development here in~\S{}\ref{sec:general} specializes that in~\S{}\ref{sec:ip}--\ref{sec:ip_for_clat}, but it is still in abstract categorical terms. The theory in~\S{}\ref{sec:general}  paves the way to the concrete applications in~\S{}\ref{sec:live}--\ref{sec:treeUnified}.

\subsection{Coincidence for \texorpdfstring{$\Omega$}{Omega}-Valued Predicates, the Pure Setting} \label{sec:general1}
We first focus on the domain fibration $d^\Omega: \Set/\Omega \to \Set$ (Example~\ref{eg:clat}), where 1) a complete lattice $\Omega$ is regarded as a truth value domain, and 2) the fibration is regarded as that of $\Omega$-valued predicates.
If $\Omega$ is the two-element lattice $\mathbf{2}=\{\bot, \top\}$, then $d^\mathbf{2}: \Set/\mathbf{2} \to \Set$ is isomorphic to $\mathbf{Pred} \to \Set$.

Towards the \cost{} for the fibration $d^\Omega$, we first need to describe a fibred lifting $\dot{F}$ of  $F$. It is induced by an $F$-algebra over $\Omega$ that is equipped with a suitable order structure.

\begin{mydefinition}[monotone algebra \cite{AguirreK20}] \label{def:mono}
  Let $F: \Set \to \Set$ be a functor and $\Omega$ be a complete lattice.
  We call $\sigma: F\Omega \to \Omega$ a \emph{monotone $F$-algebra} over $\Omega$
  if $i \leq_X i' \Rightarrow \sigma \circ Fi \leq_{FX} \sigma \circ Fi'$ holds
      for all $X\in\Set$ and all $i, i' \in \Set(X, \Omega)$.
\end{mydefinition}

\begin{mylemma}[{from \cite{BonchiKP18, AguirreK20}}] \label{lem:mono_alg}
  Let $F: \Set \to \Set$ be a functor, and $\Omega$ be a complete lattice.
  There is a bijective correspondence between monotone $F$-algebras
  $\sigma: F\Omega \to \Omega$ and fibred liftings $\dot{F}$ of $F$
  along $d^{\Omega}$. Specifically, $\sigma$ gives rise to the
  lifting $\dot{F}$ given by
  $\dot{F}(X \xrightarrow{x} \Omega) = (FX \xrightarrow{Fx} F\Omega
  \xrightarrow{\sigma} \Omega)$; conversely, $\dot{F}$ gives rise to
  $(F\Omega\xrightarrow{\sigma}\Omega) =
  \dot{F}(\Omega\xrightarrow{\mathrm{id}_{\Omega}}\Omega)$. \myqed
\end{mylemma}

Application of \S{}\ref{sec:ip_for_clat} to a domain fibration is then easy.
\begin{mytheorem}[coincidence for $\Omega$-valued predicates, pure] \label{thm:general1}
  In the setting of Def.~\ref{def:mono}, let $\sigma: F\Omega \to \Omega$ be a monotone $F$-algebra.
  By Lem.~\ref{lem:mono_alg}, $\sigma$ gives rise to a fibred lifting $\dot{F}$ of $F$ along $d^\Omega$.
  If there exists an initial $F$-algebra then $(d^\Omega, F, \dot{F})$ satisfies the \cost{}.
  \myqed
\end{mytheorem}

\subsection{Coincidence for \texorpdfstring{$\Omega$}{Omega}-Valued Predicates, Effectful}\label{subsec:effectful} \label{sec:general2}
In order to accommodate some concrete examples (those in~\S{}\ref{sec:treeUnified} to be specific), we extend the above material to the setting with monadic effects.

We aim at the situation in~(\ref{eq:fibKleisliEmbedding}), where the domain fibration $d^\Omega$ is Kleisli-embedded in the fibration
$d^{\Omega}_{\mathcal{T}, \dot{\mathcal{T}}}\colon\Kl{\dot{\mathcal{T}}}
\rightarrow
\Kl{\mathcal{T}}$
on the right. The latter is the desired fibration of effectful computations and $\Omega$-valued predicates; moreover, we extend a functor $F$ and its lifting $\dot{F}$ for the Kleisli fibration, too.
\begin{equation} \label{eq:fibKleisliEmbedding}
  \vcenter{\xymatrix@R=1em@C-1em{
     \Set/\Omega \ar[d]^{d^{\Omega}} \lloop{\dot{F}} \ar[r]^{\dot{L}} &\Kl{\dot{\mathcal{T}}} \rloop{\dot{F}_{\dot{\mathcal{T}}}} \ar[d]^{d^{\Omega}_{\mathcal{T}, \dot{\mathcal{T}}}} \\
      \Set \lloop{F} \ar[r]^L &\Kl{\mathcal{T}} \rloop{F_{\mathcal{T}}}
  }}
\end{equation}

The construction of the Kleisli fibration
$d^{\Omega}_{\mathcal{T}, \dot{\mathcal{T}}}$ is  via a {\em
  cartesian lifting} of the monad $\mT$. It is defined to be a monad
$(\dT,\dot\eta,\dot\mu)$ on $\Set/\Omega$ such that 1) $\dT$ (as a functor) is a fibred
lifting of the functor $\mT$, and 2) $\dot\eta,\dot\mu$ are componentwise cartesian
morphisms above $\eta,\mu$, respectively. Then
$d^\Omega_{\mT,\dT}:\Kl{\dT}\to\Kl{\mT}$ is defined to be the evident
extension of $d^\Omega$ to Kleisli categories, and is a
fibration \cite{AguirreK20}. Cartesian liftings of $\mathcal{T}$ from $\Set$ to
$\Set/\Omega$ bijectively correspond to Eilenberg-Moore (EM) $\mathcal{T}$-algebras, much like in Lem.~\ref{lem:mono_alg}.

\begin{mydefinition}[EM monotone algebra \cite{AguirreK20}] \label{def:em}
  Let $\mathcal{T}: \Set \to \Set$ be a monad and $\Omega$ be a complete lattice.
  A monotone $\mathcal{T}$-algebra $\tau: \mathcal{T}\Omega \to \Omega$ (where $\mathcal{T}$ is identified with its underlying functor) is called an \emph{Eilenberg-Moore (EM) monotone $\mathcal{T}$-algebra}
  if $\mathrm{id}_{\Omega} = \tau \circ \eta_{\Omega}$
    and $\tau \circ \mathcal{T}\tau = \tau \circ \mu_{\Omega}$. Here $\eta$ and $\mu$ are the unit and multiplication of the monad $\mathcal{T}$.
\end{mydefinition}
\begin{mylemma}[{from \cite[Thm.~4.4]{AguirreK20}}] \label{lem:em_mono_alg}
  Let $\mathcal{T}: \Set \to \Set$ be a monad and $\Omega$ be a complete lattice.
There is a bijective correspondence between
  \begin{itemize}
  \item EM monotone $\mathcal{T}$-algebras $\tau$, and
  \item Cartesian liftings $\dot{\mathcal{T}}$ of $\mathcal{T}$ that is itself a monad on $\Set/\Omega$.
  \end{itemize}
Specifically, $\tau$ gives rise to the
  lifting $\dT$ given by
  $\dT(X \xrightarrow{x} \Omega) = (\T X \xrightarrow{\T x} \T \Omega
  \xrightarrow{\tau} \Omega)$; conversely, $\dT$ gives rise to
  $(\T \Omega\xrightarrow{\tau}\Omega) =
  \dT(\Omega\xrightarrow{\mathrm{id}_{\Omega}}\Omega)$.
  \myqed
\end{mylemma}

Let us now describe the fibration $d^\Omega_{\mathcal{T}, \dot{\mathcal{T}}}$ between Kleisli categories---it is the one on the right in~(\ref{eq:fibKleisliEmbedding}).
Recall that the \emph{Kleisli category} $\Kl{\mathcal{T}}$ of a monad $\mathcal{T}$ on $\mathbb{C}$ has the same objects as $\mathbb{C}$, and its morphisms from $C$ to $D$ are $\mathbb{C}$-morphisms $C\to \mathcal{T}D$ (often denoted by $C \relto D$). In view of Lem.~\ref{lem:em_mono_alg}, the Kleisli category $\Kl{\dot{\mathcal{T}}}$ is described as follows:
\begin{itemize}
  \item its objects are pairs $(X,f: X\to\Omega)$ where the latter is an $\Omega$-valued predicate;
  \item
      \begin{minipage}[t]{.75\textwidth}
its morphisms from $(X,f: X\to\Omega)$ to $(Y,g: Y\to\Omega)$ are $h: X\to \mathcal{T}Y$ such that $f \leq_X \dT g \circ h$ as shown in the right,
where $\tau$ is the EM monotone $\mathcal{T}$-algebra that corresponds to the lifting $\dot{\mathcal{T}}$ (Lem.~\ref{lem:em_mono_alg}).
      \end{minipage}
      \begin{minipage}[t]{.25\textwidth}
  \begin{math}
  \xymatrix@R=1em@C=.8em{
  {X} \ar[rr]^-{h}
      \ar[rd]_-{f}
  &\ar@{}[d]|-{\leq_X}
  &{\mathcal{T}Y}
      \ar[ld]^-{\dT g=\tau \circ \T g}
  \\
  &\Omega&
}
  \end{math}
      \end{minipage}
\end{itemize}

\begin{mylemma}[the fibration $d^\Omega_{\mathcal{T}, \dot{\mathcal{T}}}\colon \Kl{\dot{\mathcal{T}}} \to \Kl{\mathcal{T}}$~{\cite[Cor.~3.5]{AguirreK20}}] \label{lem:kleisli_fib}
  Let $\mathcal{T}: \Set \to \Set$ be a monad,
  $\Omega$ be a complete lattice,
  and $\tau$ be an EM monotone $\mathcal{T}$-algebra.
  By Lem.~\ref{lem:em_mono_alg},
  $\tau$ gives the fibred lifting $\dot{\mathcal{T}}$ of $\mathcal{T}$ such that $\dot{\mathcal{T}}$ is a monad.
  \begin{enumerate}
    \item
      The functor $d^{\Omega}_{\mathcal{T}, \dot{\mathcal{T}}}: \Kl{\dot{\mathcal{T}}} \to \Kl{\mathcal{T}}$, defined as follows, is    a posetal fibration:
\begin{math}
 (X \to \Omega) \longmapsto X
\end{math} on objects, and
\begin{math}
 \bigl(f: (X \to \Omega) \relto (Y \to \Omega)\bigr) \longmapsto f: X \relto Y
\end{math} on morphisms.
\item \label{item:fibre}
  For each $X$ in $\Set$, we have the isomorphism
$(\Set/\Omega)_X \cong \Kl{\dot{\mathcal{T}}}_{LX}$ between fibers. Here $L: \Set \to \Kl{\mathcal{T}}$ is the Kleisli left adjoint that carries each object $X$ to $X$.
\myqed
  \end{enumerate}
\end{mylemma}

Now that we have described the fibration
$d^{\Omega}_{\mathcal{T}, \dot{\mathcal{T}}}\colon\Kl{\dot{\mathcal{T}}}
\rightarrow
\Kl{\mathcal{T}}$,
let us extend the functors $F,\dot{F}$ to $F_{\T}, \dot{F}_{\dotT}$ (cf.~(\ref{eq:fibKleisliEmbedding})).
We can do so by specifying how $F$ and $\T$ interact.
\begin{mydefinition}[distributive law~\cite{Mulry93}]
  Let $F: \Set \to \Set$ be a functor and $\mathcal{T}:\Set\to\Set$ be
  a monad with unit $\eta$ and multiplication $\mu$. A
  \emph{distributive law} of $F$ over $\mathcal{T}$ is a natural
  transformation $\lambda: F\mathcal{T} \Rightarrow \mathcal{T}F$ that
  makes the following diagrams commute.

  \[
  \begin{tikzcd}
    FX \arrow[dr, "\eta_{FX}"'] \arrow[r, "F\eta_X"] &F\mathcal{T}X \arrow[d, "\lambda_X"] \\
                            &\mathcal{T}FX
  \end{tikzcd}
  \qquad
  \begin{tikzcd}
    F\mathcal{T}^2X \arrow[r, "\lambda_{\mathcal{T}X}"] \arrow[d, "F\mu_X"] &\mathcal{T}F\mathcal{T}X \arrow[r, "\mathcal{T}\lambda_X"] &\mathcal{T}^2FX \arrow[d, "\mu_{FX}"] \\
    F\mathcal{T}X \arrow[rr, "\lambda_X"] & &\mathcal{T}FX
  \end{tikzcd}
  \]
\end{mydefinition}
\begin{mylemma}[{from~\cite{Mulry93}}]
  \label{lem:dist_law1}
  Let $F: \Set \to \Set$ be a functor, $\mathcal{T}:\Set\to\Set$ be a
  monad and $L:\Set\to\Kl{\T}$ be the left adjoint to the Kleisli
  category of $\mathcal T$. There is
  a bijective correspondence between distributive laws
  $\lambda: F\mathcal{T} \Rightarrow \mathcal{T}F$ and extensions
  $F_\mathcal{T}: \Kl{\mathcal{T}} \to \Kl{\mathcal{T}}$ of $F$ along
  $L$ (that is, $F_\T\circ L=L\circ F$).
  \myqed
\end{mylemma}
The next lemma tells how to lift a distributive
law $\lambda$ of $F$ over $\mT$
to that of $\dot F$ over $\dT$.
It follows from~\cite[Thm. 4.4]{AguirreK20}.
\begin{mylemma} \label{lem:dist_law2} Let $F: \Set \to \Set$ be a
  functor, $\mathcal{T}:\Set\to\Set$ be a monad, and $\Omega$ be a
  complete lattice. Consider a fibred lifting $\dot{F}$ of $F$
  corresponding to a monotone $F$-algebra $\sigma: F\Omega \to \Omega$
  and a Cartesian lifting $\dT$ of $\mT$ corresponding to an EM
  monotone $\T$-algebra $\tau: \T \Omega \to \Omega$ (see
  Lem.~\ref{lem:mono_alg} and~\ref{lem:em_mono_alg}). Assume further
  that a distributive law $\lambda: F\T \Rightarrow \T F$ is
  compatible with $\sigma$ and $\tau$, in the sense that
  $\sigma \circ F \tau \leq \tau \circ \mathcal{T} \sigma \circ
  \lambda_{\Omega}$. Then this $\lambda$ induces a distributive law
  $\dot{\lambda}: \dot{F}\dot{\mathcal{T}} \Rightarrow
  \dot{\mathcal{T}}\dot{F}$ of $\dot{F}$ over $\dot{\mathcal{T}}$
  above $\lambda$. \myqed
\end{mylemma}

Finally, we obtain the fibrations and functors shown in~(\ref{eq:fibKleisliEmbedding}).
\begin{mydefinition} \label{def:kleisli_setting}
  Assume the setting of Thm.~\ref{thm:general1}.
  Let $\mathcal{T}:\Set\to\Set$ be a monad;
  $\tau$ be an EM monotone $\mathcal{T}$-algebra on $\Omega$;
  and $\lambda$ be a distributive law
  satisfying $\sigma \circ F \tau \leq \tau \circ \mathcal{T} \sigma \circ \lambda_{\Omega}$.
  We define
  $(d^\Omega_{\mathcal{T}, \dot{\mathcal{T}}}, F_\mathcal{T}, \dot{F}_{\dot{\mathcal{T}}})$ as follows.
\begin{itemize}
  \item The EM monotone $\mathcal{T}$-algebra $\tau: \mathcal{T}\Omega \to \Omega$ gives rise to a Cartesian monad lifting $\dot{\mathcal{T}}$ of $\mathcal{T}$ along $d^{\Omega}$ (Lem.~\ref{lem:em_mono_alg}) and a fibration $d^{\Omega}_{\mathcal{T}, \dot{\mathcal{T}}}: \Kl{\dot{\mathcal{T}}} \to \Kl{\mathcal{T}}$ (Lem.~\ref{lem:kleisli_fib}).
  \item The distributive law $\lambda: F\mathcal{T} \Rightarrow \mathcal{T}F$ induces $F_\mathcal{T}: \Kl{\mathcal{T}} \to \Kl{\mathcal{T}}$ such that $F_\T$ is an extension of $F$ (in the sense of $F_\T\circ L=L\circ F$, Lem.~\ref{lem:dist_law1}).
  \item
    Because $\lambda$ satisfies $\sigma \circ F \tau \leq \tau \circ \mathcal{T} \sigma \circ \lambda_{\Omega}$,
    Lem.~\ref{lem:dist_law2} canonically induces a distributive law $\dot{\lambda}: \dot{F}\dot{\mathcal{T}} \Rightarrow \dot{\mathcal{T}}\dot{F}$.
  \item This distributive law $\dot{\lambda}$ induces an extension $\dot{F}_{\dot{\mathcal{T}}}: \Kl{\dot{\mathcal{T}}} \to \Kl{\dot{\mathcal{T}}}$ of $\dot F$ (Lem.~\ref{lem:dist_law1}), which is also a lifting of $\dot F$
    along $d^{\Omega}_{\mathcal{T}, \dot{\mathcal{T}}}$.
  \item
  (Optional) If $\lambda$ satisfies the equality $\sigma \circ F \tau = \tau \circ \mathcal{T} \sigma \circ \lambda_{\Omega}$ (instead of the inequality $\le$ required in the above), then $\dot{F}_{\dot{\mathcal{T}}}$ is a fibred lifting of $F_\mathcal{T}$.
\end{itemize}
\end{mydefinition}
The above technical material (mainly  from~\cite{AguirreK20}) allows us to state this section's main result.
\begin{mytheorem}[coincidence for $\Omega$-valued predicates, effectful] \label{thm:general2}
  In the setting of Def.~\ref{def:kleisli_setting},
  if there exists an initial $F$-algebra then
  $(d^\Omega_{\mathcal{T}, \dot{\mathcal{T}}}, F_\mathcal{T}, \dot{F}_{\dot{\mathcal{T}}})$ satisfies the \cost{}.
  \myqed
\end{mytheorem}

The proof of Thm.~\ref{thm:general2} is \emph{not} a straightforward application  of the general results in~\S{}\ref{sec:ip_for_clat} to the fibration $d^\Omega_{\mathcal{T}, \dot{\mathcal{T}}}$. Notice, for example, that fibredness of the lifting $\dot{F}_{\dot{\mathcal{T}}}$ is not mandatory in Def.~\ref{def:kleisli_setting}, while it is required in the general IF/I coincidence result (Thm.~\ref{thm:cost_clat}). Indeed, the lifting $\dot{F}_{\dot{\mathcal{T}}}$ is not fibred in our application in~\S{}\ref{sec:treeUnified}, so Thm.~\ref{thm:cost_clat} does not apply to it.

Our proof of Thm.~\ref{thm:general2} instead goes via the ``pure'' fibration $d^\Omega$ (on the left in~(\ref{eq:fibKleisliEmbedding})): using the fact that the left adjoint $L$ preserves initial chains, we essentially lift the IF/I coincidence from pure ($d^\Omega$) to effectful ($d^\Omega_{\mathcal{T}, \dot{\mathcal{T}}}$). We count this proof (in Appendix~\ref{ap:general2}) as one of our main contributions.

\section{Application 1: Probabilistic Liveness by Submartingales \conf{50}} \label{sec:live}
 We use the IF/I coincidence results in~\S{}\ref{sec:ip}--\ref{sec:general} to derive a new proof method for probabilistic liveness---more concretely, we derive the method as an instance of Thm.~\ref{thm:general1}.
Liveness properties are usually witnessed by \emph{ranking supermartingales}; see e.g.~\cite{TakisakaOUH18,ChakarovS13}. Restricting to finite trees, we show that probabilistic liveness can also be witnessed by an invariant-like \emph{sub}martingale (as opposed to \emph{super}martingale) notion.

Here is the class of probabilistic systems that we analyze. It is restricted for the simplicity of presentation; accommodating more expressive formalisms is easy by changing a functor.
\begin{mydefinition}[finite probabilistic binary tree]\label{def:finiteProbTree}
  A \emph{finite probabilistic binary tree} is a finite binary tree such that
  \begin{itemize}
  \item each internal node $n$ is labeled with either $\ok$ or $\notok$; and
  \item each edge is labeled with a real number $p\in [0,1]$, in such a way that two outgoing edge-labels sum to $1$.
  See~(\ref{eq:probBinTree}).
  \begin{equation} \label{eq:probBinTree}
       \vcenter{
      \xymatrix@R=.5em@C=.3em{
        &n \ar@{->}[dl]_-{p} \ar@{->}[dr]^-{1-p} \\
        n_1 & &n_2
      }}
  \end{equation}
  \end{itemize}
\end{mydefinition}

We restrict to \emph{finite} trees; here is one application scenario that justifies it.
We think of those probabilistic trees as models of systems with internal coin toss. We assume that there is some timeout mechanism that forces the termination of those systems, that is,  that termination of the target system is guaranteed by some external means. Such mechanism  forcing finiteness is common in real-world systems.

The liveness property we are interested in is
 eventually reaching a state labeled with $\ok$. More precisely, we are interested in the probability of eventually seeing $\ok$. The following invariant-like witness notion gives a guaranteed lower bound for the probability in question. It is derived from the IF/I coincidence; unlike ranking supermartingales, it does not  use natural numbers or ordinals.

\begin{mydefinition}[IF/I submartingale] \label{def:sub}
  Let $t$ be a finite probabilistic binary tree; the set of its nodes is denoted by $N$.
  We say $f: N \to [0, 1]$ is an \emph{IF/I submartingale} if it satisfies the following.
  \begin{enumerate}
    \item $f(n) = 0$ for each leaf node $n$.
    \item For each internal node $n$ labeled with $\notok$,
   let its children and their edge labels be as shown in~(\ref{eq:probBinTree}). Then we have
      \[
      f(n) \leq p\cdot f(n_1)+(1-p)\cdot f(n_2).
      \]
  \end{enumerate}
\end{mydefinition}
 The direction of the inequality is indeed that of a \emph{sub}martingale: the current value is a lower bound of the expected next value.
Note that there is no condition for $f(n)$ if $n$ is an internal node labeled with $\ok$. In this case, $f(n)$ can be set to $1$ to improve the lower bound.

\begin{mytheorem}\label{thm:probLivenessBySubmartingale}
  In the setting of Def.~\ref{def:sub},
  assume $f$ is an IF/I submartingale.
  Then, identifying the tree $t$ with the Markov chain with suitable probabilistic branching, the probability of eventually reaching a node labeled with $\ok$ from the root is at least $f(r_t)$ where $r_t$ is the root node of $t$.
  \myqed
\end{mytheorem}
\noindent
The proof of Thm.~\ref{thm:probLivenessBySubmartingale}	is in Appendix~\ref{appendix:prf:thm:probLivenessBySubmartingale}.
The main step is to apply the following to Thm.~\ref{thm:general1} in order to obtain a categorical data $(d^{[0, 1]}, F^\ptr, \dot{F}^\ptr)$ satisfying the \cost{}:
\begin{itemize}
  \item a complete lattice $\Omega$ is $[0, 1]$ with the usual order between real numbers;
  \item a set functor $F$ is $F^{\ptr}=\mathbf{1}+\{\ok, \notok\} \times [0, 1] \times (-)^2$; and
  \item a monotone $F^\ptr$-algebra $\sigma: F^\Sigma [0, 1] \to [0, 1]$
    is $\sigma^{\ptr}$ defined as follows:
    \begin{align*}
      \sigma^\ptr(x) &=
      \begin{cases}
        0 & \text{if } x=* \in \mathbf{1} \\
        1 & \text{if } x=(\ok, p, a, b) \\
        pa + (1-p)b & \text{if } x=(\notok, p, a, b).
      \end{cases}
    \end{align*}
\end{itemize}

\section{Application 2:  Witnesses for Bottom-Up Tree Automata} \label{sec:treeUnified}
We present an application of the \cost{} in~\S{}\ref{sec:ip} to tree automata, using the results in~\S{}\ref{sec:general} as an interface.
In this paper we restrict to \emph{bottom-up} tree automata, although a similar theory can be developed for
 top-down ones.

We restrict the ranked alphabet $\Sigma$ used for trees to
  $\Sigma=\Sigma_0\cup\Sigma_2$, where operations in $\Sigma_0$ are $0$-ary and those in $\Sigma_2$ are binary. This restriction is for simplicity and not essential.

\begin{mydefinition}[(finite) $\Sigma$-trees]\label{def:alphabet}

  A \emph{$\Sigma$-tree} $t$ is a tuple $(N, r_t, c_t)$
  where $N$ is a set of nodes, $r_t \in N$ is a root node and $c_t: N \to \Sigma_0 + \Sigma_2 \times N \times N$ is a function which determines labels and next nodes:
  if $c_t(n) = s \in \Sigma_0$ then $n$ is a leaf node labeled with $s \in \Sigma_0$,
  and if $c_t(n)=(s, n_1, n_2)$ then $n$ is an internal node labeled with $s \in \Sigma_2$ and the next nodes of $n$ are $n_1$ and $n_2$.
A \emph{finite $\Sigma$-tree} is a $\Sigma$-tree which has only finitely many nodes.
\end{mydefinition}
\begin{mydefinition}[bottom-up tree automaton] \label{def:bottomUpTreeAutom}
  A \emph{bottom-up tree automaton} is a quadruple
$\mathcal{A}=(\Sigma_0\cup\Sigma_2,Q,\delta,q_\mathsf{F})$,
  where
  1) $\Sigma_0\cup\Sigma_2$ is a ranked alphabet;
  2)  $Q$ is a set of states;
  3)  $\delta\colon \Sigma_0 + \Sigma_2\times Q\times Q\to \Pf Q$
    is a transition function (note the nondeterminism modeled by the powerset $\Pf Q$); and
  4) $q_\mathsf{F} \in Q$ is an accepting state.

A \emph{run} of $\mathcal{A}$ over a $\Sigma$-tree $t$ is a function $\rho$
from nodes $n$ of $t$ to states $\rho(n)\in Q$
such that 1)
$\rho(n) \in \delta(s)$ for each leaf node $n$ with $c_t(n)=s$, and 2)
 $\rho(n) \in \delta\bigl(s, \rho(n_1), \rho(n_2)\bigr)$
 for each internal node $n$ with $c_t(n)=(s, n_1, n_2)$.

A finite $\Sigma$-tree $t$ is \emph{accepted} by $\mathcal{A}$ if there is a run $\rho$ of $\mathcal{A}$ over $t$ such that $\rho(r_t) = q_\mathsf{F}$.
\end{mydefinition}
\noindent Note that allowing multiple accepting states does not change the theory because of the nondeterminism in transition functions.

\vspace{.3em}
\myparagraph{Upside-Down Witness for Acceptance}
For an acceptance of a single $\Sigma$-tree by a bottom-up tree automaton, the \cost{} in~\S{}\ref{sec:ip} and \S\ref{sec:general1} (the pure setting) yields the following (invariant-like, top-down) witness notion.
\begin{mydefinition}%
\label{def:ac_inv}
  Let $\mathcal{A} = (\Sigma_0 \cup \Sigma_2, Q, \delta, q_\mathsf{F})$ be a bottom-up tree automaton, and $t=(N, r_t, c_t)$ be a finite $\Sigma$-tree.
  We say $f: N \to \Pf Q$ is an \emph{acceptance invariant} if
  \begin{enumerate}
    \item \label{item:buta_tree_1}
      for each leaf node $n$ with $c_t(n)=s$, we have $f(n)\subseteq \delta(s)$;
   \item \label{item:buta_tree_2}
     for each internal node $n$ with $c_t(n)=(s, n_1, n_2)$,
     we have $f(n)\subseteq\bigcup_{q_1 \in f(n_1), q_2 \in f(n_2)} \delta\bigl(s, q_1, q_2\bigr)$;
    \item \label{item:buta_tree_3} for the root node $r_t$ of $t$, we have
      $q_\mathsf{F} \in f(r_t)$.
  \end{enumerate}
\end{mydefinition}
\noindent An acceptance invariant assigns a \emph{predicate} $f(n)$ to each node $n$, and the constraints on $f$ runs in the top-down manner. The proof of Thm.~\ref{thm:buta_tree} is in Appendix~\ref{ap:buta_tree}, where we identify suitable categorical constructs (a fibration and functors) and apply the results in \S\ref{sec:general1}.
\begin{mytheorem}[acceptance witness for a finite tree] \label{thm:buta_tree}
  In the setting of Def.~\ref{def:ac_inv},
  if there exists an acceptance invariant $f: N \to \Pf Q$,
  then $\mathcal{A}$ accepts the finite $\Sigma$-tree $t$. \myqed
\end{mytheorem}

\vspace{.3em}
\myparagraph{Upside-Down Witness for Model Checking} We extend the above theory from acceptance (of a single tree) to \emph{model checking} (whether every tree generated by a system is accepted). Besides its practical relevance, the model checking problem is categorically interesting. Specifically, for the results here, we use the extended categorical framework in \S\ref{sec:general2} (\cost{} in presence of \emph{effects}) and Thm.~\ref{thm:cost_lift} (coincidence lifting).

\begin{mydefinition}[generative tree automaton $\mathcal{C}$, its language $L^{\mathrm{fin}}_{\mathcal{C}}$, and model checking]\label{def:generativeTreeAutom}
  A \emph{generative tree automaton} is
  $\mathcal{C}=(\Sigma_0\cup\Sigma_2,X,c,x_0)$, where
 1)
 $\Sigma_0\cup\Sigma_2$ is a ranked alphabet;
 2)
 $X$ is a set of states;
 3)
 $c\colon X\to \Pf(\Sigma_0 + \Sigma_2\times X\times X)$ is a transition function (note the powerset operator $\Pf$); and
 4)
 $x_0\in X$ is an initial state.

Let $t=(N, r_t, c_t)$ be a (possibly infinite) $\Sigma$-tree.
A \emph{run} of $\mathcal{C}$ over $t$ is a function $\rho\colon N\to X$, assigning a state to each node,
such that
1) $\rho(r_t)=x_0$ for the root node $r_t$;
2) $s\in c(\rho(n))$ for each leaf node $n$ with $c_t(n)=s$; and
3)
 $\bigl(s, \rho(n_{1}), \rho(n_{2})\bigr)\in c(\rho(n))$ for each internal node $n$ with $c_t(n)=(s, n_1, n_2)$.

We say that a $\Sigma$-tree $t$ is \emph{generated} by $\mathcal{C}$ if there is a run $\rho$ of $\mathcal{C}$ over $t$. The set of all $\Sigma$-trees generated by $\mathcal{C}$ is denoted by $L_{\mathcal{C}}$; the set of all \emph{finite}  $\Sigma$-trees generated by $\mathcal{C}$ is $L^{\mathrm{fin}}_{\mathcal{C}}$.

The \emph{model checking problem} takes
 a generative tree automaton $\mathcal{C}$ and
 a bottom-up tree automaton  $\mathcal{A}$  (Def.~\ref{def:bottomUpTreeAutom}) as input, and asks if every finite $\Sigma$-tree in $L^{\mathrm{fin}}_{\mathcal{C}}$ is accepted by $\mathcal{A}$.
\end{mydefinition}
Note that we restrict to finite trees here. One possible justification is an external mechanism that forces termination, much like in~\S\ref{sec:live}.

Our general theory of the \cost{} derives the following (invariant-like, top-down) witness notion for model checking (where the specification is a bottom-up tree automaton).
\begin{mydefinition}[model checking invariant] \label{def:mo_inv}
  Let $\mathcal{A} = (\Sigma_0 \cup \Sigma_2, Q, \delta, q_\mathsf{F})$ be a bottom-up tree automaton,
  and let $\mathcal{C}=(\Sigma_0\cup\Sigma_2,X,c,x_0)$ be a generative tree automaton.
  We say $f: X \to \Pf Q$ is a \emph{model checking invariant} if it satisfies the following.
  \begin{enumerate}
    \item \label{item:buta_trees_1}
      $f(x) \subseteq \bigcap_{a \in c(x)} \delta_f(a)$  for each $x\in X$. Here
     $\delta_f: \Sigma_0+\Sigma_2 \times X\times X \to \mathcal{P}Q$ is
   defined by
1)
$\delta_f(s) = \delta(s)$ for $s\in\Sigma_0$;
2)
$\delta_f(s, x_1, x_2) = \textstyle\bigcup_{q_1 \in f(x_1), q_2 \in f(x_2)} \delta(s, q_1, q_2)$
for $s\in\Sigma_2$.
    \item \label{item:buta_trees_2}
      $q_\mathsf{F} \in f(x_{0})$.
  \end{enumerate}
\end{mydefinition}
\begin{mytheorem}%
\label{thm:buta_trees}
  In the setting of Def.~\ref{def:mo_inv},
  assume that there exists a model checking invariant $f: X \to \Pf Q$.
  Then, $\mathcal{A}$ accepts every finite $\Sigma$-tree $t\in L^{\mathrm{fin}}_{\mathcal{C}}$ generated by $\mathcal{C}$.
  \qed
\end{mytheorem}
The proof is in Appendix~\ref{ap:buta_trees}. The nondeterminism on the system side ($\mathcal{C}$ in Def.~\ref{def:generativeTreeAutom}) requires  to work in the effectful setting (\S{}\ref{sec:general2}). Another challenge is that the relevant functor lifting is not fibred (cf.\ the last item in Def.~\ref{def:kleisli_setting}); we use  the coincidence lifting (Thm.~\ref{thm:cost_lift}) to deal with it, where the required base coincidence comes from coalgebraic trace semantics~\cite{HasuoJS07b}.

  \section{Conclusions and Future Work\conf{70}}
  We presented our \cost{}, which is a general categorical framework for the coincidence of initial algebras and final coalgebras, a classic topic in computer science.
  The \cost{} is formulated in fibrational terms, and this occurs in the fiber over an initial algebra; it is therefore understood as the coincidence of logical lfp and gfp specifications. Relying on mild assumptions of fibred liftings and stable chain colimits, the \cost{} accommodates many examples. As applications, we derived seemingly new verification methods for probabilistic liveness and tree automata.

  The proofs in~\S{}\ref{sec:live}--\ref{sec:treeUnified} suggest the possibility of a structural theory of the IA-FC coincidence, where unique fixed points are pulled back along coalgebra homomorphisms. We will pursue this structural theory, together with its practical consequences.

  Another direction of future work is to formalize the relationship between the current fibrational approach to the IA-FC coincidence, and the homset enrichment approach in~\cite{SmythP82,Freyd91,Fiore96b,Barr92,HasuoJS07b}.

\bibliography{mybib}

\begin{thebibliography}{10}

\bibitem{Adamek74}
Ji{\v{r}}{\'\i} Ad{\'a}mek.
\newblock Free algebras and automata realizations in the language of
  categories.
\newblock {\em Commentationes Mathematicae Universitatis Carolinae},
  15(4):589--602, 1974.

\bibitem{Adamek94}
Ji{\v{r}}{\'\i} Ad{\'a}mek, J~Adamek, J~Rosicky, et~al.
\newblock {\em Locally presentable and accessible categories}, volume 189.
\newblock Cambridge University Press, 1994.

\bibitem{AdamekMM18}
Jir{\'{\i}} Ad{\'{a}}mek, Stefan Milius, and Lawrence~S. Moss.
\newblock Fixed points of functors.
\newblock {\em J. Log. Algebraic Methods Program.}, 95:41--81, 2018.
\newblock \href {https://doi.org/10.1016/j.jlamp.2017.11.003}
  {\path{doi:10.1016/j.jlamp.2017.11.003}}.

\bibitem{AguirreK20}
Alejandro Aguirre and Shin{-}ya Katsumata.
\newblock Weakest preconditions in fibrations.
\newblock {\em Electronic Notes in Theoretical Computer Science}, 352:5 -- 27,
  2020.
\newblock The 36th Mathematical Foundations of Programming Semantics
  Conference, 2020.
\newblock URL:
  \url{http://www.sciencedirect.com/science/article/pii/S1571066120300487},
  \href {https://doi.org/https://doi.org/10.1016/j.entcs.2020.09.002}
  {\path{doi:https://doi.org/10.1016/j.entcs.2020.09.002}}.

\bibitem{Barr92}
Michael Barr.
\newblock Algebraically compact functors.
\newblock {\em Journal of Pure and Applied Algebra}, 82(3):211--231, 1992.

\bibitem{BirdM97}
Richard~S. Bird and Oege de~Moor.
\newblock {\em Algebra of programming}.
\newblock Prentice Hall International series in computer science. Prentice
  Hall, 1997.

\bibitem{BonchiKP18}
Filippo Bonchi, Barbara K{\"{o}}nig, and Daniela Petrisan.
\newblock Up-to techniques for behavioural metrics via fibrations.
\newblock {\em CoRR}, abs/1806.11064, 2018.
\newblock URL: \url{http://arxiv.org/abs/1806.11064}, \href
  {http://arxiv.org/abs/1806.11064} {\path{arXiv:1806.11064}}.

\bibitem{CaprettaUV09}
Venanzio Capretta, Tarmo Uustalu, and Varmo Vene.
\newblock Corecursive algebras: {A} study of general structured corecursion.
\newblock In Marcel Vin{\'{\i}}cius~Medeiros Oliveira and Jim Woodcock,
  editors, {\em Formal Methods: Foundations and Applications, 12th Brazilian
  Symposium on Formal Methods, {SBMF} 2009, Gramado, Brazil, August 19-21,
  2009, Revised Selected Papers}, volume 5902 of {\em Lecture Notes in Computer
  Science}, pages 84--100. Springer, 2009.
\newblock \href {https://doi.org/10.1007/978-3-642-10452-7\_7}
  {\path{doi:10.1007/978-3-642-10452-7\_7}}.

\bibitem{CarboniLW93}
Aurelio Carboni, Stephen Lack, and R.F.C. Walters.
\newblock Introduction to extensive and distributive categories.
\newblock {\em Journal of Pure and Applied Algebra}, 84(2):145 -- 158, 1993.
\newblock URL:
  \url{http://www.sciencedirect.com/science/article/pii/002240499390035R},
  \href {https://doi.org/https://doi.org/10.1016/0022-4049(93)90035-R}
  {\path{doi:https://doi.org/10.1016/0022-4049(93)90035-R}}.

\bibitem{ChakarovS13}
Aleksandar Chakarov and Sriram Sankaranarayanan.
\newblock Probabilistic program analysis with martingales.
\newblock In Natasha Sharygina and Helmut Veith, editors, {\em Computer Aided
  Verification - 25th International Conference, {CAV} 2013, Saint Petersburg,
  Russia, July 13-19, 2013. Proceedings}, volume 8044 of {\em Lecture Notes in
  Computer Science}, pages 511--526. Springer, 2013.
\newblock \href {https://doi.org/10.1007/978-3-642-39799-8\_34}
  {\path{doi:10.1007/978-3-642-39799-8\_34}}.

\bibitem{Fiore96b}
Marcelo~P. Fiore.
\newblock {\em Axiomatic Domain Theory in Categories of Partial Maps}.
\newblock Distinguished Dissertations in Computer Science. Cambridge University
  Press, 1996.
\newblock \href {https://doi.org/10.1017/CBO9780511526565}
  {\path{doi:10.1017/CBO9780511526565}}.

\bibitem{Freyd91}
Peter Freyd.
\newblock Algebraically complete categories.
\newblock In Aurelio Carboni, Maria~Cristina Pedicchio, and Guiseppe Rosolini,
  editors, {\em Category Theory}, pages 95--104, Berlin, Heidelberg, 1991.
  Springer Berlin Heidelberg.

\bibitem{Freyd90}
Peter~J. Freyd.
\newblock Recursive types reduced to inductive types.
\newblock In {\em Proceedings of the Fifth Annual Symposium on Logic in
  Computer Science {(LICS} '90), Philadelphia, Pennsylvania, USA, June 4-7,
  1990}, pages 498--507. {IEEE} Computer Society, 1990.
\newblock \href {https://doi.org/10.1109/LICS.1990.113772}
  {\path{doi:10.1109/LICS.1990.113772}}.

\bibitem{HasuoCKJ13}
Ichiro Hasuo, Kenta Cho, Toshiki Kataoka, and Bart Jacobs.
\newblock Coinductive predicates and final sequences in a fibration.
\newblock In Dexter Kozen and Michael~W. Mislove, editors, {\em Proceedings of
  the Twenty-ninth Conference on the Mathematical Foundations of Programming
  Semantics, {MFPS} 2013, New Orleans, LA, USA, June 23-25, 2013}, volume 298
  of {\em Electronic Notes in Theoretical Computer Science}, pages 197--214.
  Elsevier, 2013.
\newblock \href {https://doi.org/10.1016/j.entcs.2013.09.014}
  {\path{doi:10.1016/j.entcs.2013.09.014}}.

\bibitem{HasuoJS07b}
Ichiro Hasuo, Bart Jacobs, and Ana Sokolova.
\newblock Generic trace semantics via coinduction.
\newblock {\em Log. Methods Comput. Sci.}, 3(4), 2007.
\newblock \href {https://doi.org/10.2168/LMCS-3(4:11)2007}
  {\path{doi:10.2168/LMCS-3(4:11)2007}}.

\bibitem{HeindelS09}
Tobias Heindel and Pawel Sobocinski.
\newblock Van kampen colimits as bicolimits in span.
\newblock In Alexander Kurz, Marina Lenisa, and Andrzej Tarlecki, editors, {\em
  Algebra and Coalgebra in Computer Science, Third International Conference,
  {CALCO} 2009, Udine, Italy, September 7-10, 2009. Proceedings}, volume 5728
  of {\em Lecture Notes in Computer Science}, pages 335--349. Springer, 2009.
\newblock \href {https://doi.org/10.1007/978-3-642-03741-2\_23}
  {\path{doi:10.1007/978-3-642-03741-2\_23}}.

\bibitem{Herrlich74}
Horst Herrlich.
\newblock Topological functors.
\newblock {\em General Topology and its Applications}, 4(2):125--142, 1974.

\bibitem{Jacobs16coalgBook}
Bart Jacobs.
\newblock {\em Introduction to Coalgebra: Towards Mathematics of States and
  Observation}, volume~59 of {\em Cambridge Tracts in Theoretical Computer
  Science}.
\newblock Cambridge University Press, 2016.
\newblock \href {https://doi.org/10.1017/CBO9781316823187}
  {\path{doi:10.1017/CBO9781316823187}}.

\bibitem{CLTT}
Bart~P. Jacobs.
\newblock {\em Categorical Logic and Type Theory}, volume 141 of {\em Studies
  in logic and the foundations of mathematics}.
\newblock North-Holland, 2001.
\newblock URL:
  \url{http://www.elsevierdirect.com/product.jsp?isbn=9780444508539}.

\bibitem{Komorida19}
Yuichi Komorida, Shin{-}ya Katsumata, Nick Hu, Bartek Klin, and Ichiro Hasuo.
\newblock Codensity games for bisimilarity.
\newblock In {\em 34th Annual {ACM/IEEE} Symposium on Logic in Computer
  Science, {LICS} 2019, Vancouver, BC, Canada, June 24-27, 2019}, pages 1--13.
  {IEEE}, 2019.
\newblock \href {https://doi.org/10.1109/LICS.2019.8785691}
  {\path{doi:10.1109/LICS.2019.8785691}}.

\bibitem{Lurie09}
Jacob Lurie.
\newblock {\em Higher Topos Theory (AM-170)}.
\newblock Princeton University Press, 2009.
\newblock URL: \url{http://www.jstor.org/stable/j.ctt7s47v}.

\bibitem{Moggi91a}
E.~Moggi.
\newblock Notions of computation and monads.
\newblock {\em Inf. \& Comp.}, 93(1):55--92, 1991.

\bibitem{Mulry93}
Philip~S. Mulry.
\newblock Lifting theorems for kleisli categories.
\newblock In Stephen~D. Brookes, Michael~G. Main, Austin Melton, Michael~W.
  Mislove, and David~A. Schmidt, editors, {\em Mathematical Foundations of
  Programming Semantics, 9th International Conference, New Orleans, LA, USA,
  April 7-10, 1993, Proceedings}, volume 802 of {\em Lecture Notes in Computer
  Science}, pages 304--319. Springer, 1993.
\newblock \href {https://doi.org/10.1007/3-540-58027-1\_15}
  {\path{doi:10.1007/3-540-58027-1\_15}}.

\bibitem{Pitts96}
Andrew~M. Pitts.
\newblock Relational properties of domains.
\newblock {\em Inf. Comput.}, 127(2):66--90, 1996.
\newblock \href {https://doi.org/10.1006/inco.1996.0052}
  {\path{doi:10.1006/inco.1996.0052}}.

\bibitem{Rutten00a}
J.~J. M.~M. Rutten.
\newblock Universal coalgebra: a theory of systems.
\newblock {\em Theoretical Computer Science}, 249:3--80, 2000.

\bibitem{SmythP82}
Michael~B. Smyth and Gordon~D. Plotkin.
\newblock The category-theoretic solution of recursive domain equations.
\newblock {\em {SIAM} J. Comput.}, 11(4):761--783, 1982.
\newblock \href {https://doi.org/10.1137/0211062} {\path{doi:10.1137/0211062}}.

\bibitem{SprungerKDH18}
David Sprunger, Shin{-}ya Katsumata, J{\'{e}}r{\'{e}}my Dubut, and Ichiro
  Hasuo.
\newblock Fibrational bisimulations and quantitative reasoning.
\newblock In Corina C{\^{\i}}rstea, editor, {\em Coalgebraic Methods in
  Computer Science - 14th {IFIP} {WG} 1.3 International Workshop, {CMCS} 2018,
  Colocated with {ETAPS} 2018, Thessaloniki, Greece, April 14-15, 2018, Revised
  Selected Papers}, volume 11202 of {\em Lecture Notes in Computer Science},
  pages 190--213. Springer, 2018.
\newblock \href {https://doi.org/10.1007/978-3-030-00389-0\_11}
  {\path{doi:10.1007/978-3-030-00389-0\_11}}.

\bibitem{TakisakaOUH18}
Toru Takisaka, Yuichiro Oyabu, Natsuki Urabe, and Ichiro Hasuo.
\newblock Ranking and repulsing supermartingales for reachability in
  probabilistic programs.
\newblock In {\em Automated Technology for Verification and Analysis - 16th
  International Symposium, {ATVA} 2018, Los Angeles, CA, USA, October 7-10,
  2018, Proceedings}, pages 476--493, 2018.
\newblock \href {https://doi.org/10.1007/978-3-030-01090-4\_28}
  {\path{doi:10.1007/978-3-030-01090-4\_28}}.

\bibitem{Trnkova75}
V{\v{e}}ra Trnkov{\'a}, Ji{\v{r}}{\'\i} Ad{\'a}mek, V{\'a}clav Koubek, and Jan
  Reiterman.
\newblock Free algebras, input processes and free monads.
\newblock {\em Commentationes Mathematicae Universitatis Carolinae},
  16(2):339--351, 1975.

\bibitem{UrabeHH17}
Natsuki Urabe, Masaki Hara, and Ichiro Hasuo.
\newblock Categorical liveness checking by corecursive algebras.
\newblock In {\em 32nd Annual {ACM/IEEE} Symposium on Logic in Computer
  Science, {LICS} 2017, Reykjavik, Iceland, June 20-23, 2017}, pages 1--12.
  {IEEE} Computer Society, 2017.
\newblock \href {https://doi.org/10.1109/LICS.2017.8005151}
  {\path{doi:10.1109/LICS.2017.8005151}}.

\bibitem{Zamdzhiev19}
Vladimir Zamdzhiev.
\newblock Reflecting algebraically compact functors.
\newblock In John Baez and Bob Coecke, editors, {\em Proceedings Applied
  Category Theory 2019, {ACT} 2019, University of Oxford, UK, 15-19 July 2019},
  volume 323 of {\em {EPTCS}}, pages 15--23, 2019.
\newblock \href {https://doi.org/10.4204/EPTCS.323.2}
  {\path{doi:10.4204/EPTCS.323.2}}.

\end{thebibliography}

\appendix

\section{\CoSt{} for general fibrations} \label{subsec:ip}
Here we show the following theorem similar to Thm.~\ref{thm:cost_clat} for general fibrations.
\begin{mytheorem}[\cost{} for general fibrations] \label{thm:cost} %
  Let $p: \mathbb{E} \to \mathbb{B}$ be a fibration; assume that
  $\mathbb{E}$ and $\mathbb{B}$ are chain-cocomplete.
   Let $\dot{F}$ be a lifting of $F$ along $p$.
   Assume further the following conditions:
  \begin{enumerate}
    \item \label{item:converge}
      The initial chain of $\dot{F}$ converges.
    \item $\dot{F}$ is a fibred lifting of $F$.
    \item \label{item:stability} $p$ has stable chain colimits.
   \item  \label{item:pPreservesColimits} $p$ strictly preserves chain colimits.
    \item \label{item:substPreservesColimits} Substitution  in $p$ preserves chain colimits in fibers.
  \end{enumerate}
  Then $(p, F, \dot{F})$ satisfies the \cost{}.
\end{mytheorem}
\noindent
The theorem follows from the next technical observation.
\begin{myproposition} \label{prop:main}
  In the setting of Thm.~\ref{thm:cost},
  assume further that the initial chain of $\dot{F}$ converges in $\lambda$ steps.
  Consider the initial chains:
\begin{equation*}
 \vcenter{\xymatrix@R=.8em@C-.5em{
  {\mathbb{E}}
  \ar[d]_{p}
  &
  {0}
  \ar[r]^-{\dot{\alpha}_{0,1}}
  &
  {\dot{F}0}
  \ar[r]^-{\dot{\alpha}_{1,2}}
  &
  \cdots
   \ar[r]
  &
  {\dot{F}^{\lambda}0}
  \ar@{->}[r]_-{\cong}^-{\dot{\alpha}_{\lambda,\lambda+1}}
  &
  {\dot{F}^{\lambda+1}0}
  \ar[r]
  &
  \cdots
  \\
  {\mathbb{B}}
  &
  {0}
  \ar[r]^-{\alpha_{0,1}}
  &
  {F0}
  \ar[r]^-{\alpha_{1,2}}
  &
  \cdots
   \ar[r]
  &
  {F^{\lambda}0}
  \ar@{->}[r]_-{\cong}^-{\alpha_{\lambda,\lambda+1}}
  &
  {\dot{F}^{\lambda+1}0}
  \ar[r]
  &
  \cdots
 }}
\end{equation*}
where $0$ in the two lines denote initial objects in $\mathbb{E}$ and in $\mathbb{B}$, respectively.
Note that both $\dot{\alpha}_{\lambda, \lambda+1}$ and $\alpha_{\lambda, \lambda+1}$ are isomorphisms: the former is  by the convergence assumption; the latter is by the assumption and Lem.~\ref{lem:init}. Therefore their inverses are initial algebras by Prop.~\ref{prop:converge_to_initial}.

 In this setting,  $\dot{\alpha}_{\lambda, \lambda+1}$ is a final object of $\Coalg{\dot{F}}_{\alpha_{\lambda, \lambda+1}}$ (cf.\ Def.~\ref{def:cost}).
\end{myproposition}

From now on, we aim to prove Prop.~\ref{prop:main}.

To show finality of $\dot{\alpha}_{\lambda, \lambda+1}$ in $\Coalg{\dot{F}}_{\alpha_{\lambda, \lambda+1}}$,
we first claim the existence of a morphism from an arbitrary $\dot{F}$-coalgebra $\gamma: P \to \dot{F}P$ in $\Coalg{\dot{F}}_{\alpha_{\lambda, \lambda+1}}$ to $\dot{\alpha}_{\lambda, \lambda+1}$.
The next lemma shows a construction of such a morphism $p_\lambda$
by transfinite induction along initial chains.
\begin{displaymath}
 \vcenter{\xymatrix@R=.6em@C-.4em{
    \mathbb{E} \ar[dd]^p
    &\dot{F}^\lambda0 \ar[r]^{\dot{\alpha}_{\lambda, \lambda+1}} &\dot{F}^{\lambda+1}0 \\
    &P \ar[u]^{p_\lambda} \ar[r]^\gamma &\dot{F}P \ar[u]_{\dot{F}p_\lambda} \\
    \mathbb{B}
    &F^\lambda0 \ar[r]^{\alpha_{\lambda, \lambda+1}} &F^{\lambda+1}0 \\
  }}
\end{displaymath}
This construction
exploits
the singleton property of $\mathbb{E}_0$ (Lem.~\ref{lem:init_obj}) in the base case,
fibredness of the lifting $\dot{F}$ in the step case,
and stability of chain colimits in the limit case.
\begin{mylemma} \label{lem:exist}
  Let $p: \mathbb{E} \to \mathbb{B}$ be a fibration with stable chain colimits.
  Assume that $\mathbb{E}$ and $\mathbb{B}$ are chain-cocomplete and $p$ strictly preserves chain colimits.
   Let $\dot{F}$ be a fibred lifting of $F$ along $p$.

  For each ordinal $\lambda$ and each coalgebra $\gamma: P \to \dot{F}P$ above $\alpha_{\lambda, \lambda+1}$ (or equivalently, $\gamma \in \Coalg{\dot{F}}_{\alpha_{\lambda, \lambda+1}}$),
  there exists a morphism $p_\lambda: \gamma \to \dot{\alpha}_{\lambda, \lambda+1}$ in $\Coalg{\dot{F}}_{\alpha_{\lambda, \lambda+1}}$.
\end{mylemma}
\begin{myproof}
  Let $(k_{j, i}: \alpha_{j, \lambda}^*P \to \alpha_{i, \lambda}^*P)_{j \leq i \leq \lambda}$ denote those morphisms induced from $(\alpha_{i, \lambda}^*P \to P)_{i \leq \lambda}$ via their universality as cartesian liftings.
  We construct vertical morphisms $(p_i: \alpha_{i, \lambda}^*P \to \dot{F}^i0)_{i \leq \lambda}$
  such that $p_i \circ k_{j, i} = \dot{\alpha}_{j, i} \circ p_j$
  for all $i, j$ with $j \leq i \leq \lambda$.
  Such a $(p_i)_{i \leq \lambda}$ makes the diagram below commute.
  (The rightmost square commutes, see Appendix~\ref{ap:exist}.\ref{item:exist2} for details.)
\begin{displaymath}
  \xymatrix@R=.8em@C-.9em{
  \mathbb{E} \ar[dd]^{p} &0 \ar[r]^{{\dot{\alpha}_{0, 1}}} &\dot{F}0 \ar[r]^{{\dot{\alpha}_{1, 2}}} &\cdots \ar[r] &\dot{F}^\lambda 0 \ar[r]^{\dot{\alpha}_{\lambda, \lambda+1}} &\dot{F}^{\lambda+1}0 \\
                         &\alpha_{0, \lambda}^*P \ar[u]^{p_0} \ar[r]^{{k_{0, 1}}} &\alpha_{1, \lambda}^*P \ar[u]^{p_1} \ar[r]^{{k_{1, 2}}}&\cdots \ar[r] &{\alpha_{\lambda, \lambda}^*P = P} \ar[u]^{p_\lambda} \ar[r]^(.6){\gamma} &\dot{F}P \ar[u]^{\dot{F}p_\lambda}\\
  \mathbb{B} &0 \ar[r]^{{\alpha_{0, 1}}} &F0 \ar[r]^{{\alpha_{1, 2}}} &\cdots \ar[r] &F^\lambda 0 \ar[r]^{\alpha_{\lambda, \lambda+1}} &F^{\lambda+1}0
 }
\end{displaymath}
Then this $p_\lambda$ is what we want.
  The construction of $(p_i)_{i \leq \lambda}$ is  by the following transfinite induction on $i$.
  \begin{itemize}
    \item (Base case)
      Lem.~\ref{lem:init_obj} says $\alpha_{0, \lambda}^*P \cong 0$ in $\mathbb{E}$
      and this isomorphism is vertical
      because both $p\alpha^*_{0, \lambda}P$ and $p0$ are $0$ in $\mathbb{B}$.
      We define $p_0$ as this isomorphism.

    \item (Step case)
      If $i$ is a successor ordinal, we define $p_i$ by
      \begin{equation*}
        \alpha_{i, \lambda}^*P
          \xrightarrow{(\star)} (F\alpha_{{i-1}, \lambda})^*\dot{F}P
          \xrightarrow[\cong]{\xi} \dot{F}\alpha_{{i-1}, \lambda}^*P
          \xrightarrow{\dot{F}p_{i-1}} \dot{F}^i0
      \end{equation*}
      where $\xi$ is from fibredness of $\dot{F}$ and $(\star)$ is induced as follows by universality of a cartesian lifting.

\begin{displaymath}
        \xymatrix@R=.8em@C-1em{
           \alpha_{i, \lambda}^*P \ar[r]_-{\overline{\alpha_{i, \lambda}}} \ar@{.>}[d]^{(\star)} &P \ar@/^0.5pc/[dr]^{\gamma} \\
            (F\alpha_{{i-1}, \lambda})^*\dot{F}P \ar[rr]^-{\overline{F\alpha_{i-1, \lambda}}} & &\dot{F}P \\
            F^i0 \ar[r]^{\alpha_{i, \lambda}} &F^\lambda 0 \ar[r]^(.4){\alpha_{\lambda, \lambda+1}} &F^{\lambda+1}0.
      }
\end{displaymath}
Note that $F\alpha_{i-1, \lambda} = \alpha_{\lambda, \lambda+1} \circ \alpha_{i, \lambda}$ by the definition of $\alpha$.
      We can prove $\dot{\alpha}_{j, i} \circ p_j = p_i \circ k_{j, i}$ for all $j \leq i$ by transfinite induction on $j$.
      See Appendix~\ref{ap:exist}.\ref{item:exist1}.
    \item (Limit case)
      If $i$ is a limit ordinal,
      we define $p_i$ by the stability of chain colimits.
      By applying chain colimit stability of $p$ to $\alpha_{i, \lambda}^*P$ above $F^i0 = \colim_{j < i} F^j 0$ (by Def.~\ref{def:init_chain}, see below),
      we have
      $\alpha_{i, \lambda}^*P \cong \colim_{j < i} \alpha_{j, i}^*(\alpha_{i, \lambda}^*P) \cong \colim_{j < i} \alpha_{j, \lambda}^*P$.
      For all $l,j$ with $l < j < i$, by the induction hypothesis, we have
      $\dot{\alpha}_{l, i} \circ p_l = \dot{\alpha}_{j, i} \circ \dot{\alpha}_{l, j} \circ p_l = \dot{\alpha}_{j, i} \circ p_j \circ k_{l, j}$.
      Hence $(\dot{F}^i0, (\alpha_{j, i} \circ p_j)_{j < i})$ is a cocone over $(j \mapsto \alpha_{j, \lambda}^*P)$, as shown below.
      We define $p_i$ as the mediating morphism from a colimit, as in the following diagram.
      \begin{displaymath}
        \xymatrix@R=.8em@C-.8em{
            &&&\dot{F}^i 0 \mathrlap{\text{ (colim.)}}\\
            \cdots \ar[r] &\dot{F}^l0 \ar@/^/[urr]^{\dot{\alpha}_{l, i}} \ar[r] &\dot{F}^j0 \ar@/^/[ur]_{\dot{\alpha}_{j, i}} \ar[r] &\cdots
       \\
            \cdots \ar[r] &\alpha_{l, \lambda}^*P \ar[u]^{p_l} \ar@/_/[drr]_{k_{l, i}} \ar[r] &\alpha_{j, \lambda}^*P \ar[u]^{p_j} \ar@/_/[dr]^{k_{j, i}} \ar[r] &\cdots
       \\
            &&&\alpha_{i, \lambda}^*P \mathrlap{\text{ (colim.)}} \ar@/_2em/@{.>}[uuu]_{p_i}
           \\ \cdots \ar[r]
        & F^{l}0 \ar[r]^-{\alpha_{l,j}}
        & F^{j}0 \ar[r]^-{\alpha_{j,i}}
        & F^{i}0 \ar[r]
        &\cdots
      }
    \end{displaymath}
  \end{itemize}
 This concludes the proof. \myqed
\end{myproof}

For Prop.~\ref{prop:main}, it remains to show the uniqueness of $p_\lambda: \gamma \to \dot{\alpha}_{\lambda, \lambda+1}$. The uniqueness does not immediately follow from the construction of $p_{\lambda}$ in Lem.~\ref{lem:exist}.

Our uniqueness proof (in the proof of Prop.~\ref{prop:main} shown later), we work on a suitable chain in the fiber $\mathbb{E}_{F^{\lambda}0}$, defined as follows.

  The following fact (cf.~\cite[Prop.~9.2.2 and Exercise~9.2.4]{CLTT}) shows $\clat$-fibrations have properties suitable for colimits.

\begin{myproposition} \label{prop:fiberwiseBaseAndTotalCoLimits}
  Let $p: \mathbb{E} \to \mathbb{B}$ be an opfibration.
  Assume the base category $\mathbb{B}$ has colimits of shape $\mathbb{I}$.
  Then the following statements are equivalent.
  \begin{enumerate}
    \item Each fiber of the opfibration $p$ has colimits of shape $\mathbb{I}$.
    \item The total category $\mathbb{E}$ has colimits of shape $\mathbb{I}$ and $p$ strictly preserves them.
  \myqed
  \end{enumerate}
\end{myproposition}

\begin{mynotation}[$\dot{\alpha}, \alpha$]\label{notation:alpha}
    In the setting of Lem.~\ref{lem:exist},
  let us fix $\lambda$ to be a converging ordinal of the initial $\dot{F}$-chain, in the sense that $\dot{\alpha}_{\lambda, \lambda+1}\colon \dot{F}^{\lambda}0\iso\dot{F}^{\lambda+1}0$ is an isomorphism.
  In the rest of the section, we write $\dot{\alpha}, \alpha$ for $\dot{\alpha}_{\lambda, \lambda+1}, \alpha_{\lambda, \lambda+1}$, respectively. Then $\alpha\colon F^{\lambda}0\iso F^{\lambda+1}0$ is an isomorphism, too.
\end{mynotation}

\begin{mydefinition}\label{def:chainV}
   In Prop.~\ref{prop:main},
   we define a chain
\begin{displaymath}
   P\xrightarrow{\beta_{0,1}} \asdf P\xrightarrow{\beta_{1,2}} (\asdf)^{2} P\xrightarrow{\beta_{2,3}}\cdots
\end{displaymath}
by repeated application of  $\asdf$. The whole chain resides in the fiber $\mathbb{E}_{F^{\lambda}0}$, as shown in~(\ref{eq:diagramASDF}).
 \begin{equation}\label{eq:diagramASDF}
 \vcenter{\xymatrix@R=.8em@C-1em{
        \vdots &\vdots \\
        (\asdf)^2P \ar[dr]|{\overline{\alpha}} \ar[u] &\dot{F}(\asdf)^2P \ar[u]  \\
        \asdf P \ar[u]^{\beta_{1, 2}} \ar[dr]|{\overline{\alpha}} &\dot{F}\asdf P \ar[u]_{\dot{F}\beta_{1, 2}} \\
        P \ar[u]^{\beta_{0, 1}} \ar[r]_\gamma &\dot{F}P \ar[u]_{\dot{F}\beta_{0, 1}} \\
        F^\lambda 0 \ar[r]^\alpha &F^{\lambda+1}0
    }}
\end{equation}

The precise definition is as follows. It is similar to Def.~\ref{def:init_chain}, but starting from $P$ (instead of from  $0$) calls for some care.
  \begin{itemize}
    \item (Objects)   $(\asdf)^{i}P$ is given for each $i\in\mathrm{Ord}$: $(\asdf)^{0}P = P$, $(\asdf)^{i+1}P = \asdf((\asdf)^{i}P)$, and $(\asdf)^{i}0 = \colim_{j < i} (\asdf)^{j}0$ for a limit ordinal $i$.
    \item (Morphisms) The morphism $\beta_{i,i+1}\colon (\asdf)^{i}0\to (\asdf)^{i+1}0$ for each ordinal $i$ is defined as follows.
      \begin{itemize}
        \item (Base case) $\beta_{0,1}: P \to \asdf P$ is induced from $\gamma\colon P\to \dot{F}P$ by universality of the cartesian lifting $\bar{\alpha}: \asdf P \to \dot{F}P$. See~(\ref{eq:diagramASDF}).
       \item (Step case) $\beta_{i+1,i+2}$ is defined by $\asdf\beta_{i,i+1}$.
       \item (Limit case) $\beta_{i,i+1}: (\asdf)^{i}P\to (\asdf)^{i+1}P$ for a limit ordinal $i$ is  induced by universality of $(\asdf)^i P= \colim_{j<i}(\asdf)^j P$.
         Prop.~\ref{prop:fiberwiseBaseAndTotalCoLimits} ensures this colimit vertex is above $F^\lambda 0$.
      \end{itemize}
      We have defined $\beta_{j,j+1}$ for each ordinal $j$. This induces morphisms  $\beta_{i,j}\colon (\asdf)^{i}P\to  (\asdf)^{j}P$ for each $i<j$ in a straight-forward manner: one repeats the step and limit cases; when $j$ is a limit ordinal, $\beta_{i,j}$ is the  cocone component to $(\asdf)^j P= \colim_{k<j}(\asdf)^k P$.
  \end{itemize}
\end{mydefinition}

\begin{mylemma} \label{lem:pii_iso}
In the setting of Prop.~\ref{prop:main}, let us
 assume that $\lambda$ is a converging ordinal in the initial chain of $\dot{F}$, and adopt Notation~\ref{notation:alpha}.

 Fig.~\ref{fig:lem} shows the following constructs.
\begin{itemize}
 \item
       The morphism $\gamma: P \to \dot{F}P$ above $\alpha$ induces the chain $P \xrightarrow{\beta_{0, 1}} \asdf P \xrightarrow{\beta_{1, 2}} (\asdf)^2P \to \cdots$ as in Def.~\ref{def:chainV}.
 \item
The last chain induces, for each ordinal $l$ such that $l<\lambda$, the chain
       \begin{displaymath}
   \alpha^{*}_{l,\lambda}P\xrightarrow{\alpha^{*}_{l,\lambda}\beta_{0,1}} \alpha^{*}_{l,\lambda}\asdf P\xrightarrow{\alpha^{*}_{l,\lambda}\beta_{1,2}} \alpha^{*}_{l,\lambda}(\asdf)^{2} P\rightarrow\cdots
       \end{displaymath}
       above $F^{l}0$, via the substitution along $\alpha_{l,\lambda}$.
 \item
  For each ordinal $i$,
  we obtain a $\dot{F}$-coalgebra as follows. It is above $\alpha$; it is denoted by  $\gamma_{i}$.
\begin{equation}\label{eq:gammai}
 \vcenter{\xymatrix@R=.8em@C+3em{
       (\asdf)^i P  \ar[r]^-{\gamma_{i}\; := \;\overline{\alpha} \circ \beta_{i, i+1}}
       &
       \dot{F}(\asdf)^i P
       \\
       F^{\lambda}0\ar[r]^-{\alpha}_-{\cong}
       &
F^{\lambda +1}0
    }}
  \end{equation}
 \item We
  apply Lem.~\ref{lem:exist} to the last coalgebras $\gamma_{i}$, using each of them  in place of the coalgebra $\gamma$ in Lem.~\ref{lem:exist}. Following the proof of Lem.~\ref{lem:exist}, we obtain vertical morphisms
  $(p_j^i: \alpha_{j, \lambda}^*(\asdf)^iP \to \dot{F}^j 0)_{j \leq \lambda}$ for each ordinal $i$.
\end{itemize}
 In this case, for each $i$ such that $i \leq \lambda$,
  (i) $p_i^i$ is an isomorphism;
  (ii) $p_i^l = p_i^m \circ \alpha_{i, \lambda}^*\beta_{l, m}$ for all $l, m$ with $l \leq m$.
  \myqed
\end{mylemma}
See Appendix~\ref{ap:pii_iso} for the proof.
\begin{figure*}[h]
  {\footnotesize
    \begin{displaymath}
    \xymatrix@R=2em@C=1.5em{
        &0 \ar[r] &\dot{F}0 \ar[r] &\dot{F}^20 \ar[r] &\cdots \ar[r] &\dot{F}^\lambda 0 \ar[r]^{\dot{\alpha}} &\dot{F}^{\lambda+1}0 \\
        &\vdots &\vdots &\vdots & &\vdots &\vdots \\
        \mathbb{E}\ar[ddd]^{p}&\alpha_{0, \lambda}^*(\asdf)^2P \ar[r] \ar[u] \ar@/^1pc/[uu]|{p_0^2} &\alpha_{1, \lambda}^*(\asdf)^2P \ar[r] \ar[u] \ar@/^1pc/[uu]|{p_1^2} &\alpha_{2, \lambda}^*(\asdf)^2P \ar[r] \ar[u] \ar@/^1pc/[uu]|{p_2^2} &\cdots \ar[r] &(\asdf)^2P \ar[dr]|{\overline{\alpha}} \ar[u] \ar@/^1pc/[uu]|{p_\lambda^2} &\dot{F}(\asdf)^2P \ar[u] \ar@/_1pc/[uu]|{\dot{F}p_\lambda^2} \\
        &
        \alpha_{0, \lambda}^*\asdf P \ar[r] \ar[u]_{\alpha_{0, \lambda}^*\beta_{1, 2}} \ar@/^2pc/[uuu]|{p_0^1} &\alpha_{1, \lambda}^*\asdf P \ar[r] \ar[u]_{\alpha_{1, \lambda}^*\beta_{1, 2}} \ar@/^2pc/[uuu]|{p_1^1} &\alpha_{2, \lambda}^*\asdf P \ar[r] \ar[u]_{\alpha_{2, \lambda}^*\beta_{1, 2}} \ar@/^2pc/[uuu]|{p_2^1} &\cdots \ar[r] &\asdf P \ar[u]|{\beta_{1, 2}} \ar@/^2pc/[uuu]|{p_\lambda^1} \ar[dr]|{\overline{\alpha}} &\dot{F}\asdf P \ar[u]|{\dot{F}\beta_{1, 2}} \ar@/_2pc/[uuu]|{\dot{F}p_\lambda^1} \\
        &
        \alpha_{0, \lambda}^*P \ar[r] \ar[u]_{\alpha_{0, \lambda}^*\beta_{0, 1}} \ar@/^3pc/[uuuu]|(.4){p_0^0} &\alpha_{1, \lambda}^*P \ar[r] \ar[u]_{\alpha_{1, \lambda}^*\beta_{0, 1}} \ar@/^3pc/[uuuu]|(.4){p_1^0} &\alpha_{2, \lambda}^*P \ar[r] \ar[u]_{\alpha_{2, \lambda}^*\beta_{0, 1}} \ar@/^3pc/[uuuu]|(.4){p_2^0} &\cdots \ar[r] &P \ar[u]|{\beta_{0, 1}} \ar[r]^\gamma \ar@/^3pc/[uuuu]|(.4){p_\lambda^0} &\dot{F}P \ar[u]|{\dot{F}\beta_{0, 1}} \ar@/_3pc/[uuuu]|(.4){\dot{F}p_\lambda^0} \\
        \mathbb{B}&0 \ar[r] &F0 \ar[r] &F^20 \ar[r] &\cdots \ar[r] &F^\lambda 0 \ar[r]^\alpha &F^{\lambda+1}0
    }
    \end{displaymath}
  }
  \caption{A diagram for Lem.~\ref{lem:pii_iso}.}
  \label{fig:lem}
\end{figure*}

The last lemma shows that the $\dot{F}$-coalgebra
$\gamma_{i}
: (\asdf)^i P \to \dot{F}(\asdf)^i P$ gets closer to $\dot{\alpha}$ as $i$ gets larger, with a particular consequence that
$\gamma_\lambda$ is isomorphic to $\dot{\alpha}$ (via $p_\lambda^\lambda$). This is used in the following proof of
 Prop.~\ref{prop:main}.
\begin{myproof}[Prop.~\ref{prop:main}]
  Let $\gamma$ be an arbitrary coalgebra $P \to \dot{F}P$ above $\alpha=\alpha_{\lambda, \lambda+1}$ (Notation~\ref{notation:alpha}).
  Lem.~\ref{lem:exist} shows the existence of a vertical morphism from $\gamma$ to $\dot{\alpha}_{\lambda, \lambda+1}$.
  We only need to show the uniqueness of morphisms.
  Let $f$ be an arbitrary vertical morphism from $\gamma$ to $\dot{\alpha}_{\lambda, \lambda+1}$.
  The isomorphic correspondence in
  Lem.~\ref{lem:iso_alg_coalg}
  carries
  $f: (P \xrightarrow{\gamma} \dot{F}P) \to (\dot{F}^\lambda 0 \xrightarrow{\dot{\alpha}} \dot{F}^{\lambda+1}0)$ in $\Coalg{\dot{F}}_\alpha$
  to $f: (P \xrightarrow{\beta_{0, 1}} \alpha^*\dot{F}P) \to (\dot{F}^\lambda 0 \xrightarrow{\delta} \alpha^*\dot{F}^{\lambda+1}0)$ in $\Coalg{\asdf}$,
  where $\delta$ is the mediating morphism from $\dot{\alpha}$ by universality of the cartesian lifting $\overline{\alpha}: \alpha^*\dot{F}^{\lambda+1}0 \to \dot{F}^{\lambda+1}0$.

  Using the above $f$ in $\Coalg{\asdf}$, we consider the following two chains and a morphism between them. Everything here is above $F^\lambda 0$; cf.~(\ref{eq:diagramASDF}). $\delta$ is an isomorphism since the initial chain of $\dot{F}$ converges in $\lambda$ steps.
  \begin{equation} \label{eq:rectangle}
      \vcenter{\xymatrix@R=.8em@C-.3em{
          \dot{F}^\lambda 0 \ar[r]^(.4){\cong}|(.4){\delta} &(\asdf)\dot{F}^\lambda 0 \ar[r]^{\cong} &\cdots \ar[r]^(.3){\cong} &(\asdf)^\lambda \dot{F}^\lambda 0 \ar[r]^(.45){\cong} &(\asdf)^{\lambda+1} \dot{F}^{\lambda} 0 \\
          P \ar[r]_(.4){\beta_{0, 1}} \ar[u]^{f} &(\asdf)P \ar[u]_{(\asdf)f} \ar[r] &\cdots \ar[r]_(.4){\beta_{\lambda-1, \lambda}} &(\asdf)^\lambda P \ar[u]^{(\asdf)^\lambda f} \ar[r]_(.45){\beta_{\lambda, \lambda+1}} &(\asdf)^{\lambda+1} P \ar[u]^{(\asdf)^{\lambda+1} f}
    }}
  \end{equation}

  It follows easily that
 $\beta_{\lambda, \lambda+1}$ is the inverse of an initial $\asdf$-algebra. This is essentially because 1) $\gamma_{\lambda}$ is isomorphic to $\dot{\alpha}$ (see the paragraph that follows Lem.~\ref{lem:pii_iso}); and 2) (the inverse of) $\gamma_{\lambda}=\overline{\alpha} \circ \beta_{\lambda, \lambda+1}$ corresponds to (the inverse of) $\beta_{\lambda, \lambda+1}$ in the isomorphic correspondence $\Alg{\dot{F}}_{\alpha^{-1}} \cong \Alg{\asdf}$ in Lem.~\ref{lem:iso_alg_coalg}.

  Consider the rightmost square in~(\ref{eq:rectangle}).
  By universality of the initial $\asdf$-algebra $(\beta_{\lambda, \lambda+1})^{-1}$,
  $(\asdf)^\lambda f$ is unique; therefore the composite $(\asdf)^\lambda f\circ \beta_{\lambda-1,\lambda}\circ\cdots\circ \beta_{0,1}$ (on the left in~(\ref{eq:rectangle})) is uniquely determined. By the commutativity of~(\ref{eq:rectangle}) and the fact that all the morphisms in the first row are isomorphisms, this uniquely determines $f$, too.
  \myqed
\end{myproof}

\section{Non-Example for Stable Chain Colimits} \label{ap:non_example}
Here we present $\clat$-fibrations which do not have stable chain colimits.
\begin{myproposition}
  For any complete lattice $L$
  and a $\clat$-fibration $p : \mathbb{E} \rightarrow
  \mathbb{B}$ with stable chain colimits,
  the composite
  \[ p \circ \pi_2 : L \times \mathbb{E} \rightarrow \mathbb{E} \rightarrow
     \mathbb{B} \]
  is a $\clat$-fibration.
\end{myproposition}
\begin{myproof}
  The substitution is given by $f^{\ast} (l, X) = (l, f^{\ast} X)$. This
  evidently preserves all meets.
  \myqed
\end{myproof}
The fibre over $0$ is $L \times \mathbb{E}_0$ and an initial object of $L \times \mathbb{E}$ is $(\bot, 0)$ where $\bot$ is the bottom element of $L$.
This and Lem.~\ref{lem:init_obj} lead that the $\clat$-fibration $p \circ \pi_2$ does not have stable chain colimits
if $L$ is not trivial.

\begin{myproposition}
  Let $h : L \rightarrow L$ be a monotone function. Define $\dot{I} : L \times
  \mathbb{E} \rightarrow L \times \mathbb{E}$ by
  \[ \dot{I} (l, X) = (h l, X) . \]
  Then $\dot{I}$ is a fibred lifting of $\mathrm{id}_{\mathbb{B}}$ along $p
  \circ \pi_2 : L \times \mathbb{E} \rightarrow \mathbb{B}$.
\end{myproposition}
\begin{myproof}
  It suffices to show the fibredness.
  The following equations conclude the proof:
  $\dot{I} (f^{\ast} (l, X)) = \dot{I} (l, f^{\ast} X) = (h l, f^{\ast} X) =
     f^{\ast} (h l, X) = f^{\ast} (\dot{I} (l, X))$.
     \myqed
\end{myproof}

The tuple $(p \circ \pi_2, \mathrm{id}_\mathbb{B}, \dot{I})$ is a non-example of Cor.~\ref{cor:same_converge}.
The initial chain of $\mathrm{id}_{\mathbb{B}}$ stabilizes from the beginning,
while the initial chain of $\dot{I}$ corresponds to that of $h \times
\mathrm{id}_{\mathbb{E}_0} : L \times \mathbb{E}_0 \rightarrow L \times \mathbb{E}_0$, whose
convergence depends on $h$
(e.g. $L=\{\bot, \top\}$ and $h(\bot) = \top, h(\top)=\top$).

\section{Omitted Proofs}
\subsection{Proof of Lem.~\ref{lem:init}}
\begin{myproof}
  \begin{enumerate}
    \item \label{item:init}
      We prove $\alpha_{i, i+1} = p\dot{\alpha}_{i, i+1}$ for each ordinal $i$ by transfinite induction on $i$.
      This statement also shows $\alpha_{i, j} = p \dot{\alpha}_{i, j}$ for all ordinals $i, j$ with $i < j$.
      For objects, it is easy to check that $p$ sends all objects of the initial chain of $\dot{F}$ to the corresponding objects of the initial chain of $F$
      because $Fp = p\dot{F}$ and $p$ strictly preserves chain colimits including initial objects.
      \begin{itemize}
        \item (Base case) There exists only one morphism $0 \to F0$ thus $\alpha_{0, 1} = p\dot{\alpha}_{0, 1}$.
        \item (Step case) Assume $\alpha_{i, i+1} = p\dot{\alpha}_{i, i+1}$.
          Then $\alpha_{i+1, i+2} = F\alpha_{i, i+1} = Fp\dot{\alpha}_{i, i+1} = p\dot{F}\dot{\alpha}_{i, i+1} = p\dot{\alpha}_{i+1, i+2}$.

        \item (Limit case)
          Because $\dot{F}^i0 \xrightarrow{\dot{\alpha}_{i, i+1}} \dot{F}^{i+1}0$ is a coprojection of the colimit $\dot{F}^{i+1}0 = \colim_{j < i+1} F^j0$ and $p$ strictly preserves colimits,
          $\alpha_{i, i+1} = p\dot{\alpha}_{i, i+1}$.
      \end{itemize}
    \item If the initial chain of $\dot{F}$ converges in $\lambda$ steps,
      the initial chain of $F$ also converges by Lem.~\ref{lem:init}.\ref{item:init}.
      Therefore, by Prop.~\ref{prop:converge_to_initial}, $\alpha_{\lambda, \lambda+1}^{-1}$ is an initial $F$-algebra.
      \myqed
  \end{enumerate}
\end{myproof}

\subsection{Proof of Prop.~\ref{prop:cost_fibre}} \label{ap:cost_fibre}
  \begin{myproof}
   (2 iff 2'): see Prop.~\ref{prop:fibred_initalg_finalcoalg},
   (3 iff 3'): by definition.
       \myqed
  \end{myproof}

  \begin{myremark}
     We note that the choice of $\beta$ is irrelevant in Prop.~\ref{prop:cost_fibre}: if one initial algebra $\beta$ satisfies Cond.~\ref{item:totalInitAlg}'--\ref{item:totalFinalCoalgToo}', then any initial algebra does.
     Indeed, for any $F$-algebra $\gamma$ isomorphic to $\beta$,
     there are isomorphisms of fibres
     $\Alg{\dot{F}}_\beta \cong \Alg{\dot{F}}_\gamma$ and $\Coalg{\dot{F}}_{\beta^{-1}} \cong \Coalg{\dot{F}}_{\gamma^{-1}}$
     because cartesian liftings of isomorphisms are also isomorphisms.
 \end{myremark}

\subsection{Proof of Lem.~\ref{lem:iso_alg_coalg}}
This holds because cartesian liftings of isomorphisms are isomorphisms.
See the following diagrams.
\[
  \xymatrix@R=.8em{
    \mathbb{E} \ar[dd]^p
    &P \ar[r]^-c \ar[d]_-{c^\dagger} &\dot{F}P &\dot{F}P \ar[r]^-a &P \\
    &\alpha^*\dot{F}P \ar[ur]_-{\overline{\alpha}}|\cong &&&\alpha^*\dot{F}P \ar[ul]^-{\overline{\alpha}}|\cong \ar[u]_-{a^\dagger}
    \\
    \mathbb{B}
    &X \ar[r]_-\cong^-\alpha &FX &FX \ar[r]_-\cong^-{\alpha^{-1}} &X
  }
\]

\subsection{Proof of Thm.~\ref{thm:cost_lift}} \label{ap:cost_lift}
\begin{mylemma} \label{lem:fibred_init_final}
  \begin{enumerate}
    \item \label{item:fibred_init_final_1}
  Let $p: \mathbb{E} \to \mathbb{B}$ be a fibration,
  $0$ be an initial object in $\mathbb{B}$
  and $\bot$ be an initial object in $\mathbb{E}_0$.
  Then $\bot$ is also initial in $\mathbb{E}$.
    \item
      Let $p: \mathbb{E} \to \mathbb{B}$ be an opfibration,
      $1$ be a final object in $\mathbb{B}$
      and $\top$ be a final object in $\mathbb{E}_1$.
      Then $\top$ is also final in $\mathbb{E}$.
  \myqed
  \end{enumerate}
\end{mylemma}
\begin{myproof}
  We only give a proof for \ref{item:fibred_init_final_1}:
  For each object $X \in \mathbb{E}$,
  $\mathbb{E}(\bot, X)
  \cong \mathbb{E}_0(\bot, (!_{pX})^*X)
  \cong 1$
  where $!_{pX}: 0 \to pX$ is the unique morphism.
  \[
  \vcenter{\xymatrix@R=1em{
      \mathbb{E} \ar[dd]^p &\bot \ar@{.>}[d] \ar[r] &X \\
                           &(!_{pX})^*X \ar[ur]_{\overline{!_{pX}}} \\
      \mathbb{B} &0 \ar[r]^{!_{pX}} &pX
  }}
  \]
  This concludes the proof.
  \myqed
\end{myproof}

\begin{myproposition} \label{prop:fibred_initalg_finalcoalg}
  \begin{enumerate}
    \item
      Let $p: \mathbb{E} \to \mathbb{B}$ be a fibration and $\dot{F}: \mathbb{E} \to \mathbb{E}$ be a lifting of $F: \mathbb{B} \to \mathbb{B}$.
      Consider an initial $F$-algebra $\beta$ and a $\dot{F}$-algebra $\dot{\beta}$ above $\beta$.
      Then $\dot{\beta}$ is an initial $\dot{F}$-algebra
      iff $\dot{\beta}$ is initial in $\Alg{\dot{F}}_\beta$.
    \item
      Let $p: \mathbb{E} \to \mathbb{B}$ be a opfibration and $\dot{F}: \mathbb{E} \to \mathbb{E}$ be a lifting of $F: \mathbb{B} \to \mathbb{B}$.
      Consider a final $F$-coalgebra $\alpha$ and a $\dot{F}$-coalgebra $\dot{\alpha}$ above $\alpha$.
      Then $\dot{\alpha}$ is a final $\dot{F}$-coalgebra
      iff $\dot{\alpha}$ is final in $\Coalg{\dot{F}}_\alpha$.
  \end{enumerate}
\end{myproposition}
\begin{myproof}
  Apply Lem.~\ref{lem:fibred_init_final} to Prop.~\ref{prop:alg_coalg}.
  \myqed
\end{myproof}

\begin{myproof}[Thm.~\ref{thm:cost_lift}]
  Since $p$ is an opfibration,
  Prop.~\ref{prop:fibred_initalg_finalcoalg} says that Cond.~\ref{item:totalFinalCoalgToo}' in Prop.~\ref{prop:cost_fibre} is equivalent to the following condition:
  \ref{item:totalFinalCoalgToo}'')
           $\dot{\beta}^{-1}$ is a final $\dot{F}$-coalgebra.
   This concludes the proof.
   \myqed
\end{myproof}

\subsection{Proof of Example~\ref{eg:stable chain colimits}}
\begin{myproof}
  In each case,
  let us consider an arbitrary diagram $D: \mathrm{Ord}_{<\lambda} \to \Set$.
  Then there exists a colimit of $D$ since $\Set$ is cocomplete.
  We write $\kappa_i: D i \to \colim D$ for the $i$-th cocone component.
  In order to show that a fibration $p$ has stable chain colimits,
  we prove $P \cong \colim_{\mathrm{Ord}_{<\lambda}} \kappa_{(-)}^*P$ for each $P \in \mathbb{E}_{\colim D}$.
  The case $\lambda=0$ is easily shown for each example,
  so we give the proof for the case $\lambda > 0$.

  \[
  \xymatrix{
        \mathbb{E} \ar[d]^p
        &\kappa_0^*P \ar[r] &\kappa_1^*P \ar[r] &\cdots \ar[r] &P \\
        \Set
        &D0 \ar[r] &D1 \ar[r] &\cdots \ar[r] &\colim D
  }
\]
  \begin{itemize}
    \item []
    \item $\mathbf{Pred} \to \Set$:
      Let $(Q \subseteq X, (f_i: \kappa_i^*P \to Q)_{i < \lambda})$ be an arbitrary cocone over $\kappa_{(-)}^*P$.
      Note that $f_i$ is a function $D i \to X$ satisfying $f_i(\kappa_i^*P) \subseteq Q$.
      This cocone induces a unique morphism $h: \colim D \to X$ in $\Set$ by universality of $\colim D$.
      It is sufficient to prove that $h$ is also a morphism from $(P \subseteq \colim D)$ to $(Q \subseteq X)$ in $\mathbf{Pred}$.
      For each $a \in P$, because $a \in \colim D$,
      there exists $n \in Di$ for some $i (< \lambda)$ such that $a = \kappa_i n$.
      Then $n \in \kappa_i^*P = \kappa_i^{-1}P$.
      Since $f_i(\kappa_i^*P) \subseteq Q$ and $f_i = h \circ \kappa_i$,
      $h a = h (\kappa_i n) = f_i n \in Q$.
      Therefore, $\mathbf{Pred} \to \Set$ has stable chain colimits.
    \item $(\mathbf{Pre}, \mathbf{ERel}) \to \Set$:
      Here we give the proof only for $\mathbf{ERel} \to \Set$.
      Let $(Q \subseteq X^2, (f_i: \kappa_i^*P \to Q)_{i < \lambda})$ be an arbitrary cocone over $\kappa_{(-)}^*P$.
      Note that $f_i$ is a function $D i \to X$ satisfying $(f_i x, f_i y) \in Q$ for all $(x, y) \in \kappa_i^*P \subseteq (Di)^2$.
      This cocone induces a unique morphism $h: \colim D \to X$ in $\Set$ by universality of $\colim D$.
      It is sufficient to prove $h$ is also a morphism from $(P \subseteq (\colim D)^2)$ to $(Q \subseteq X^2)$ in $\mathbf{ERel}$.
      For each $(a, b) \in P$, because $a, b \in \colim D$,
      there exists $n \in Di$ and $m \in Dj$ for some $i, j (< \lambda)$ such that $a = \kappa_i n$ and $b = \kappa_j m$.
      Without loss of generality, we can assume $i \leq j$.
      Then $(D(i, j)n, m) \in \kappa_j^*P$ because $(\kappa_j \circ D(i, j)n, \kappa_j m) = (\kappa_i n, \kappa_j m) = (a, b) \in P$.
      Since $f_j: \kappa_j^*P \to Q$ in $\mathbf{ERel}$ and $f_j = h \circ \kappa_j$,
      $(h a, h b) = (h (\kappa_i n), h (\kappa_j m))
      = (h (\kappa_j \circ D(i, j) n), h (\kappa_j m))
      = (f_j \circ D(i, j) n, f_j m) \in Q$.
      Therefore, $\mathbf{ERel} \to \Set$ has stable chain colimits.
    \item $d^\Omega: \Set/\Omega \to \Set$:
      In this case, $P$ is a function from $\colim D$ to $\Omega$.
      Let $((X \xrightarrow{Q} \Omega) , (f_i: \kappa_i^*P \to Q)_{i < \lambda})$ be an arbitrary cocone over $\kappa_{(-)}^*P$.
      Note that $f_i$ is a function $D i \to X$ satisfying $(\kappa_i^*P)(x) = P (\kappa_i x) \leq Q (f_ix)$ for each $x \in D i$.
      This cocone induces a unique morphism $h: \colim D \to X$ in $\Set$ by universality of $\colim D$.
      It is sufficient to prove that $h$ is also a morphism from $\colim D \xrightarrow{P} \Omega$ to $X \xrightarrow{Q} \Omega$.
      For each $a \in \colim D$,
      there exists an ordinal $i<\lambda$ and $b \in Di$ such that $a = \kappa_i b$.
      Since $f_i: \kappa_i^*P \to Q$ in $\Set/\Omega$, $P (\kappa_i b) \leq Q (f_i b)$.
      Because $f_i = h \circ \kappa_i$,
      $P a = P (\kappa_i b) \leq Q (f_i b) = Q (h(\kappa_i b)) = Q h a$.
      Therefore, $d^\Omega$ has stable chain colimits.
      \myqed
  \end{itemize}
\end{myproof}

\subsection{Proof of Prop.~\ref{prop:coinci_clat}} \label{ap:clat_ip}
We use the following lemma.
\begin{mylemma}[{\cite[Thm. 1.6]{Barr92}}] \label{lem:barr}
  Let $F$ be an endofunctor on $\mathbb{C}$ and
  $(\alpha_{i, j})_{i, j \in \mathrm{Ord}}, (\beta_{i, j})_{i, j \in \mathrm{Ord}}$ be the initial and final chain morphisms of $F$ respectively.
  Then there exists a unique family of morphisms $\{F^i 0 \xrightarrow{f_i} F^i 1\}_{i \in \mathrm{Ord}}$ such that $f_{i+1} = Ff_i$ and $f_i = \beta_{i, j} \circ f_j \circ \alpha_{i, j}$
  for each $i, j$ with $i \leq j$.

  \[
  \xymatrix{
        1 &F1 \ar[l]^{\beta_{0, 1}} &\cdots \ar[l]^-{\beta_{1, 2}} F^i 1 \ar[l] &F^j 1 \ar[l]^{\beta_{i, j}} &\cdots \ar[l] \\
        0 \ar[u]^{f_0} \ar[r]^{\alpha_{0, 1}} &F0 \ar[u]^{f_1} \ar[r]^-{\alpha_{1, 2}} &\cdots F^i 0 \ar[u]^{f_i} \ar[r]^{\alpha_{i, j}} &F^j 0 \ar[u]^{f_j} \ar[r] &\cdots
  }
\]
\myqed
\end{mylemma}

\begin{myproof}[Prop.~\ref{prop:coinci_clat}]
\begin{enumerate}
  \item \label{item:coinci_clat_proof_1}
    We prove $\alpha_{i, \lambda}^*(\alpha^*\dot{F})^i \bot = \dot{F}^i 0$ by transfinite induction on $i$.
  \begin{itemize}
    \item (Base case) $\alpha_{0, \lambda}^*\bot = 0$ by Lem.~\ref{lem:init_obj}.
    \item (Step case)
    Assume $\alpha_{i, \lambda}^*(\alpha^*\dot{F})^i \bot = \dot{F}^i 0$.
    \begin{align*}
      \alpha_{i+1, \lambda}^*(\alpha^*\dot{F})^{i+1} \bot
        &= \alpha_{i+1, \lambda}^*\alpha^*\dot{F}(\alpha^*\dot{F})^{i} \bot \\
        &= (F\alpha_{i, \lambda})^*\dot{F}(\alpha^*\dot{F})^{i} \bot &\text{since $\alpha \circ \alpha_{i+1, \lambda} = \alpha_{i+1, \lambda+1} = F\alpha_{i, \lambda}$} \\
        &= \dot{F}\alpha_{i, \lambda}^* (\alpha^*\dot{F})^{i} \bot &\text{by fibredness of $\dot{F}$} \\
        &= \dot{F}^{i+1}0
        &\text{by assumption}
    \end{align*}
    \item (Limit case) Assume $\alpha_{j, \lambda}^*(\alpha^*\dot{F})^j \bot = \dot{F}^j 0$ for each $j < i$.
      \begin{align*}
        \alpha_{i, \lambda}^*(\alpha^*\dot{F})^i \bot
        &= \colim_{j < i} \alpha_{j, \lambda}^*(\alpha^*\dot{F})^i \bot &\text{since $p$ has stable chain colimits}\\
        &\geq \colim_{j < i} \alpha_{j, \lambda}^*(\alpha^*\dot{F})^j \bot &\text{since $(\alpha^*\dot{F})^i \bot \geq (\alpha^*\dot{F})^j \bot$} \\
        &= \colim_{j < i} \dot{F}^j 0 &\text{by assumption} \\
        &= \dot{F}^i 0
      \end{align*}
      The inverse inequality $\alpha_{i, \lambda}^*(\alpha^*\dot{F})^i \bot \leq \dot{F}^i 0$ is got by Lem.~\ref{lem:exist}:
      By Prop.~\ref{prop:clat}, both $\mathbb{E}$ and $\Set$ are cocomplete and $p$ strictly preserves colimits.
      From the proof of Lem.~\ref{lem:exist} and $(\alpha^*\dot{F})^i \bot \leq (\alpha^*\dot{F})^{i+1} \bot$, $\alpha_{j, \lambda}^*(\alpha^*\dot{F})^i \bot \leq \dot{F}^j 0$ for each $j \leq \lambda$.
      By considering $j=i$, $\alpha_{i, \lambda}^*(\alpha^*\dot{F})^i \bot \leq \dot{F}^i 0$.
      Hence, $\alpha_{i, \lambda}^*(\alpha^*\dot{F})^i \bot = \dot{F}^i 0$.
  \end{itemize}

  \item We prove $\alpha_{i, \lambda}^*(\alpha^*\dot{F})^i \top = \dot{F}^i 0$ by transfinite induction on $i$.
  \begin{itemize}
    \item (Base case), (Step case)
      The same discussions as in~\ref{item:coinci_clat_proof_1} can be applied.
    \item (Limit case) Assume $\alpha_{j, \lambda}^*(\alpha^*\dot{F})^j \top = \dot{F}^j 0$ for each $j < i$.
    \begin{align*}
      \alpha_{i, \lambda}^*(\alpha^*\dot{F})^i \top
      &= \colim_{j < i} \alpha_{j, \lambda}^*(\alpha^*\dot{F})^i \top &\text{since $p$ has stable colimits} \\
      &\leq \colim_{j < i} \alpha_{j, \lambda}^*(\alpha^*\dot{F})^j \top &\text{since $(\alpha^*\dot{F})^i \top \leq (\alpha^*\dot{F})^j \top$}\\
      &= \colim_{j < i} \dot{F}^j 0 &\text{by assumption}\\
      &= \dot{F}^i 0
    \end{align*}
    Lem.~\ref{lem:barr} relates the initial and final chain of $\asdf$ on $\mathbb{E}_{F^\lambda 0}$,
    and in particular this shows $(\alpha^*\dot{F})^i \bot \leq (\alpha^*\dot{F})^i \top$.
    Because $\dot{F}^i0 = \alpha_{i, \lambda}^*(\alpha^*\dot{F})^i \bot$
    by the first item of this proof,
    $\dot{F}^i 0 \leq \alpha_{i, \lambda}^*(\alpha^*\dot{F})^i \top$ holds.
    This concludes the proof.
  \end{itemize}
  \myqed
\end{enumerate}
\end{myproof}

\subsection{Proof of Cor.~\ref{cor:same_converge}}
\begin{myproof}
  The (only if)-part is obvious (since $p$ preserves isomorphisms).

  (if) By Prop.~\ref{prop:coinci_clat}, $\dot{F}^\lambda 0 = \alpha_{\lambda, \lambda+1}^*\dot{F}^{\lambda+1}0$
  so $\dot{\alpha}_{\lambda, \lambda+1}$ is the cartesian lifting $\dot{F}^\lambda 0 \xrightarrow{\overline{\alpha_{\lambda, \lambda+1}}} \dot{F}^{\lambda+1}0$.
  Because $\alpha_{\lambda, \lambda+1}$ is an isomorphism,
  $\overline{\alpha_{\lambda, \lambda+1}}$ is also an isomorphism.
  Therefore, the initial chain of $\dot{F}$ converges in $\lambda$ steps.
  \myqed
\end{myproof}

\subsection{Proof of Thm.~\ref{thm:general1}}
\begin{myproof}
  The domain fibration $d^\Omega$ is a $\clat$-fibration as in Example~\ref{eg:clat}.
  By Example~\ref{eg:stable chain colimits}, $d^\Omega$ has stable chain colimits.
  Thm.~\ref{thm:cost_clat} concludes that $(d^\Omega, F, \dot{F})$ satisfies the \cost{}
  if $F$ has an initial algebra. \myqed
\end{myproof}

\subsection{Proof of Lem.~\ref{lem:dist_law2}}
\begin{myproof}
      Consider an arbitrary object $(X \xrightarrow{x} \Omega) \in \Set/\Omega$.
      We prove that $\lambda_X: F\T X \to \T F X$ in $\Set$ is also
      a morphism $\dot{F}\dotT x \to \dotT \dot{F} x$ in $\Set/\Omega$.
      The following diagram includes $\dot{F}\dotT x \ (= \sigma \circ F\tau \circ F\T x) $ and $\dotT \dot{F} x \ (= \tau \circ \T \sigma \circ \T F x)$.
      The top square commutes by naturality of $\lambda: F\T \Rightarrow \T F$.
      Because $\sigma \circ F\tau \leq \tau \circ \T\sigma \circ \lambda_\Omega$ by assumption, the pentagon has an inequality as the following diagram.
      \[
        \xymatrix@R=1.5em@C=1em{
          F\T X \ar@{}[rrd]|{=} \ar[d]^{F\T x} \ar[rr]^{\lambda_X} &&\T F X \ar[d]^{\T Fx}\\
          F\T \Omega \ar@{}[rrd]|{\leq} \ar[d]^{F \tau} \ar[rr]^{\lambda_\Omega} &&\T F \Omega \ar[d]^{\T \sigma} \\
          F \Omega \ar[dr]_{\sigma}  &&\T \Omega \ar[dl]^{\tau} \\
                    &\Omega \\
        }
      \]
      Therefore, $\lambda_X$ is a morphism $\dot{F}\dotT x \to \dotT \dot{F} x$ in $\Set/\Omega$.
      \myqed
\end{myproof}

\subsection{Proof of Thm.~\ref{thm:general2}} \label{ap:general2}
We use the following lemma.
\begin{mylemma} \label{lem:general2_coalg}
  Consider in the setting of Def.~\ref{def:kleisli_setting},
  Let $\beta: F\mu F \iso \mu F$ be an initial $F$-algebra.

  Then $\Coalg{\dot{F}}_{\beta^{-1}} \cong \Coalg{\dot{F}_{\dot{\mathcal{T}}}}_{L\beta^{-1}}$.
\end{mylemma}
\begin{myproof}
  We show that
  the following conditions are equivalent for each $f: \mu F \to \Omega$ in $\Set/\Omega$:
  \begin{enumerate}
    \item $\beta^{-1}: f \to \dot{F}f$ in $\Set/\Omega$
      (i.e. $f \leq_{\mu F} (\beta^{-1})^*\dot{F}f$);
    \item $L\beta^{-1}: f \to \dot{F}_{\dT} f$ in $\Kl{\dot{\mathcal{T}}}$
      (i.e. $f \leq_{L\mu F} (L\beta^{-1})^*\dot{F}_{\dT}f$).
  \end{enumerate}
  The first condition means that $f$ yields a $\dot{F}$-coalgebra above $\beta^{-1}$
  and the second condition means that $f$ yields a $\dot{F}_{\dot{\mathcal{T}}}$-coalgebra above $L\beta^{-1}$.

  \begin{align*}
    &(\dot{L}\beta^{-1})^*\dot{F}_{\dot{\mathcal{T}}}f \\
    &= \left(\begin{array}{r}
         \mu F \xrightarrow{\beta^{-1}} F\mu F \xrightarrow{\eta_{F\mu F}} \mathcal{T}F\mu F \xrightarrow{\mathcal{T}Ff} \mathcal{T}F\Omega \\
         \xrightarrow{\mathcal{T}\sigma} \mathcal{T}\Omega \xrightarrow{\tau} \Omega
      \end{array}
      \right)
    &\text{by definition} \\
    &= \left(\begin{array}{r}
         \mu F \xrightarrow{\beta^{-1}} F\mu F \xrightarrow{Ff} F\Omega \xrightarrow{\sigma} \Omega \xrightarrow{\eta_\Omega} \mathcal{T}\Omega \xrightarrow{\tau} \Omega
      \end{array}
      \right)
    &\text{by naturality of the unit $\eta$} \\
    &= \left(\begin{array}{r}
         \mu F \xrightarrow{\beta^{-1}} F\mu F \xrightarrow{Ff} F\Omega \xrightarrow{\sigma} \Omega \\
      \end{array}
      \right)
    &\text{because $\tau$ is an EM monotone algebra} \\
    &= (\beta^{-1})^*\dot{F}f.
  \end{align*}
  Therefore, (2) is equivalent to (1).
  Lem.~\ref{lem:kleisli_fib}.\ref{item:fibre} concludes this lemma.
  \myqed
\end{myproof}

\begin{myproof}[Thm.~\ref{thm:general2}]
  Let $\beta: F\mu F \iso \mu F$ be an initial $F$-algebra.
  By Thm.~\ref{thm:general1}, we have an initial $\dot{F}$-algebra $\dot{\beta}$ above $\beta$.
  Because the initial chains are defined by initial objects and colimits,
  Kleisli left adjoints $L, \dot{L}$ preserve them.
  Thus, $L\beta$ is an initial $F_\mathcal{T}$-algebra
  and $\dot{L}\dot{\beta}$ is also an initial $\dot{F}_{\dot{\mathcal{T}}}$-algebra above $L\beta$.

  By Thm.~\ref{thm:general1}, $\dot{\beta}^{-1}$ is a final $\dot{F}$-coalgebra over $\beta^{-1}$.
  Lem.~\ref{lem:general2_coalg} says that
  $\dot{L}\dot{\beta}^{-1}$ is also a final $\dot{F}_{\dot{\mathcal{T}}}$-coalgebra over $L\beta^{-1}$.
  Therefore, $(d^\Omega_{\mathcal{T}, \dot{\mathcal{T}}}, F_\mathcal{T}, \dot{F}_{\dot{\mathcal{T}}})$ satisfies the \cost{}.
  \myqed
\end{myproof}

\subsection{Proof of Thm.~\ref{thm:probLivenessBySubmartingale}}
\label{appendix:prf:thm:probLivenessBySubmartingale}
\begin{wrapfigure}[3]{r}{0pt}
\begin{math}
  \vcenter{\xymatrix@R=1.5em{
      \Set/[0, 1] \ar[d]^{d^{[0, 1]}} \lloop{\dot{F}^\ptr}  \\
      \Set \lloop{F^\ptr}
  }}
\end{math}
\end{wrapfigure}
The following tuple satisfying the \cost{} shows the theorem:
$(d^{[0, 1]}, F^\ptr, \dot{F}^\ptr)$ defined by applying Thm.~\ref{thm:general1} to the following.
\begin{itemize}
  \item a complete lattice $\Omega$ is $[0, 1]$ with the usual order between real numbers;
  \item a set functor $F$ is $F^{\ptr}=\mathbf{1}+\{\ok, \notok\} \times [0, 1] \times (-)^2$;
  \item a monotone $F^\ptr$-algebra $\sigma: F^\Sigma [0, 1] \to [0, 1]$
    is $\sigma^{\ptr}$ defined as follows:
    \begin{align*}
      \sigma^\ptr(x) &=
      \begin{cases}
        0 & \text{if } x=* \in \mathbf{1} \\
        1 & \text{if } x=(\ok, p, a, b) \\
        pa + (1-p)b & \text{if } x=(\notok, p, a, b)
      \end{cases}
    \end{align*}
\end{itemize}
\begin{mylemma} \label{lem:probLivenessAsInitialAlg}
  $(d^{[0, 1]}: \Set/[0, 1] \to \Set, F^\ptr, \dot{F}^\ptr)$ satisfies the \cost{}.

  The coincidence occurs for initial algebras $\beta: F^{\ptr} \mu F^{\ptr} \iso \mu F^{\ptr}$ in $\Set$ and $\dot{\beta}: \dot{F}^\ptr \mu \dot{F}^\ptr \iso \mu \dot{F}^\ptr$ in $\Set/[0, 1]$
  such that $\mu F^{\ptr}$ is the set of all finite probabilistic binary trees (Def.~\ref{def:finiteProbTree})
  and $\mu \dot{F}^{\ptr} = (\mu F^{\ptr} \to [0, 1])$ in $\Set/[0, 1]$ carries a finite probabilistic binary tree $t \in \mu F^{\ptr}$ to the probability of eventually reaching an $\ok$ node.
  \myqed
\end{mylemma}

\begin{myproof}[Thm.~\ref{thm:probLivenessBySubmartingale}]
  We note that the given tree $t$ is naturally identified with a subcoalgebra of $\beta^{-1}\colon \mu F^\ptr\iso F^\ptr (\mu F^\ptr)$: it has the set $N$ of nodes of $t$ as a carrier; its coalgebra structure $c\colon N\to F^\ptr N$ is a natural ``decomposition'' function; and the embedding $\iota\colon N\to \mu F^\ptr$, carrying a state to the subtree below it, obviously preserves the coalgebra structures.

  If $f$ satisfies the conditions in the statement then
  there exists a $\dot{F}^\ptr$-coalgebra $f \to \dot{F}^\ptr f$ above $c$.

  The domain fibration $d^{[0, 1]}$ is a bifibration
(Example~\ref{eg:clat}, Prop.~\ref{prop:clat}).
  Because the lifting $\dot{F}^{\ptr}$ of $F^{\ptr}$ is fibred (Lem.~\ref{lem:mono_alg}),
  Prop.~\ref{prop:alg_coalg} shows that $\Coalg{d^{[0, 1]}}: \Coalg{\dot{F}^\ptr} \to \Coalg{F^\ptr}$ is a bifibration, too. Consider the substitution
  $\iota^*: \Coalg{\dot{F}^{\ptr}}_{\beta^{-1}} \to \Coalg{\dot{F}^{\ptr}}_c$ along the coalgebra morphism $\iota$ we discussed; since each substitution in a bifibration is a right adjoint,
  $\iota^*$ preserves a final object.
  By Lem.~\ref{lem:probLivenessAsInitialAlg}, $\dot{\beta}^{-1}$ is final object in $\Coalg{\dot{F}^{\ptr}}_{\beta^{-1}}$ so
  $\iota^*(\dot{\beta}^{-1})$ is also final in $\Coalg{\dot{F}^{\ptr}}_c$.

  Therefore, there exists a vertical morphism from $f\colon N\to [0,1]$ to $\mu \dot{F}^{\ptr} \circ \iota\colon N\to [0,1]$, that is, that $f$ is below $\mu \dot{F}^{\ptr} \circ \iota$ in the pointwise order. Lem.~\ref{lem:probLivenessAsInitialAlg} yields the claim.
  \myqed
\end{myproof}
\subsection{Proof of Thm.~\ref{thm:buta_tree}} \label{ap:buta_tree}

\begin{wrapfigure}[2]{r}{0pt}
\begin{math}
  \vcenter{\xymatrix@R=1.5em{
      \Set/\Pf Q \ar[d]^{d^{\Pf Q}} \lloop{\dot{F}^\tr}  \\
      \Set \lloop{F^\tr}
  }}
\end{math}
\end{wrapfigure}
The following tuple satisfying the \cost{} shows the theorem:
$(d^{\mathcal{P}Q}, F^\tr, \dot{F}^\tr)$ defined by applying Thm.~\ref{thm:general1} to the following.
\begin{itemize}
  \item a complete lattice $\Omega$ is $\Pf Q$ ordered by inclusion;
  \item a set functor $F$ is $F^{\tr}=\Sigma_0+\Sigma_2 \times (-)^2$;
  \item a monotone $F^\tr$-algebra $\sigma: F^\Sigma \Pf Q \to \Pf Q$
    is $\sigma^{\mathrm{bu}}$ defined as follows:
    \begin{align*}
      \sigma^\mathrm{bu}(x) &=
      \begin{cases}
        \delta(s) & \text{if } x=s \in \Sigma_0 \\
        \bigcup_{q_1 \in Q_1, q_2 \in Q_2} \delta(s, q_1, q_2) & \text{if } x=(s, Q_1, Q_2) \in \Sigma_2 \times (\Pf Q)^2
      \end{cases}
    \end{align*}
\end{itemize}
\begin{mylemma} \label{lem:dpq_cost}
  $(d^{\mathcal{P}Q}: \Set/\Pf Q \to \Set, F^\tr, \dot{F}^\tr)$ satisfies the \cost{}.

  The coincidence occurs for initial algebras $\beta: F^{\tr} \mu F^{\tr} \iso \mu F^{\tr}$ in $\Set$ and $\dot{\beta}: \dot{F}^\tr \mu \dot{F}^\tr \iso \mu \dot{F}^\tr$ in $\Set/\Pf Q$
  such that $\mu F^{\tr}$ is the set of all finite $\Sigma$-trees
  and $\mu \dot{F}^{\tr} = (\mu F^{\tr} \to \Pf Q)$ in $\Set/\Pf Q$ carries a finite $\Sigma$-tree $t \in \mu F^{\tr}$ to the set of states $\{\rho(r_t) \in Q \mid \rho \text{ is a run of $\mathcal{A}$ over $t$} \}$.
  \myqed
\end{mylemma}
\begin{myproof}[Proof of Thm.~\ref{thm:buta_tree}]
  If $f$ satisfies Cond.~\ref{item:buta_tree_1} and~\ref{item:buta_tree_2}
  then there exists a $\dot{F}^\tr$-coalgebra $f \to \dot{F}^\tr f$ above $c_t$.
  Since $d^{\Pf Q}$ is a bifibration and $\dot{F}^\tr$ is a fibred lifting of $F^\tr$,
  Prop.~\ref{prop:alg_coalg} says that there is a right adjoint $\mathrm{tr}^*: \Coalg{\dot{F}^\tr}_{\beta^{-1}} \to \Coalg{\dot{F}^\tr}_{c_t}$
  where
  $\mathrm{tr}: c_t \to \beta$ mapping a node $n \in N$ to the subtree under $n$ of $t$.
  By Lem.~\ref{lem:dpq_cost},
  $\dot{\beta}^{-1}$ is final in $\Coalg{\dot{F}^\tr}_{\beta^{-1}}$
  so $\mathrm{tr}^*\dot{\beta}^{-1}$ is also final in $\Coalg{\dot{F}^\tr}_{c_t}$.
  Therefore, $f \leq \mathrm{tr}^*\mu \dot{F}$ holds.
  Then Cond.~\ref{item:buta_tree_3} induces $q_\mathsf{F} \in f(r_t) \subseteq \mu \dot{F}^\tr(t)$.
  \myqed
\end{myproof}

\subsection{Proof of Thm.~\ref{thm:buta_trees}} \label{ap:buta_trees}

\begin{wrapfigure}[4]{r}{0pt}
\begin{math}\footnotesize
  \vcenter{\xymatrix@R=1em@C-1em{
     \Set/\Pf Q \ar[d]^{d^{\Pf}} \lloop{\dot{F}^\tr} \ar[r]^{\dot{L}} &\Kl{\dot{\Pf}} \rloop{\dot{F}^\tr_{\dot{\Pf}}} \ar[d]^{d^{\Pf}_{\Pf, \dot{\Pf}}} \\
      \Set \lloop{F^\tr} \ar[r]^L &\Kl{\Pf} \rloop{F^\tr_{\Pf}}
  }}
\end{math}
\end{wrapfigure}

We use the ``effectful'' results in~\S{}\ref{subsec:effectful} to accommodate the nondeterminism effect appearing in the transition $c$ of a generative tree automaton $\mathcal{C}$ (Def.~\ref{def:bottomUpTreeAutom}).
For a generative tree automaton $\mathcal{C}$,
we define $(d^{\mathcal{P}Q}_{\Pf, \dotP}, F^\tr_\Pf, \dot{F}^\tr_{\dotP})$ with the \cost{} by applying Thm.~\ref{thm:general2} to the following and items defined before Lem.~\ref{lem:dpq_cost}.
\begin{itemize}
  \item a set monad $\mathcal{T}$ is the powerset monad $\mathcal{P}$ on $\Set$;
  \item an EM monotone $\Pf$-algebra $\tau$ is
    $\tau_\cap: \Pf \Pf Q \to \Pf Q$ mapping $A$ to $\bigcap_{a \in A} a$;
  \item a distributive law $\lambda$ is
    $\lambda^{\tr}: F^{\tr}\Pf \Rightarrow \Pf F^{\tr}$ defined by
    \begin{align*}
      \lambda^{\tr}_X(x) &=
      \begin{cases}
        \{ s \} & \text{if } x=s \in \Sigma_0 \\
        \{ (s, x_1, x_2) \mid x_1 \in X_1, x_2 \in X_2 \} & \text{if } x=(s, X_1, X_2) \in \Sigma_2 \times (\Pf X)^2.
      \end{cases}
    \end{align*}
\end{itemize}

\begin{mylemma}
  $(d^{\mathcal{P}Q}_{\Pf, \dotP}: \Kl{\dotP} \to \Kl{\Pf}, F^\tr_\Pf, \dot{F}^\tr_{\dotP})$
satisfies the \cost{}.

The coincidence occurs for initial algebras $L\beta: F^\tr_\Pf \mu F^{\tr} \relto \mu F^{\tr}$ in $\Kl{\Pf}$ and $\dot{L}\dot{\beta}: \dot{F}^\tr_{\dotP} \mu \dot{F}^\tr \relto \mu \dot{F}^\tr$ in $\Kl{\dotP}$
  where $\beta$ and $\dot{\beta}$ are initial algebras defined in Lem.~\ref{lem:dpq_cost}, and $L: \Set \to \Kl{\Pf}$ and $\dot{L}: \Set/{\Pf Q} \to \Kl{\dotP}$ are Kleisli left adjoints.
\myqed
\end{mylemma}
\begin{auxproof}
  We use the initial algebras $\beta$ and $\dot{\beta}$ used in Prop.~\ref{prop:acceptanceByInitAlg}: the former is carried by the set of finite $\Sigma$-trees; and the latter characterizes acceptance in the sense of Prop.~\ref{prop:acceptanceByInitAlg}.
  We write $L: \Set \to \Kl{\Pf}$ and $\dot{L}: \Set/{\Pf Q} \to \Kl{\dotP}$ for Kleisli left adjoints.
  Because the initial chains are defined by initial objects and colimits and left adjoints preserve them,
  $\dot{L}\dot{\beta}$ is an initial $\dot{F}^\Sigma_{\dotP}$-algebra above an initial $F^\Sigma_\Pf$-algebra $L\beta$.
\end{auxproof}
The next lemma proved by {\cite[Prop.~3.2 and Thm.~3.8]{HasuoJS07b}}
shows the lifted coincidence in $\Kl{\Pf}$.
\begin{mylemma} \label{lem:co_klp}
  The initial $F^\Sigma$-algebra $\beta: F^\Sigma \mu F^\Sigma \iso \mu F^\Sigma$ in $\Set$
  gives rise to both an initial $F^\Sigma_\Pf$-algebra $L\beta$ and a final $F^\Sigma_\Pf$-coalgebra $L\beta^{-1}$
  where $L: \Set \to \Kl{\Pf}$ is the Kleisli left adjoint.
\end{mylemma}
\begin{auxproof}
  Thm.~3.8 in HasuoJS07b:
  $F^{\Sigma} = \Sigma_0 + \Sigma_2 \times (-)^2$ preserves weak pullbacks,
  and $\lambda$ is given by the $F$-relation lifting (Lem.~2.3).
\end{auxproof}

\begin{myproof}[Thm.~\ref{thm:buta_trees}]
  It is easy to see that $d^{\Pf Q}_{\Pf, \dotP}\colon\Kl{\dotP} \to \Kl{\Pf}$ is a $\clat$-fibration: each fibre is a complete lattice by definition; $\bigwedge$ in a fibre is preserved by substitution since $\tau_\cap$ preserves $\bigwedge=\bigcap$.
  Therefore, Lem.~\ref{lem:co_klp} and Thm.~\ref{thm:cost_lift} yield coincidence lifting,
  so that $L\beta^{-1}$ and $\dot{L}\dot{\beta}^{-1}$ are both initial algebras and final coalgebras.

  Because $f$ satisfies Cond.~\ref{item:buta_trees_1},
  there is a ${F}^{\tr}_{\dot{\Pf}}$-coalgebra $c: f \relto \dot{F}^{\tr}_{\dot{\mathcal{P}}} f$.
  Consider the diagram below:
  \[
    \xymatrix@R-1em{
      \Kl{\dotP} \ar[ddd]^{d^{\Pf Q}_{\Pf, \dot{\Pf}}} \lloop{\dot{F}^\Sigma_{\dotP}} &L\mu \dot{F}^\Sigma \ar[r]^-{\dot{L}\dot{\beta}^{-1}}_-\cong &\dot{F}^\Sigma_{\dotP} (L\mu \dot{F}^\Sigma) \mathrlap{\text{ (init alg and final coalg in $\Kl{\dotP}$)}} \\
                                                         &f \ar[r]^-c \ar@{.>}[u]^{L^{\mathrm{fin}}_{\mathcal{C}}} &\dot{F}^\Sigma_{\dotP} f \ar@{.>}[u]_{\dot{F}^\Sigma_{\dotP}L^{\mathrm{fin}}_{\mathcal{C}}} \\
                          &L\mu F^\Sigma \ar[r]^-{L\beta^{-1}}_-\cong &F^\Sigma_{\Pf} (L\mu F^\Sigma) \mathrlap{\text{ (init alg and final coalg in $\Kl{\Pf}$)}} \\
      \Kl{\Pf} \lloop{F^\Sigma_{\Pf}} &X \ar[r]^-c \ar@{.>}[u]^{L^{\mathrm{fin}}_{\mathcal{C}}} &F^\Sigma_{\Pf} X \ar@{.>}[u]_{F^\Sigma_{\Pf}L^{\mathrm{fin}}_{\mathcal{C}}}
    }
  \]

  The unique morphism from $c$ to the final coalgebra $L\beta^{-1}$ in $\Kl{\Pf}$
  is a function $L^{\mathrm{fin}}_{\mathcal{C}}\colon X\to \Pf(\mu F^{\tr})$ (this is the trace semantics for $c$, see \cite[Cor.~4.1]{HasuoJS07b}) defined by
  \[
  L^{\mathrm{fin}}_{\mathcal{C}}\colon x\longmapsto
  \bigl\{\,t\in \mu F^{\tr}\,\big|\,\text{$t$ is generated by $\mathcal{C}_{x}$}\,\bigr\}
  \]
  where $\mathcal{C}_{x}$ is obtained from $\mathcal{C}$ by replacing the initial state $x_{0}$ by $x$.
  This overrides the notation of the set $L^{\mathrm{fin}}_{\mathcal{C}}\in \Pf(\mu F^{\tr})$ in Def.~\ref{def:generativeTreeAutom} for the sake of argument.
  By finality of $\dot{L}\dot{\beta}^{-1}$ that we established in the above, we have
   $f \leq (L^{\mathrm{fin}}_{\mathcal{C}})^*(L\mu \dot{F}^{\tr})
   = \tau_\cap \circ \mathcal{P}(\mu\dot{F}^{\tr}) \circ L^{\mathrm{fin}}_{\mathcal{C}}$
  in $\Kl{\dotP}_{X}$. The latter equality follows from the definition of the fibration $d^{\Pf Q}_{\Pf, \dotP}\colon\Kl{\dotP} \to \Kl{\Pf}$.
  Therefore, if $f$ satisfies Cond.~\ref{item:buta_trees_2}, then we have the following.
  \begin{align*}
    q_\mathsf{F} \in
  f(x_{0})
  &\subseteq \bigl(\tau_\cap \circ \mathcal{P}(\mu\dot{F}^{\tr}) \circ L^{\mathrm{fin}}_{\mathcal{C}}\bigr)(x_{0})
                     = \bigcap_{t \in L^{\mathrm{fin}}_{\mathcal{C}}(x_{0})} \{\rho(r_t) \in Q \mid \rho \text{ is a run of $\mathcal{A}$ over $t$} \}.
  \end{align*}
  The last equality crucially relies on Prop.~\ref{lem:dpq_cost}. The above means that,
  each tree $t\in L^{\mathrm{fin}}_{\mathcal{C}}(x_{0})$ (i.e.\ one generated by $\mathcal{C}$) has a run $\rho$ that assigns $q_\mathsf{F}$ to the root $r_t$ (i.e.\ $\rho(r_t) = q_\mathsf{F}$).
  Therefore all the trees generated by $\mathcal{C}$ are accepted by $\mathcal{A}$.
  \myqed
\end{myproof}

\subsection{Additional Proof of Lem.~\ref{lem:exist}} \label{ap:exist}
\begin{enumerate}
  \item \label{item:exist1} %
  In the step case of the transfinite induction on $i$,
  we prove $\dot{\alpha}_{j, i} \circ p_j = p_i \circ k_{j, i}$ for each $j$ with $j \leq i$ by transfinite induction on $j$.
    \[
      \xymatrix@R-1em@C-1em{
        \dot{F}^j 0 \ar[r]^{\dot{\alpha}_{j, i}} &\dot{F}^i0 \\
        \alpha^*_{j, \lambda}P \ar[r]^{k_{j, i}} \ar[u]^{p_j} &\alpha^*_{i, \lambda}P \ar[u]^{p_i}
    }
    \]

  \begin{itemize}
    \item (Base case) $\dot{\alpha}_{0, i} \circ p_0 = p_i \circ k_{0, i}$ by initiality of $0 \in \mathbb{E}$.
  \item (Step case)
    In this case, $p_i$ and $p_j$ are defined to be
    \begin{align*}
      p_i &= (\alpha^*_{i, \lambda}P \xrightarrow{(\star)}
      (F\alpha_{i-1, \lambda})^*\dot{F}P
          \xrightarrow[\cong]{\xi_0} \dot{F}\alpha^*_{i-1, \lambda}P \xrightarrow{\dot{F}p_{i-1}} \dot{F}^i 0) \\
      p_j &= (\alpha^*_{j, \lambda}P \xrightarrow{(\dagger)}
      (F\alpha_{j-1, \lambda})^*\dot{F}P
          \xrightarrow[\cong]{\xi_1} \dot{F}\alpha^*_{j-1, \lambda}P \xrightarrow{\dot{F}p_{j-1}} \dot{F}^j 0)
    \end{align*}
    where both $\xi_0$ and $\xi_1$ are from fibredness of $\dot{F}$;
    $(\star)$ and $(\dagger)$ are induced as follows by universality of cartesian liftings.

\[
\scalebox{.8}{     \xymatrix@C-.5em{
        \alpha_{j, \lambda}^*P \ar@{.>}[d]^{(\dagger)} \ar[r]^{k_{j, i} = \overline{\alpha_{j, i}}} &\alpha_{i, \lambda}^*P \ar@{.>}[d]^{(\star)} \ar[r]^-{\overline{\alpha_{i, \lambda}}} &P \ar@/^1pc/[dr]^\gamma \\
        (F\alpha_{j-1, \lambda})^*\dot{F}P \ar[r]^{\overline{\alpha_{j, i}}} &(F\alpha_{i-1, \lambda})^*\dot{F}P \ar[rr]^-{\overline{F\alpha_{i-1, \lambda}}} & &\dot{F}P \\
        F^j 0 \ar[r]^{\alpha_{j, i}} &F^i0 \ar[r]^{\alpha_{i, \lambda}} &F^\lambda 0 \ar[r]^{\alpha_{\lambda, \lambda+1}} &F^{\lambda+1}0
    }
}\]
This implies in particular that the square on the left commutes.

    Fibredness of $\dot{F}$ makes the following diagram commute.

    \[
    \xymatrix{
        (F\alpha_{j-1, \lambda})^*\dot{F}P \ar[r]^{\overline{\alpha_{j, i}}} \ar@{.>}[d]^{\xi_1} &(F\alpha_{i-1, \lambda})^*\dot{F}P \ar@{.>}[d]^{\xi_0} \ar@/^1pc/[dr] \\
        \dot{F}\alpha_{j-1, \lambda}^*P \ar[r]^{\dot{F}k_{j-1, i-1}} &\dot{F}\alpha_{i-1, \lambda}^*P \ar[r] &\dot{F}P \\
        F^j 0 \ar[r]^{\alpha_{j, i}} &F^i 0 \ar[r]^{\alpha_{i, \lambda+1}} &F^{\lambda+1}0
    }
    \]

    Therefore, the following diagram commutes, where the last two diagrams establish the commutativity of the top two squares.
    The bottom square commutes because of the induction hypothesis ($\dot{\alpha}_{j-1, i-1} \circ p_{j-1} = p_{i-1} \circ k_{j-1, i-1}$) and that $\dot{\alpha}_{j,i}=\dot{F}\dot{\alpha}_{j-1, i-1}$.

    \[
    \xymatrix{
        \alpha^*_{j, \lambda}P \ar[r]^{k_{j, i}} \ar[d]^{(\dagger)} &\alpha^*_{i, \lambda}P \ar[d]^{(\star)} \\
        (F\alpha_{j-1, \lambda})^*\dot{F}P \ar[r]^{\overline{\alpha_{j, i}}} \ar@{.>}[d]^{\xi_1} &(F\alpha_{i-1, \lambda})^*\dot{F}P \ar@{.>}[d]^{\xi_0} \\
        \dot{F}\alpha_{j-1, \lambda}^*P \ar[r]^{\dot{F}k_{j-1, i-1}} \ar[d]^{\dot{F}p_{j-1}} &\dot{F}\alpha_{i-1, \lambda}^*P \ar[d]^{\dot{F}p_{i-1}} \\
        \dot{F}^j0 \ar[r]^{\dot{\alpha}_{j, i}} &\dot{F}^i0
    }
    \]

    By the definition of $p_i$ and $p_j$,
    the vertical composite on the left is $p_j$ and that on the right is $p_i$.

    Hence we conclude that $\dot{\alpha}_{j, i} \circ p_j = p_i \circ k_{j, i}$.

  \item (Limit case)
    We use universality of $\alpha_{j, \lambda}^*P \cong \colim_{l < j}\alpha_{l, \lambda}^*P$.
    For all $l, m$ with $l \leq m < j$,
      \begin{align*}
        \dot{\alpha}_{l, i} \circ p_l &= \dot{\alpha}_{m, i} \circ \dot{\alpha}_{l, m} \circ p_l \\
        &= \dot{\alpha}_{m, i} \circ p_m \circ k_{l, m}
      \end{align*}
      by induction hypothesis.
      Hence $(\dot{F}^i0, (\alpha_{l, i} \circ p_l)_{l < j})$ is a cocone.

      By the definition of $p_j$, $p_j$ is the mediating morphism by universality of the colimit $\alpha^*_{j, \lambda}P \cong \colim_{l < j} \alpha_{l, \lambda}^*P$.

      \begin{equation} \label{eq:term}
        \xymatrix@R=.8em{
            &&&\dot{F}^j 0 \mathrlap{\text{ (colim.)}}\\
            \cdots \ar[r] &\dot{F}^l0 \ar@/^/[urr]^{\dot{\alpha}_{l, j}} \ar[r] &\dot{F}^m0 \ar@/^/[ur]_{\dot{\alpha}_{m, j}} \ar[r] &\cdots
       \\
            \cdots \ar[r] &\alpha_{l, \lambda}^*P \ar[u]^{p_l} \ar@/_/[drr]_{k_{l, j}} \ar[r] &\alpha_{m, \lambda}^*P \ar[u]^{p_m} \ar@/_/[dr]^{k_{m, j}} \ar[r] &\cdots
       \\
            &&&\alpha_{j, \lambda}^*P \mathrlap{\text{ (colim.)}} \ar@/_2em/@{.>}[uuu]_{p_j}
           \\ \cdots \ar[r]
        & F^{l}0 \ar[r]^-{\alpha_{l,m}}
        & F^{m}0 \ar[r]^-{\alpha_{m,j}}
        & F^{j}0 \ar[r]
        &\cdots
      }
    \end{equation}
    By the diagram above, $\dot{\alpha}_{j, i} \circ p_j$ is also a mediating morphism in:

    \[
      \xymatrix@R=.8em@C-.8em{
          &&&\dot{F}^j 0 \ar[r]^{\dot{\alpha}_{j, i}} &\dot{F}^i 0 \mathrlap{\text{ (colim.)}}\\
          \cdots \ar[r] &\dot{F}^l0 \ar@/^/[urr]^{\dot{\alpha}_{l, j}} \ar[r] &\dot{F}^m0 \ar@/^/[ur]_{\dot{\alpha}_{m, j}} \ar[r] &\cdots
     \\
          \cdots \ar[r] &\alpha_{l, \lambda}^*P \ar[u]^{p_l} \ar@/_/[drr]_{k_{l, j}} \ar[r] &\alpha_{m, \lambda}^*P \ar[u]^{p_m} \ar@/_/[dr]^{k_{m, j}} \ar[r] &\cdots
     \\
                        &&&\alpha_{j, \lambda}^*P \mathrlap{\text{ (colim.)}} \ar@/_2em/@{.>}[uuur]_{\dot{\alpha}_{j, i} \circ p_j}
         \\ \cdots \ar[r]
      & F^{l}0 \ar[r]^-{\alpha_{l,m}}
      & F^{m}0 \ar[r]^-{\alpha_{m,j}}
      & F^{j}0 \ar[r]^-{\alpha_{j,i}}
      & F^{i}0 \ar[r]
      &\cdots
    }
  \]
  Therefore,
  if we show
  $\dot{\alpha}_{l, i} \circ p_l = (p_i \circ k_{j, i}) \circ k_{l, j}$
  for each $l$ with $l < j$ then
  $\dot{\alpha}_{j, i} \circ p_j = p_i \circ k_{j, i}$
  by universality of the colimit
  $\alpha^*_{j, \lambda}P \cong \colim_{l < j} \alpha_{l, \lambda}^*P$.

  Let $l$ be an arbitrary ordinal satisfying $l < j$.
  By the induction hypothesis ($\dot{\alpha}_{l, i} \circ p_l = p_i \circ k_{l, i}$),
  $\dot{\alpha}_{l, i} \circ p_l = p_i \circ k_{l, i} = p_i \circ k_{j, i} \circ k_{l, j}$.
  This concludes the proof.
      \myqed
\end{itemize}
\item \label{item:exist2}
  After defining $(p_i)_{i \leq \lambda}$,
  we show $\dot{\alpha}_{\lambda, \lambda+1} \circ p_\lambda = \dot{F}p_\lambda \circ \gamma$
by transfinite induction on $\lambda$.
    \[
      \xymatrix@R-1em@C-.5em{
        \dot{F}^\lambda 0 \ar[r]^{\dot{\alpha}_{\lambda, \lambda+1}} &\dot{F}^{\lambda+1}0 \\
        P \ar[r]^{\gamma} \ar[u]^{p_\lambda} &\dot{F}P \ar[u]_{\dot{F}p_\lambda}
    }
    \]

  \begin{itemize}
    \item (Base case) If $\lambda$=0 then it is true by initiality.
    \item (Step case) Assume $\lambda$ is a successor ordinal.

      By definition, $p_\lambda$ is equal to
      \begin{equation} \label{eq:p lambda}
         P \xrightarrow{(\star)} (F\alpha_{{\lambda-1}, \lambda})^*\dot{F}P
          \xrightarrow[\cong]{\xi_2} \dot{F}\alpha_{{\lambda-1}, \lambda}^*P
        \xrightarrow{\dot{F}p_{\lambda-1}} \dot{F}^\lambda0
      \end{equation}
       where $\xi_2$ is from fibredness of $\dot{F}$ and $(\star)$ is induced as follows by universality of a cartesian lifting.
       \begin{equation} \label{di:ast}
         \begin{gathered}
           \xymatrix@R-.5pc{
               P \ar@{.>}[d]^{(\star)} \ar@/^1pc/[dr]^{\gamma} \\
               (F\alpha_{{\lambda-1}, \lambda})^*\dot{F}P \ar[r]_(.65){\overline{F\alpha_{\lambda-1, \lambda}}} &\dot{F}P \\
               F^\lambda 0 \ar[r]^{\alpha_{\lambda, \lambda+1}} &F^{\lambda+1}0.
          }
         \end{gathered}
       \end{equation}
       Then by definition, $\xi_2$ makes the diagram below commute:
       \begin{equation} \label{di:xi}
         \begin{gathered}
           \xymatrix@R-.5pc{
             \dot{F}\alpha_{\lambda-1, \lambda}^*P \ar@{.>}[d]^{\cong}_{\xi_2^{-1}} \ar@/^1pc/[dr]^{\dot{F}k_{\lambda-1, \lambda}} \\
             (F\alpha_{\lambda-1, \lambda})^*\dot{F}P \ar[r]_(.65){\overline{F\alpha_{\lambda-1, \lambda}}} &\dot{F}P \\
             F^\lambda 0 \ar[r]^{\alpha_{\lambda, \lambda+1}} &F^{\lambda+1}0
           }
         \end{gathered}
       \end{equation}
       Therefore, $\dot{\alpha_{\lambda, \lambda+1}} \circ p_\lambda = \dot{F}p_\lambda \circ \gamma$ by the following equations.
      \begin{align*}
        \dot{\alpha}_{\lambda, \lambda+1} \circ p_\lambda
          &= \dot{F}\alpha_{\lambda-1, \lambda} \circ \dot{F}p_{\lambda-1} \circ \xi_2 \circ (\star) &\text{by (\ref{eq:p lambda}) and $\alpha_{\lambda, \lambda+1} = \dot{F}\alpha_{\lambda-1, \lambda}$} \\
          &= \dot{F}p_\lambda \circ \dot{F}k_{\lambda-1, \lambda} \circ \xi_2 \circ (\star) &\text{since $\dot{\alpha}_{\lambda-1, \lambda} \circ p_{\lambda-1} = p_\lambda \circ k_{\lambda-1, \lambda}$}\\
          &= \dot{F}p_\lambda \circ \overline{F\alpha_{\lambda-1, \lambda}} \circ (\star) &\text{by (\ref{di:xi})}\\
          &= \dot{F}p_\lambda \circ \gamma &\text{by (\ref{di:ast})}
      \end{align*}

    \item (Limit case) Assume $\lambda$ is a limit ordinal.
      In this case, $p_\lambda$ is defined as the mediating morphism below by universality of $P = \colim_{i < \lambda} \alpha_{i, \lambda}^*P$:

      \[
        \xymatrix@R=.8em@C-.8em{
            &&&\dot{F}^\lambda 0 \mathrlap{\text{ (colim.)}}\\
            \cdots \ar[r] &\dot{F}^i0 \ar@/^/[urr]^{\dot{\alpha}_{i, \lambda}} \ar[r] &\dot{F}^j0 \ar@/^/[ur]_{\dot{\alpha}_{j, \lambda}} \ar[r] &\cdots
       \\
            \cdots \ar[r] &\alpha_{i, \lambda}^*P \ar[u]^{p_i} \ar@/_/[drr]_{k_{i, \lambda}} \ar[r] &\alpha_{j, \lambda}^*P \ar[u]^{p_j} \ar@/_/[dr]^{k_{j, \lambda}} \ar[r] &\cdots
       \\
            &&&P \mathrlap{\text{ (colim.)}} \ar@/_2em/@{.>}[uuu]_{p_\lambda}
           \\ \cdots \ar[r]
        & F^{i}0 \ar[r]^-{\alpha_{i,j}}
        & F^{j}0 \ar[r]^-{\alpha_{j,\lambda}}
        & F^{\lambda}0 \ar[r]
        &\cdots
      }
    \]

      By the diagram above, $\dot{\alpha}_{\lambda, \lambda+1} \circ p_\lambda$ is also a mediating morphism in:

      \[
        \xymatrix@R=.8em@C-.8em{
            &&&\dot{F}^\lambda 0 \ar[r]^{\dot{\alpha}_{\lambda, \lambda+1}} &\dot{F}^{\lambda+1} 0 \mathrlap{\text{ (colim.)}}\\
            \cdots \ar[r] &\dot{F}^i0 \ar@/^/[urr]^{\dot{\alpha}_{i, \lambda}} \ar[r] &\dot{F}^j0 \ar@/^/[ur]_{\dot{\alpha}_{j, \lambda}} \ar[r] &\cdots
       \\
            \cdots \ar[r] &\alpha_{i, \lambda}^*P \ar[u]^{p_i} \ar@/_/[drr]_{k_{i, \lambda}} \ar[r] &\alpha_{j, \lambda}^*P \ar[u]^{p_j} \ar@/_/[dr]^{k_{j, \lambda}} \ar[r] &\cdots
       \\
                          &&&P \mathrlap{\text{ (colim.)}} \ar@/_2em/@{.>}[uuur]_{\dot{\alpha}_{\lambda, \lambda+1}\circ p_\lambda}
           \\ \cdots \ar[r]
        & F^{i}0 \ar[r]^-{\alpha_{i,j}}
        & F^{j}0 \ar[r]^-{\alpha_{j,\lambda}}
        & F^{\lambda}0 \ar[r]^-{\alpha_{\lambda, \lambda+1}}
        & F^{\lambda+1}0 \ar[r]
        &\cdots
      }
    \]
      Therefore,
      if we show $\dot{\alpha}_{i, \lambda+1} \circ p_i = (\dot{F}p_\lambda \circ \gamma) \circ k_{i, \lambda}$ for each $i$ with $i < \lambda$
      then $\dot{\alpha}_{\lambda, \lambda+1} \circ p_\lambda = \dot{F}p_\lambda \circ \gamma$
      by universality of the colimit $P = \colim_{i < \lambda} \alpha_{i, \lambda}^*P$.

      Let $i$ be an arbitrary ordinal satisfying $i < \lambda$.
      By definition, $p_{i+1}$ is equal to
      \begin{equation} \label{eq:p i1}
        \alpha^*_{i+1, \lambda}P \xrightarrow{(\dagger)} (F\alpha_{i, \lambda})^*\dot{F}P \xrightarrow[\cong]{\xi_3} \dot{F}\alpha^*_{i, \lambda}P \xrightarrow{\dot{F}p_i} \dot{F}^{i+1}0
      \end{equation}
      where $\xi_3$ is from fibredness of $\dot{F}$ and $(\dagger)$ is induced as follows by universality of a cartesian lifting.
      \begin{equation} \label{di:star}
        \begin{gathered}
          \xymatrix@R-.5pc{
            \alpha^*_{i+1, \lambda}P \ar@{.>}[d]^{(\dagger)} \ar[r]^{\overline{\alpha_{i+1, \lambda}}} &P \ar@/^1pc/[dr]^{\gamma} \\
            (F\alpha_{i, \lambda})^*\dot{F}P \ar[rr]^{\overline{F\alpha_{i, \lambda}}} & &\dot{F}P \\
            F^{i+1}0 \ar[r]^{\alpha_{i+1, \lambda}} &F^\lambda 0 \ar[r]^{\alpha_{\lambda, \lambda+1}} &F^{\lambda+1}0.
         }
        \end{gathered}
      \end{equation}
       Then by definition, $\xi_3$ makes the diagram below commute:
       \begin{equation} \label{di:xi3}
         \begin{gathered}
           \xymatrix@R-.5pc{
             \dot{F}\alpha_{i, \lambda}^*P \ar@{.>}[d]^{\cong}_{\xi_3^{-1}} \ar@/^1pc/[dr]^{\dot{F}k_{i, \lambda}} \\
             (F\alpha_{i, \lambda})^*\dot{F}P \ar[r]_(.65){\overline{F\alpha_{i, \lambda}}} &\dot{F}P \\
             F^{i+1} 0 \ar[r]^{\alpha_{i+1, \lambda+1}} &F^{\lambda+1}0
           }
         \end{gathered}
       \end{equation}
       Therefore, $\dot{\alpha}_{i, \lambda+1} \circ p_i = \dot{F}p_\lambda \circ \gamma \circ k_{i, \lambda}$ by the following equations.
       \begin{align*}
         \dot{\alpha}_{i, \lambda+1} \circ p_i
         &= \dot{F}\dot{\alpha}_{i, \lambda} \circ \dot{\alpha}_{i, i+1} \circ p_i &\text{since $\dot{\alpha}_{i, \lambda+1} = \dot{F}\dot{\alpha}_{i, \lambda} \circ \dot{\alpha}_{i, i+1}$}\\
         &= \dot{F}\dot{\alpha}_{i, \lambda} \circ p_{i+1} \circ k_{i, i+1} &\text{since $\dot{\alpha}_{i, i+1} \circ p_i = p_{i+1} \circ k_{i, i+1}$}\\
         &= \dot{F}\dot{\alpha}_{i, \lambda} \circ \dot{F}p_i \circ \xi_3 \circ (\dagger) \circ k_{i, i+1} &\text{by definition of $p_i$ $(\ref{eq:p i1})$}\\
         &= \dot{F}p_\lambda \circ \dot{F}k_{i, \lambda} \circ \xi_3 \circ (\dagger) \circ k_{i, i+1} &\text{since $\dot{\alpha}_{i, \lambda} \circ p_i = p_\lambda \circ k_{i, \lambda}$} \\
         &= \dot{F}p_\lambda \circ \overline{F\alpha_{i, \lambda}} \circ (\dagger) \circ k_{i, i+1} &\text{by (\ref{di:xi3})} \\
         &= \dot{F}p_\lambda \circ \gamma \circ \overline{\alpha_{i+1, \lambda}} \circ k_{i, i+1} &\text{by (\ref{di:star})} \\
         &= \dot{F}p_\lambda \circ \gamma \circ k_{i, \lambda} &\text{by property of cartesian liftings}
       \end{align*}
  \end{itemize}
\end{enumerate}

\subsection{Proof of Lem.~\ref{lem:pii_iso}} \label{ap:pii_iso}
We use the following lemma. Its proof is easy.
\begin{mylemma} \label{lem:cheese}
  Let $p: \mathbb{E} \to \mathbb{B}$ be a fibration;
  $\dot{F}$ be a fibred lifting of $F$ along $p$.
  For all morphisms $f: X \to Y$ in $\mathbb{B}$ and morphisms $x: P \to Q$ in $\mathbb{E}_Y$,
  $((Ff)^*\dot{F}P \iso \dot{F}f^*P \xrightarrow{\dot{F}f^*x} \dot{F}f^*Q)
  = ((Ff)^*\dot{F}P \xrightarrow{(Ff)^*\dot{F}x} (Ff)^*\dot{F}Q \iso \dot{F}f^*Q)$
  where each isomorphism is derived from fibredness.
  \myqed
\end{mylemma}

\begin{myproof}[Lem.~\ref{lem:pii_iso}]
  We show it by transfinite induction on $i$.

  \begin{itemize}
    \item (Base case)
  If $i=0$ then (i) and (ii) follow from Lem.~\ref{lem:init_obj}.
    \item (Step case)
Let us first prove
      (i). By the proof of Lem.~\ref{lem:exist},
      the morphism $p_i^i: \alpha_{i, \lambda}^*(\asdf)^iP \to \dot{F}^i0$ is defined to be the following composite:
      \begin{equation}\label{eq:piiStep}
\begin{aligned}
         \alpha^*_{i, \lambda}(\asdf)^iP &\xrightarrow{(\star)}
        (F\alpha_{i-1, \lambda})^*\dot{F}(\asdf)^iP \\
                                                  &\xrightarrow[\cong]{\xi_0} \dot{F}\alpha^*_{i-1, \lambda}(\asdf)^iP \xrightarrow{\dot{F}p_{i-1}^i} \dot{F}^i 0
\end{aligned}
      \end{equation}
      where $\xi_0$ is from fibredness of $\dot{F}$ and $(\star)$ is induced as follows by universality of the cartesian lifting $\overline{F\alpha_{i-1, \lambda}}$.

      \begin{equation}\label{eq:pentagonTriangle}
\vcenter{      \xymatrix@C=1em{
          \alpha^*_{i, \lambda}(\asdf)^i P \ar@{.>}[dd]^{(\star)} \ar[r]^{\overline{\alpha_{i, \lambda}}} &(\asdf)^iP \ar[d]^{\beta_{i, i+1}} \\
                                                                               &(\asdf)^{i+1}P \ar@/^1pc/[dr]^{\overline{\alpha}} \\
          (F\alpha_{i-1, \lambda})^*\dot{F}(\asdf)^iP \ar[rr]^{\overline{F\alpha_{i-1, \lambda}}} & &\dot{F}(\asdf)^iP \\
                                                                        &F^\lambda 0 \ar@/^1pc/[rd]^{\alpha = \alpha_{\lambda, \lambda+1}} \\
          F^i0 \ar@/^1pc/[ru]^{\alpha_{i, \lambda}} \ar[rr]^{F\alpha_{i-1, \lambda} = \alpha_{i, \lambda+1}} & &F^{\lambda+1}0
      }
}    \end{equation}
In the top pentagon, three morphisms (the overlined ones) are cartesian liftings; the vertical morphism $\beta_{i,i+1}$ is from Def.~\ref{def:chainV}.
   For the bottom triangle, we note that $\alpha_{i,\lambda +1}=F\alpha_{i-1,\lambda}$ by the definition of $\alpha_{i,j}$, which shows the commutativity of the triangle.

    We first establish some properties of the morphism $(\star)$.
     Note that we have
\begin{align*}
& \Big( (\asdf)^iP \xrightarrow{\beta_{i, i+1}} (\asdf)^{i+1}P \xrightarrow{\overline{\alpha}} \dot{F}(\asdf)^i P \Big) \\&= \Big((\asdf)^iP \xrightarrow{\overline{\alpha}} \dot{F}(\asdf)^{i-1}P \xrightarrow{\dot{F}\beta_{i-1, i}} \dot{F}(\asdf)^iP\Big);
\end{align*}
  this follows from the definition of $\beta_{i, i+1}$ by the action of $\alpha^{*}$ (Def.~\ref{def:chainV}). Therefore
$(\star)$ is equal to the vertical morphism on the left in:
      \[
      \xymatrix@R=1em@C=1em{
          \alpha^*_{i, \lambda}(\asdf)^i P \ar@{.>}[dd]^{(\star)} \ar[r]^{\overline{\alpha_{i, \lambda}}} &(\asdf)^iP \ar@/^1pc/[dr]^-{\overline{\alpha}} \\
                                                                                                          & &\dot{F}(\asdf)^{i-1}P \ar[d]^{\dot{F}\beta_{i-1, i}} \\
          (F\alpha_{i-1, \lambda})^*\dot{F}(\asdf)^iP \ar[rr]^-{\overline{F\alpha_{i-1, \lambda}}} & &\dot{F}(\asdf)^iP\mathrlap{.}
      }
      \]
  Moreover, the morphism $(\star)$
       can be factored as follows.
      \begin{equation}\label{eq:starStepConcl}
      \xymatrix@C=1em{
          \alpha^*_{i, \lambda}(\asdf)^i P \ar@/_2pc/@{.>}[dd]_(.3){(\star)} \ar@{.>}[d]^\cong_{(\ddagger)} \ar[r]^-{\overline{\alpha_{i, \lambda}}} &(\asdf)^iP \ar@/^1pc/[dr]^-{\overline{\alpha}} \\
          (F\alpha_{i-1, \lambda})^*\dot{F}(\asdf)^{i-1}P \ar[rr]^-{\overline{F\alpha_{i-1, \lambda}}} \ar@{.>}[d]^{(F\alpha_{i-1, \lambda})^*\dot{F}\beta_{i-1, i}} & &\dot{F}(\asdf)^{i-1}P \ar[d]^{\dot{F}\beta_{i-1, i}} \\
          (F\alpha_{i-1, \lambda})^*\dot{F}(\asdf)^iP \ar[rr]^-{\overline{F\alpha_{i-1, \lambda}}} & &\dot{F}(\asdf)^iP
       \\
       F^{\lambda}0 \ar[rr]^-{F\alpha_{i-1, \lambda}}
       &&
       F^{\lambda +1}0
      }
      \end{equation}
  Indeed, the bottom component $(F\alpha_{i-1, \lambda})^*\dot{F}\beta_{i-1, i}$ arises from universality of the cartesian lifting $\overline{F\alpha_{i-1, \lambda}}$; the top component (an isomorphism $(\ddagger)$) is obtained from $\alpha\circ \alpha_{i,\lambda}=F\alpha_{i-1, \lambda}$, the bottom triangle in~(\ref{eq:pentagonTriangle}).

Let us get back to the morphism $p_i^i: \alpha_{i, \lambda}^*(\asdf)^iP \to \dot{F}^i0$ in question, defined by the composite in~(\ref{eq:piiStep}).
\begin{align*}
& p_i^i
\\
&=
\left(
         \alpha^*_{i, \lambda}(\asdf)^iP \xrightarrow{(\star)}
        (F\alpha_{i-1, \lambda})^*\dot{F}(\asdf)^iP
                                                  \xrightarrow[\cong]{\xi_0} \dot{F}\alpha^*_{i-1, \lambda}(\asdf)^iP \xrightarrow{\dot{F}p_{i-1}^i} \dot{F}^i 0
      \right)
  \quad\text{by~(\ref{eq:piiStep})}
\\
&=
\left(\begin{array}{r}
         \alpha^*_{i, \lambda}(\asdf)^iP \xrightarrow[\cong]{(\ddagger)}
	 \bullet \xrightarrow{(F\alpha_{i-1, \lambda})^*\dot{F}\beta_{i-1, i}}
        (F\alpha_{i-1, \lambda})^*\dot{F}(\asdf)^iP \\
                                                  \xrightarrow[\cong]{\xi_0} \dot{F}\alpha^*_{i-1, \lambda}(\asdf)^iP
       \xrightarrow{\dot{F}p_{i-1}^i} \dot{F}^i 0
\end{array}      \right) \quad\text{by~(\ref{eq:starStepConcl})}
\\
&=
\left(\begin{array}{r}
         \alpha^*_{i, \lambda}(\asdf)^iP \xrightarrow[\cong]{(\ddagger)}
	 \bullet
\xrightarrow[\cong]{\xi_1} \dot{F}\alpha_{i-1, \lambda}^*(\alpha^*\dot{F})^{i-1}P
                                                  \xrightarrow{\dot{F}\alpha_{i-1, \lambda}^*\beta_{i-1, i}}
\dot{F}\alpha^*_{i-1, \lambda}(\asdf)^iP
       \xrightarrow{\dot{F}p_{i-1}^i} \dot{F}^i 0
\end{array}      \right)
  \\&\qquad\qquad\qquad\text{by~(\ref{eq:stepCheese}) later}
\\&=
\left(
         \alpha^*_{i, \lambda}(\asdf)^iP \xrightarrow[\cong]{(\ddagger)}
	 \bullet
\xrightarrow[\cong]{\xi_1} \dot{F}\alpha_{i-1, \lambda}^*(\alpha^*\dot{F})^{i-1}P
                                                  \xrightarrow{\dot{F}\bigl(p_{i-1}^i\circ\alpha_{i-1, \lambda}^*\beta_{i-1, i}\bigr)}
\dot{F}^i 0
\right)
  \\&\qquad\qquad\qquad\text{by the functoriality of $\dot{F}$}
\\&=
\left(
         \alpha^*_{i, \lambda}(\asdf)^iP \xrightarrow[\cong]{(\ddagger)}
	 \bullet
\xrightarrow[\cong]{\xi_1} \dot{F}\alpha_{i-1, \lambda}^*(\alpha^*\dot{F})^{i-1}P
                                                  \xrightarrow[\cong]{\dot{F}p_{i-1}^{i-1}}
\dot{F}^i 0
\right)
  \\&\qquad\qquad\qquad\text{by the induction hypothesis (ii).}
\end{align*}
We used the following equality, where $\xi_1: (F\alpha_{i-1, \lambda})^*\dot{F}(\asdf)^iP \iso \dot{F}\alpha^*_{i-1, \lambda}(\asdf)^iP$ is an isomorphism induced by fibredness of $\dot{F}$.
The equality follows from Lem.~\ref{lem:cheese}.
\begin{equation}\label{eq:stepCheese}
\begin{aligned}
  &      \xi_0 \circ (F\alpha_{i-1, \lambda})^*\dot{F}\beta_{i-1, i}
= \dot{F}\alpha^*_{i-1, \lambda}\beta_{i-1, i} \circ \xi_1,
\end{aligned}
\end{equation}
Since $p_{i-1}^{i-1}$ is an isomorphism (the induction hypothesis (i)), the above equational reasoning concludes that $p_i^i$ is an isomorphism.

      We turn to (ii) in the step case. Let us consider all $l < m$ because the case of $l=m$ is easily proved.
      By Lem.~\ref{lem:exist}, the morphism $p_i^m \circ \alpha_{i, \lambda}^*\beta_{l, m}$ is equal to
      \begin{align*}
        \alpha_{i, \lambda}^*(\asdf)^lP &\xrightarrow{\alpha_{i, \lambda}^*\beta_{l, m}} \alpha^*_{i, \lambda}(\asdf)^mP \\
                                                  &\xrightarrow{(\dagger)} (F\alpha_{i-1, \lambda})^*\dot{F}(\asdf)^mP \\
                                                  &\xrightarrow[\cong]{\xi_2} \dot{F}\alpha_{i-1, \lambda}^*(\asdf)^mP \\
                                                  &\xrightarrow{\dot{F}p_{i-1}^m} \dot{F}^i0
      \end{align*}
      where $\xi_2$ is from fibredness of $\dot{F}$ and $(\dagger)$ is induced as follows by universality of the cartesian lifting $\overline{F\alpha_{i-1, \lambda}}$.

      \[
      \xymatrix@C=1em@R=1.5em{
          \alpha^*_{i, \lambda}(\asdf)^m P \ar@{.>}[dd]^{(\dagger)} \ar[r]^{\overline{\alpha_{i, \lambda}}} &(\asdf)^mP \ar[d]^{\beta_{m, m+1}} \\
                                                                               &(\asdf)^{m+1}P \ar@/^1pc/[dr]^{\overline{\alpha}} \\
          (F\alpha_{i-1, \lambda})^*\dot{F}(\asdf)^mP \ar[rr]^{\overline{F\alpha_{i-1, \lambda}}} & &\dot{F}(\asdf)^mP
      }
    \]

    Because $\Big((\asdf)^{l+1}P \xrightarrow{\beta_{l+1, m+1}} (\asdf)^{m+1}P \xrightarrow{\overline{\alpha}} \dot{F}(\asdf)^m P\Big) = \Big((\asdf)^{l+1}P \xrightarrow{\overline{\alpha}} \dot{F}(\asdf)^{l}P \xrightarrow{\dot{F}\beta_{l, m}} \dot{F}(\asdf)^mP\Big)$,
      \begin{align*}
        &\Big( (\asdf)^lP \xrightarrow{\beta_{l, m+1}} (\asdf)^{m+1}P \xrightarrow{\overline{\alpha}} \dot{F}(\asdf)^mP \Big) \\
        &= \Big( (\asdf)^lP \xrightarrow{\beta_{l, l+1}} (\asdf)^{l+1}P \\
        &\phantom{=} \xrightarrow{\beta_{l+1, m+1}} (\asdf)^{m+1}P \xrightarrow{\overline{\alpha}} \dot{F}(\asdf)^mP \Big) \\
        &= \Big( (\asdf)^lP \xrightarrow{\beta_{l, l+1}} (\asdf)^{l+1}P \\
        &\phantom{=}  \xrightarrow{\bar{\alpha}} \dot{F}(\asdf)^lP \xrightarrow{\dot{F}\beta_{l, m}} \dot{F}(\asdf)^mP \Big).
      \end{align*}
      Thus $(\dagger) \circ \alpha_{i, \lambda}^*\beta_{l, m}$ is equal to the vertical morphism on the left in the below.

      \[
      \xymatrix@C=1em@R=1.5em{
          \alpha^*_{i, \lambda}(\asdf)^l P \ar@{.>}[dd]^{(\diamondsuit)} \ar[r]^{\overline{\alpha_{i, \lambda}}} &(\asdf)^lP \ar[d]^{\beta_{l, l+1}} \\
                                                                               &(\asdf)^{l+1}P \ar@/^1pc/[dr]^{\overline{\alpha}} \\
          (F\alpha_{i-1, \lambda})^*\dot{F}(\asdf)^lP \ar@{.>}[d]^{(F\alpha_{i-1, \lambda})^*\dot{F}\beta_{l, m}} \ar[rr]^{\overline{F\alpha_{i-1, \lambda}}} & &\dot{F}(\asdf)^lP \ar[d]^{\dot{F}\beta_{l, m}} \\
          (F\alpha_{i-1, \lambda})^*\dot{F}(\asdf)^mP \ar[rr]^{\overline{F\alpha_{i-1, \lambda}}} & &\dot{F}(\asdf)^mP
      }
      \]

      By Lem.~\ref{lem:cheese},
      $\xi_2 \circ (F\alpha_{i-1, \lambda})^*\dot{F}\beta_{l, m} = \dot{F}\alpha^*_{i-1, \lambda}\beta_{l, m} \circ \xi_3$
      where
      $\xi_3: (F\alpha_{i-1, \lambda})^*\dot{F}(\asdf)^lP \iso \dot{F}\alpha^*_{i-1, \lambda}(\asdf)^lP)$ is a morphism made by fibredness of $\dot{F}$.
      Therefore, by Lem.~\ref{lem:cheese},
      $p_i^m \circ \alpha_{i, \lambda}^*\beta_{l, m}$ is equal to:
      \begin{align*}
        \alpha^*_{i, \lambda}(\asdf)^lP &\xrightarrow{(\diamondsuit)} (F\alpha_{i-1, \lambda})^*\dot{F}(\asdf)^lP \\
                                        &\xrightarrow[\cong]{\xi_3} \dot{F}\alpha^*_{i-1, \lambda}(\asdf)^lP \\
                                        &\xrightarrow{\dot{F}\alpha^*_{i-1, \lambda}\beta_{l, m}} \dot{F}\alpha^*_{i-1, \lambda}(\asdf)^mP \\
                                        &\xrightarrow{\dot{F}p_{i-1}^m} \dot{F}^i0.
      \end{align*}
      induction hypotheses say $\dot{F}p_{i-1}^m \circ \dot{F}\alpha^*_{i-1, \lambda}\beta_{l, m} = \dot{F}p_{i-1}^l$ so $p_i^m \circ \alpha_{i, \lambda}^*\beta_{l, m}$ is actually equal to $p_i^l$ by definition in Lem.~\ref{lem:exist}.
    \item (Limit case)
Let us first prove
      (i).

Firstly we note that the diagram
  \begin{equation}\label{eq:colimitAlphaii}
\vcenter{        \xymatrix@C-1em{
\alpha_{0, \lambda}^*(\asdf)^iP \ar[r]
  &\alpha_{1, \lambda}^*(\asdf)^iP \ar[r]
&\cdots \ar[r]
&\alpha_{i, \lambda}^*(\asdf)^iP
}}, \end{equation}
which is the $i$-th row of Fig.~\ref{fig:lem} but is truncated at the $i$-th column, is a colimit with $\alpha_{i, \lambda}^*(\asdf)^i P \cong \colim_{j < i} \alpha_{j, \lambda}^*(\asdf)^i P$. This fact follows from that $F^i0=\colim_{j<i} F^j0$ (recall that $i$ is a limit ordinal; Def.~\ref{def:init_chain} \& Lem.~\ref{lem:init}) and that $p$ has stable chain colimits. We also note that the top row of Fig.~\ref{fig:lem} truncated at the $i$-th column, namely
\[
\vcenter{        \xymatrix@C-1em{
  0 \ar[r] &\dot{F}0 \ar[r] &\dot{F}^20 \ar[r] &\cdots \ar[r] &\dot{F}^i 0
}},
\]
is a colimit with $\dot{F}^i 0 \cong \colim_{j < i} \dot{F}^j 0$. This is by the definition of the initial chain for $\dot{F}$ (Def.~\ref{def:init_chain}).

  We now have the following situation; it shows part of the diagram in Fig.~\ref{fig:lem}. The morphism $p_i^i$ is defined to be the mediating morphism from the colimit $\alpha_{i, \lambda}^*(\asdf)^i P\cong \colim_{j < i} \alpha_{j, \lambda}^*(\asdf)^i P$ to the cone designated in the diagram.
      \begin{equation}
       \label{eq:isoStepGrid}\footnotesize
       \xymatrix@R=1em@C=1.3em{
            &&&\dot{F}^i 0\\
            \cdots \ar[r] &\dot{F}^l0 \ar@/^/[urr]^{\dot{\alpha}_{l, i}} \ar[r] &\dot{F}^j0 \ar@/^/[ur]_{\dot{\alpha}_{j, i}} \ar[r] &\cdots
       \\
        \cdots \ar[r] &\alpha_{l, \lambda}^*(\asdf)^iP \ar[u]^{p_l^i} \ar@/_/[dddrr] \ar[r] &\alpha_{j, \lambda}^*(\asdf)^iP \ar[u]^{p_j^i} \ar@/_/[dddr] \ar[r] &\cdots
       \\
        \cdots \ar[r] &\alpha_{l, \lambda}^*(\asdf)^jP \ar[u]^{p_l^j}  \ar[r] &\alpha_{j, \lambda}^*(\asdf)^jP \ar[u]^{p_j^j}  \ar[r] &\cdots
       \\
        \cdots \ar[r] &\alpha_{l, \lambda}^*(\asdf)^lP \ar[u]^{p_l^l}  \ar[r] &\alpha_{j, \lambda}^*(\asdf)^lP \ar[u]^{p_j^l}  \ar[r] &\cdots
       \\
            &&&\alpha_{i, \lambda}^*(\asdf)^iP \ar@/_2em/@{.>}[uuuuu]_{p_i^i}
           \\ \cdots \ar[r]
        & F^{l}0 \ar[r]^-{\alpha_{l,j}}
        & F^{j}0 \ar[r]^-{\alpha_{j,i}}
        & F^{i}0 \ar[r]
        &\cdots
      }
      \end{equation}
      Our proof strategy is by showing that the colimit $\alpha_{i, \lambda}^*(\asdf)^i P$ is the colimit of the following chain~(\ref{eq:diagonalChain}) at the same time. The chain occurs as a diagonal in~(\ref{eq:isoStepGrid}).
    \begin{equation}\label{eq:diagonalChain}
\vcenter{        \xymatrix@C-1em{
\alpha_{0, \lambda}^*(\asdf)^0P \ar[r]
  &\alpha_{1, \lambda}^*(\asdf)^1P \ar[r]
&\cdots \ar[r]
&\alpha_{i, \lambda}^*(\asdf)^iP
\ar[r]
&\cdots
}}
    \end{equation}
    That $\alpha_{i, \lambda}^*(\asdf)^i P$ is the colimit of the ``diagonal'' chain~(\ref{eq:diagonalChain}) follows from these facts.
  \begin{itemize}
  \item For each $j$ such that $j<i$, the ``vertical'' chain
    \[
\vcenter{        \xymatrix@C-1em{
\alpha_{j, \lambda}^*(\asdf)^0P \ar[r]
  &\alpha_{j, \lambda}^*(\asdf)^1P \ar[r]
&\cdots \ar[r]
&\alpha_{j, \lambda}^*(\asdf)^iP
}},
    \]
	that occurs in the $j$-th column but is truncated at height $i$, is a colimit. This is because
    \begin{math}
(\asdf)^0P\to (\asdf)^1P\to\cdots\to (\asdf)^iP
    \end{math} is a colimit above $F^{j}0$ (Def.~\ref{def:chainV}), and that substitution preserves colimits (as assumed, see Prop.~\ref{prop:main}).
  \item The diagram~(\ref{eq:colimitAlphaii}) is a colimit, as we showed in the above.
  \end{itemize}
Therefore the current situation~(\ref{eq:isoStepGrid}) is reorganized as follows.
\[\footnotesize
       \xymatrix@R=1em@C=1.3em{
            &&&\dot{F}^i 0 \mathrlap{\text{ (colim.)}}\\
            \cdots \ar[r] &\dot{F}^l0 \ar@/^/[urr]^{\dot{\alpha}_{l, i}} \ar[r] &\dot{F}^j0 \ar@/^/[ur]_{\dot{\alpha}_{j, i}} \ar[r] &\cdots
       \\
        \cdots \ar[r] &\alpha_{l, \lambda}^*(\asdf)^lP \ar[u]^{p_l^l}_{\cong} \ar@/_/[drr] \ar[r] &\alpha_{j, \lambda}^*(\asdf)^jP \ar[u]^{p_j^j}_{\cong} \ar@/_/[dr] \ar[r] &\cdots
       \\
            &&&\alpha_{i, \lambda}^*(\asdf)^iP \mathrlap{\text{ (colim.)}} \ar@/_2em/@{.>}[uuu]_{p_i^i}
           \\ \cdots \ar[r]
        & F^{l}0 \ar[r]^-{\alpha_{l,j}}
        & F^{j}0 \ar[r]^-{\alpha_{j,i}}
        & F^{i}0 \ar[r]
        &\cdots
      }
\]
The commutativity of the squares between the two chains follows easily from the induction hypothesis (ii); that the vertical morphisms are isomorphisms are the induction hypothesis (i); and the upper cone is colimiting, as we observed before. Therefore the mediating morphism $p_i^i$, mediating the colimits of two isomorphic chains, is an isomorphism. This concludes (ii) for the limit case.

      We turn to (ii) in the limit case. For all ordinals $l, m$ with $l \leq m$ and $j$ with $j < i$,
      induction hypotheses make the diagram below commute:

      \[
      \xymatrix{
          \cdots \ar[r] &\dot{F}^j0 \ar[r]^{\dot{\alpha}_{j, i}} &\dot{F}^i0 \ar[r] &\cdots \\
          \cdots \ar[r] &\alpha_{j, \lambda}^*(\asdf)^mP \ar[u]_{p_j^m} \ar[r] &\alpha_{i, \lambda}^*(\asdf)^mP \ar[r] \ar[u]_{p_i^m} &\cdots \\
          \cdots \ar[r] &\alpha_{j, \lambda}^*(\asdf)^lP \ar[r] \ar[u]_{\alpha^*_{j, \lambda}\beta_{l, m}} \ar@/^1pc/[uu]^(.3){p_j^l} &\alpha_{i, \lambda}^*(\asdf)^lP \ar[r] \ar[u]_{\alpha^*_{i, \lambda}\beta_{l, m}} &\cdots \\
      }
      \]

      The morphism $p_i^l$ is defined as the mediating morphism from $(\alpha_{j, \lambda}^*(\asdf)^lP \xrightarrow{p_j^l} \dot{F}^j0 \xrightarrow{\dot{\alpha}_{j, i}} \dot{F}^i0)_{j < i}$
      by universality of the colimit $\alpha^*_{i, \lambda}(\asdf)^lP \cong \colim_{j<i}\alpha^*_{j, \lambda}(\asdf)^lP$.
      Because $p_i^m \circ \alpha^*_{i, \lambda} \beta_{l, m}$ commutes the above diagram,
      the universality of the colimit $\alpha^*_{i, \lambda}(\asdf)^lP$ concludes $p_i^l = p_i^m \circ \alpha_{i, \lambda}^*\beta_{l, m}$.
  \end{itemize}
      \myqed
\end{myproof}

\end{document}